\documentclass[10pt, a4paper, twoside, titlepages]{article}
\usepackage[utf8]{inputenc}
\usepackage{amsmath,color,oldgerm,mathrsfs}
\usepackage{amssymb}

\usepackage[utf8]{inputenc}
\usepackage[english]{babel}
 
\usepackage{amsthm}

\usepackage{systeme}

\usepackage{parskip}

\usepackage{geometry}
\usepackage{multirow}
\usepackage{array}
\usepackage{dsfont}

\usepackage{pgffor}
\usepackage{graphicx}
\usepackage{subcaption}

\usepackage[section]{placeins}
\usepackage{float}
\usepackage{python}

\usepackage{verbatim}
\usepackage[draft]{minted}
\usepackage{tikz}
\usepackage{url}

\newcolumntype{M}{>{\centering\arraybackslash}p{1.3cm}}

\makeatletter
\def\PYG@reset{\let\PYG@it=\relax \let\PYG@bf=\relax%
    \let\PYG@ul=\relax \let\PYG@tc=\relax%
    \let\PYG@bc=\relax \let\PYG@ff=\relax}
\def\PYG@tok#1{\csname PYG@tok@#1\endcsname}
\def\PYG@toks#1+{\ifx\relax#1\empty\else%
    \PYG@tok{#1}\expandafter\PYG@toks\fi}
\def\PYG@do#1{\PYG@bc{\PYG@tc{\PYG@ul{%
    \PYG@it{\PYG@bf{\PYG@ff{#1}}}}}}}
\def\PYG#1#2{\PYG@reset\PYG@toks#1+\relax+\PYG@do{#2}}

\expandafter\def\csname PYG@tok@w\endcsname{\def\PYG@tc##1{\textcolor[rgb]{0.73,0.73,0.73}{##1}}}
\expandafter\def\csname PYG@tok@c\endcsname{\let\PYG@it=\textit\def\PYG@tc##1{\textcolor[rgb]{0.25,0.50,0.50}{##1}}}
\expandafter\def\csname PYG@tok@cp\endcsname{\def\PYG@tc##1{\textcolor[rgb]{0.74,0.48,0.00}{##1}}}
\expandafter\def\csname PYG@tok@k\endcsname{\let\PYG@bf=\textbf\def\PYG@tc##1{\textcolor[rgb]{0.00,0.50,0.00}{##1}}}
\expandafter\def\csname PYG@tok@kp\endcsname{\def\PYG@tc##1{\textcolor[rgb]{0.00,0.50,0.00}{##1}}}
\expandafter\def\csname PYG@tok@kt\endcsname{\def\PYG@tc##1{\textcolor[rgb]{0.69,0.00,0.25}{##1}}}
\expandafter\def\csname PYG@tok@o\endcsname{\def\PYG@tc##1{\textcolor[rgb]{0.40,0.40,0.40}{##1}}}
\expandafter\def\csname PYG@tok@ow\endcsname{\let\PYG@bf=\textbf\def\PYG@tc##1{\textcolor[rgb]{0.67,0.13,1.00}{##1}}}
\expandafter\def\csname PYG@tok@nb\endcsname{\def\PYG@tc##1{\textcolor[rgb]{0.00,0.50,0.00}{##1}}}
\expandafter\def\csname PYG@tok@nf\endcsname{\def\PYG@tc##1{\textcolor[rgb]{0.00,0.00,1.00}{##1}}}
\expandafter\def\csname PYG@tok@nc\endcsname{\let\PYG@bf=\textbf\def\PYG@tc##1{\textcolor[rgb]{0.00,0.00,1.00}{##1}}}
\expandafter\def\csname PYG@tok@nn\endcsname{\let\PYG@bf=\textbf\def\PYG@tc##1{\textcolor[rgb]{0.00,0.00,1.00}{##1}}}
\expandafter\def\csname PYG@tok@ne\endcsname{\let\PYG@bf=\textbf\def\PYG@tc##1{\textcolor[rgb]{0.82,0.25,0.23}{##1}}}
\expandafter\def\csname PYG@tok@nv\endcsname{\def\PYG@tc##1{\textcolor[rgb]{0.10,0.09,0.49}{##1}}}
\expandafter\def\csname PYG@tok@no\endcsname{\def\PYG@tc##1{\textcolor[rgb]{0.53,0.00,0.00}{##1}}}
\expandafter\def\csname PYG@tok@nl\endcsname{\def\PYG@tc##1{\textcolor[rgb]{0.63,0.63,0.00}{##1}}}
\expandafter\def\csname PYG@tok@ni\endcsname{\let\PYG@bf=\textbf\def\PYG@tc##1{\textcolor[rgb]{0.60,0.60,0.60}{##1}}}
\expandafter\def\csname PYG@tok@na\endcsname{\def\PYG@tc##1{\textcolor[rgb]{0.49,0.56,0.16}{##1}}}
\expandafter\def\csname PYG@tok@nt\endcsname{\let\PYG@bf=\textbf\def\PYG@tc##1{\textcolor[rgb]{0.00,0.50,0.00}{##1}}}
\expandafter\def\csname PYG@tok@nd\endcsname{\def\PYG@tc##1{\textcolor[rgb]{0.67,0.13,1.00}{##1}}}
\expandafter\def\csname PYG@tok@s\endcsname{\def\PYG@tc##1{\textcolor[rgb]{0.73,0.13,0.13}{##1}}}
\expandafter\def\csname PYG@tok@sd\endcsname{\let\PYG@it=\textit\def\PYG@tc##1{\textcolor[rgb]{0.73,0.13,0.13}{##1}}}
\expandafter\def\csname PYG@tok@si\endcsname{\let\PYG@bf=\textbf\def\PYG@tc##1{\textcolor[rgb]{0.73,0.40,0.53}{##1}}}
\expandafter\def\csname PYG@tok@se\endcsname{\let\PYG@bf=\textbf\def\PYG@tc##1{\textcolor[rgb]{0.73,0.40,0.13}{##1}}}
\expandafter\def\csname PYG@tok@sr\endcsname{\def\PYG@tc##1{\textcolor[rgb]{0.73,0.40,0.53}{##1}}}
\expandafter\def\csname PYG@tok@ss\endcsname{\def\PYG@tc##1{\textcolor[rgb]{0.10,0.09,0.49}{##1}}}
\expandafter\def\csname PYG@tok@sx\endcsname{\def\PYG@tc##1{\textcolor[rgb]{0.00,0.50,0.00}{##1}}}
\expandafter\def\csname PYG@tok@m\endcsname{\def\PYG@tc##1{\textcolor[rgb]{0.40,0.40,0.40}{##1}}}
\expandafter\def\csname PYG@tok@gh\endcsname{\let\PYG@bf=\textbf\def\PYG@tc##1{\textcolor[rgb]{0.00,0.00,0.50}{##1}}}
\expandafter\def\csname PYG@tok@gu\endcsname{\let\PYG@bf=\textbf\def\PYG@tc##1{\textcolor[rgb]{0.50,0.00,0.50}{##1}}}
\expandafter\def\csname PYG@tok@gd\endcsname{\def\PYG@tc##1{\textcolor[rgb]{0.63,0.00,0.00}{##1}}}
\expandafter\def\csname PYG@tok@gi\endcsname{\def\PYG@tc##1{\textcolor[rgb]{0.00,0.63,0.00}{##1}}}
\expandafter\def\csname PYG@tok@gr\endcsname{\def\PYG@tc##1{\textcolor[rgb]{1.00,0.00,0.00}{##1}}}
\expandafter\def\csname PYG@tok@ge\endcsname{\let\PYG@it=\textit}
\expandafter\def\csname PYG@tok@gs\endcsname{\let\PYG@bf=\textbf}
\expandafter\def\csname PYG@tok@gp\endcsname{\let\PYG@bf=\textbf\def\PYG@tc##1{\textcolor[rgb]{0.00,0.00,0.50}{##1}}}
\expandafter\def\csname PYG@tok@go\endcsname{\def\PYG@tc##1{\textcolor[rgb]{0.53,0.53,0.53}{##1}}}
\expandafter\def\csname PYG@tok@gt\endcsname{\def\PYG@tc##1{\textcolor[rgb]{0.00,0.27,0.87}{##1}}}
\expandafter\def\csname PYG@tok@err\endcsname{\def\PYG@bc##1{\setlength{\fboxsep}{0pt}\fcolorbox[rgb]{1.00,0.00,0.00}{1,1,1}{\strut ##1}}}
\expandafter\def\csname PYG@tok@kc\endcsname{\let\PYG@bf=\textbf\def\PYG@tc##1{\textcolor[rgb]{0.00,0.50,0.00}{##1}}}
\expandafter\def\csname PYG@tok@kd\endcsname{\let\PYG@bf=\textbf\def\PYG@tc##1{\textcolor[rgb]{0.00,0.50,0.00}{##1}}}
\expandafter\def\csname PYG@tok@kn\endcsname{\let\PYG@bf=\textbf\def\PYG@tc##1{\textcolor[rgb]{0.00,0.50,0.00}{##1}}}
\expandafter\def\csname PYG@tok@kr\endcsname{\let\PYG@bf=\textbf\def\PYG@tc##1{\textcolor[rgb]{0.00,0.50,0.00}{##1}}}
\expandafter\def\csname PYG@tok@bp\endcsname{\def\PYG@tc##1{\textcolor[rgb]{0.00,0.50,0.00}{##1}}}
\expandafter\def\csname PYG@tok@fm\endcsname{\def\PYG@tc##1{\textcolor[rgb]{0.00,0.00,1.00}{##1}}}
\expandafter\def\csname PYG@tok@vc\endcsname{\def\PYG@tc##1{\textcolor[rgb]{0.10,0.09,0.49}{##1}}}
\expandafter\def\csname PYG@tok@vg\endcsname{\def\PYG@tc##1{\textcolor[rgb]{0.10,0.09,0.49}{##1}}}
\expandafter\def\csname PYG@tok@vi\endcsname{\def\PYG@tc##1{\textcolor[rgb]{0.10,0.09,0.49}{##1}}}
\expandafter\def\csname PYG@tok@vm\endcsname{\def\PYG@tc##1{\textcolor[rgb]{0.10,0.09,0.49}{##1}}}
\expandafter\def\csname PYG@tok@sa\endcsname{\def\PYG@tc##1{\textcolor[rgb]{0.73,0.13,0.13}{##1}}}
\expandafter\def\csname PYG@tok@sb\endcsname{\def\PYG@tc##1{\textcolor[rgb]{0.73,0.13,0.13}{##1}}}
\expandafter\def\csname PYG@tok@sc\endcsname{\def\PYG@tc##1{\textcolor[rgb]{0.73,0.13,0.13}{##1}}}
\expandafter\def\csname PYG@tok@dl\endcsname{\def\PYG@tc##1{\textcolor[rgb]{0.73,0.13,0.13}{##1}}}
\expandafter\def\csname PYG@tok@s2\endcsname{\def\PYG@tc##1{\textcolor[rgb]{0.73,0.13,0.13}{##1}}}
\expandafter\def\csname PYG@tok@sh\endcsname{\def\PYG@tc##1{\textcolor[rgb]{0.73,0.13,0.13}{##1}}}
\expandafter\def\csname PYG@tok@s1\endcsname{\def\PYG@tc##1{\textcolor[rgb]{0.73,0.13,0.13}{##1}}}
\expandafter\def\csname PYG@tok@mb\endcsname{\def\PYG@tc##1{\textcolor[rgb]{0.40,0.40,0.40}{##1}}}
\expandafter\def\csname PYG@tok@mf\endcsname{\def\PYG@tc##1{\textcolor[rgb]{0.40,0.40,0.40}{##1}}}
\expandafter\def\csname PYG@tok@mh\endcsname{\def\PYG@tc##1{\textcolor[rgb]{0.40,0.40,0.40}{##1}}}
\expandafter\def\csname PYG@tok@mi\endcsname{\def\PYG@tc##1{\textcolor[rgb]{0.40,0.40,0.40}{##1}}}
\expandafter\def\csname PYG@tok@il\endcsname{\def\PYG@tc##1{\textcolor[rgb]{0.40,0.40,0.40}{##1}}}
\expandafter\def\csname PYG@tok@mo\endcsname{\def\PYG@tc##1{\textcolor[rgb]{0.40,0.40,0.40}{##1}}}
\expandafter\def\csname PYG@tok@ch\endcsname{\let\PYG@it=\textit\def\PYG@tc##1{\textcolor[rgb]{0.25,0.50,0.50}{##1}}}
\expandafter\def\csname PYG@tok@cm\endcsname{\let\PYG@it=\textit\def\PYG@tc##1{\textcolor[rgb]{0.25,0.50,0.50}{##1}}}
\expandafter\def\csname PYG@tok@cpf\endcsname{\let\PYG@it=\textit\def\PYG@tc##1{\textcolor[rgb]{0.25,0.50,0.50}{##1}}}
\expandafter\def\csname PYG@tok@c1\endcsname{\let\PYG@it=\textit\def\PYG@tc##1{\textcolor[rgb]{0.25,0.50,0.50}{##1}}}
\expandafter\def\csname PYG@tok@cs\endcsname{\let\PYG@it=\textit\def\PYG@tc##1{\textcolor[rgb]{0.25,0.50,0.50}{##1}}}

\def\PYGZus{\char`\_}

\def\PYGZgt{\char`\>}
\def\PYGZsh{\char`\#}

\def\PYGZhy{\char`\-}

\makeatother

\makeatletter
\def\PYGdefault@reset{\let\PYGdefault@it=\relax \let\PYGdefault@bf=\relax%
    \let\PYGdefault@ul=\relax \let\PYGdefault@tc=\relax%
    \let\PYGdefault@bc=\relax \let\PYGdefault@ff=\relax}
\def\PYGdefault@tok#1{\csname PYGdefault@tok@#1\endcsname}
\def\PYGdefault@toks#1+{\ifx\relax#1\empty\else%
    \PYGdefault@tok{#1}\expandafter\PYGdefault@toks\fi}
\def\PYGdefault@do#1{\PYGdefault@bc{\PYGdefault@tc{\PYGdefault@ul{%
    \PYGdefault@it{\PYGdefault@bf{\PYGdefault@ff{#1}}}}}}}
\def\PYGdefault#1#2{\PYGdefault@reset\PYGdefault@toks#1+\relax+\PYGdefault@do{#2}}

\expandafter\def\csname PYGdefault@tok@w\endcsname{\def\PYGdefault@tc##1{\textcolor[rgb]{0.73,0.73,0.73}{##1}}}
\expandafter\def\csname PYGdefault@tok@c\endcsname{\let\PYGdefault@it=\textit\def\PYGdefault@tc##1{\textcolor[rgb]{0.25,0.50,0.50}{##1}}}
\expandafter\def\csname PYGdefault@tok@cp\endcsname{\def\PYGdefault@tc##1{\textcolor[rgb]{0.74,0.48,0.00}{##1}}}
\expandafter\def\csname PYGdefault@tok@k\endcsname{\let\PYGdefault@bf=\textbf\def\PYGdefault@tc##1{\textcolor[rgb]{0.00,0.50,0.00}{##1}}}
\expandafter\def\csname PYGdefault@tok@kp\endcsname{\def\PYGdefault@tc##1{\textcolor[rgb]{0.00,0.50,0.00}{##1}}}
\expandafter\def\csname PYGdefault@tok@kt\endcsname{\def\PYGdefault@tc##1{\textcolor[rgb]{0.69,0.00,0.25}{##1}}}
\expandafter\def\csname PYGdefault@tok@o\endcsname{\def\PYGdefault@tc##1{\textcolor[rgb]{0.40,0.40,0.40}{##1}}}
\expandafter\def\csname PYGdefault@tok@ow\endcsname{\let\PYGdefault@bf=\textbf\def\PYGdefault@tc##1{\textcolor[rgb]{0.67,0.13,1.00}{##1}}}
\expandafter\def\csname PYGdefault@tok@nb\endcsname{\def\PYGdefault@tc##1{\textcolor[rgb]{0.00,0.50,0.00}{##1}}}
\expandafter\def\csname PYGdefault@tok@nf\endcsname{\def\PYGdefault@tc##1{\textcolor[rgb]{0.00,0.00,1.00}{##1}}}
\expandafter\def\csname PYGdefault@tok@nc\endcsname{\let\PYGdefault@bf=\textbf\def\PYGdefault@tc##1{\textcolor[rgb]{0.00,0.00,1.00}{##1}}}
\expandafter\def\csname PYGdefault@tok@nn\endcsname{\let\PYGdefault@bf=\textbf\def\PYGdefault@tc##1{\textcolor[rgb]{0.00,0.00,1.00}{##1}}}
\expandafter\def\csname PYGdefault@tok@ne\endcsname{\let\PYGdefault@bf=\textbf\def\PYGdefault@tc##1{\textcolor[rgb]{0.82,0.25,0.23}{##1}}}
\expandafter\def\csname PYGdefault@tok@nv\endcsname{\def\PYGdefault@tc##1{\textcolor[rgb]{0.10,0.09,0.49}{##1}}}
\expandafter\def\csname PYGdefault@tok@no\endcsname{\def\PYGdefault@tc##1{\textcolor[rgb]{0.53,0.00,0.00}{##1}}}
\expandafter\def\csname PYGdefault@tok@nl\endcsname{\def\PYGdefault@tc##1{\textcolor[rgb]{0.63,0.63,0.00}{##1}}}
\expandafter\def\csname PYGdefault@tok@ni\endcsname{\let\PYGdefault@bf=\textbf\def\PYGdefault@tc##1{\textcolor[rgb]{0.60,0.60,0.60}{##1}}}
\expandafter\def\csname PYGdefault@tok@na\endcsname{\def\PYGdefault@tc##1{\textcolor[rgb]{0.49,0.56,0.16}{##1}}}
\expandafter\def\csname PYGdefault@tok@nt\endcsname{\let\PYGdefault@bf=\textbf\def\PYGdefault@tc##1{\textcolor[rgb]{0.00,0.50,0.00}{##1}}}
\expandafter\def\csname PYGdefault@tok@nd\endcsname{\def\PYGdefault@tc##1{\textcolor[rgb]{0.67,0.13,1.00}{##1}}}
\expandafter\def\csname PYGdefault@tok@s\endcsname{\def\PYGdefault@tc##1{\textcolor[rgb]{0.73,0.13,0.13}{##1}}}
\expandafter\def\csname PYGdefault@tok@sd\endcsname{\let\PYGdefault@it=\textit\def\PYGdefault@tc##1{\textcolor[rgb]{0.73,0.13,0.13}{##1}}}
\expandafter\def\csname PYGdefault@tok@si\endcsname{\let\PYGdefault@bf=\textbf\def\PYGdefault@tc##1{\textcolor[rgb]{0.73,0.40,0.53}{##1}}}
\expandafter\def\csname PYGdefault@tok@se\endcsname{\let\PYGdefault@bf=\textbf\def\PYGdefault@tc##1{\textcolor[rgb]{0.73,0.40,0.13}{##1}}}
\expandafter\def\csname PYGdefault@tok@sr\endcsname{\def\PYGdefault@tc##1{\textcolor[rgb]{0.73,0.40,0.53}{##1}}}
\expandafter\def\csname PYGdefault@tok@ss\endcsname{\def\PYGdefault@tc##1{\textcolor[rgb]{0.10,0.09,0.49}{##1}}}
\expandafter\def\csname PYGdefault@tok@sx\endcsname{\def\PYGdefault@tc##1{\textcolor[rgb]{0.00,0.50,0.00}{##1}}}
\expandafter\def\csname PYGdefault@tok@m\endcsname{\def\PYGdefault@tc##1{\textcolor[rgb]{0.40,0.40,0.40}{##1}}}
\expandafter\def\csname PYGdefault@tok@gh\endcsname{\let\PYGdefault@bf=\textbf\def\PYGdefault@tc##1{\textcolor[rgb]{0.00,0.00,0.50}{##1}}}
\expandafter\def\csname PYGdefault@tok@gu\endcsname{\let\PYGdefault@bf=\textbf\def\PYGdefault@tc##1{\textcolor[rgb]{0.50,0.00,0.50}{##1}}}
\expandafter\def\csname PYGdefault@tok@gd\endcsname{\def\PYGdefault@tc##1{\textcolor[rgb]{0.63,0.00,0.00}{##1}}}
\expandafter\def\csname PYGdefault@tok@gi\endcsname{\def\PYGdefault@tc##1{\textcolor[rgb]{0.00,0.63,0.00}{##1}}}
\expandafter\def\csname PYGdefault@tok@gr\endcsname{\def\PYGdefault@tc##1{\textcolor[rgb]{1.00,0.00,0.00}{##1}}}
\expandafter\def\csname PYGdefault@tok@ge\endcsname{\let\PYGdefault@it=\textit}
\expandafter\def\csname PYGdefault@tok@gs\endcsname{\let\PYGdefault@bf=\textbf}
\expandafter\def\csname PYGdefault@tok@gp\endcsname{\let\PYGdefault@bf=\textbf\def\PYGdefault@tc##1{\textcolor[rgb]{0.00,0.00,0.50}{##1}}}
\expandafter\def\csname PYGdefault@tok@go\endcsname{\def\PYGdefault@tc##1{\textcolor[rgb]{0.53,0.53,0.53}{##1}}}
\expandafter\def\csname PYGdefault@tok@gt\endcsname{\def\PYGdefault@tc##1{\textcolor[rgb]{0.00,0.27,0.87}{##1}}}
\expandafter\def\csname PYGdefault@tok@err\endcsname{\def\PYGdefault@bc##1{\setlength{\fboxsep}{0pt}\fcolorbox[rgb]{1.00,0.00,0.00}{1,1,1}{\strut ##1}}}
\expandafter\def\csname PYGdefault@tok@kc\endcsname{\let\PYGdefault@bf=\textbf\def\PYGdefault@tc##1{\textcolor[rgb]{0.00,0.50,0.00}{##1}}}
\expandafter\def\csname PYGdefault@tok@kd\endcsname{\let\PYGdefault@bf=\textbf\def\PYGdefault@tc##1{\textcolor[rgb]{0.00,0.50,0.00}{##1}}}
\expandafter\def\csname PYGdefault@tok@kn\endcsname{\let\PYGdefault@bf=\textbf\def\PYGdefault@tc##1{\textcolor[rgb]{0.00,0.50,0.00}{##1}}}
\expandafter\def\csname PYGdefault@tok@kr\endcsname{\let\PYGdefault@bf=\textbf\def\PYGdefault@tc##1{\textcolor[rgb]{0.00,0.50,0.00}{##1}}}
\expandafter\def\csname PYGdefault@tok@bp\endcsname{\def\PYGdefault@tc##1{\textcolor[rgb]{0.00,0.50,0.00}{##1}}}
\expandafter\def\csname PYGdefault@tok@fm\endcsname{\def\PYGdefault@tc##1{\textcolor[rgb]{0.00,0.00,1.00}{##1}}}
\expandafter\def\csname PYGdefault@tok@vc\endcsname{\def\PYGdefault@tc##1{\textcolor[rgb]{0.10,0.09,0.49}{##1}}}
\expandafter\def\csname PYGdefault@tok@vg\endcsname{\def\PYGdefault@tc##1{\textcolor[rgb]{0.10,0.09,0.49}{##1}}}
\expandafter\def\csname PYGdefault@tok@vi\endcsname{\def\PYGdefault@tc##1{\textcolor[rgb]{0.10,0.09,0.49}{##1}}}
\expandafter\def\csname PYGdefault@tok@vm\endcsname{\def\PYGdefault@tc##1{\textcolor[rgb]{0.10,0.09,0.49}{##1}}}
\expandafter\def\csname PYGdefault@tok@sa\endcsname{\def\PYGdefault@tc##1{\textcolor[rgb]{0.73,0.13,0.13}{##1}}}
\expandafter\def\csname PYGdefault@tok@sb\endcsname{\def\PYGdefault@tc##1{\textcolor[rgb]{0.73,0.13,0.13}{##1}}}
\expandafter\def\csname PYGdefault@tok@sc\endcsname{\def\PYGdefault@tc##1{\textcolor[rgb]{0.73,0.13,0.13}{##1}}}
\expandafter\def\csname PYGdefault@tok@dl\endcsname{\def\PYGdefault@tc##1{\textcolor[rgb]{0.73,0.13,0.13}{##1}}}
\expandafter\def\csname PYGdefault@tok@s2\endcsname{\def\PYGdefault@tc##1{\textcolor[rgb]{0.73,0.13,0.13}{##1}}}
\expandafter\def\csname PYGdefault@tok@sh\endcsname{\def\PYGdefault@tc##1{\textcolor[rgb]{0.73,0.13,0.13}{##1}}}
\expandafter\def\csname PYGdefault@tok@s1\endcsname{\def\PYGdefault@tc##1{\textcolor[rgb]{0.73,0.13,0.13}{##1}}}
\expandafter\def\csname PYGdefault@tok@mb\endcsname{\def\PYGdefault@tc##1{\textcolor[rgb]{0.40,0.40,0.40}{##1}}}
\expandafter\def\csname PYGdefault@tok@mf\endcsname{\def\PYGdefault@tc##1{\textcolor[rgb]{0.40,0.40,0.40}{##1}}}
\expandafter\def\csname PYGdefault@tok@mh\endcsname{\def\PYGdefault@tc##1{\textcolor[rgb]{0.40,0.40,0.40}{##1}}}
\expandafter\def\csname PYGdefault@tok@mi\endcsname{\def\PYGdefault@tc##1{\textcolor[rgb]{0.40,0.40,0.40}{##1}}}
\expandafter\def\csname PYGdefault@tok@il\endcsname{\def\PYGdefault@tc##1{\textcolor[rgb]{0.40,0.40,0.40}{##1}}}
\expandafter\def\csname PYGdefault@tok@mo\endcsname{\def\PYGdefault@tc##1{\textcolor[rgb]{0.40,0.40,0.40}{##1}}}
\expandafter\def\csname PYGdefault@tok@ch\endcsname{\let\PYGdefault@it=\textit\def\PYGdefault@tc##1{\textcolor[rgb]{0.25,0.50,0.50}{##1}}}
\expandafter\def\csname PYGdefault@tok@cm\endcsname{\let\PYGdefault@it=\textit\def\PYGdefault@tc##1{\textcolor[rgb]{0.25,0.50,0.50}{##1}}}
\expandafter\def\csname PYGdefault@tok@cpf\endcsname{\let\PYGdefault@it=\textit\def\PYGdefault@tc##1{\textcolor[rgb]{0.25,0.50,0.50}{##1}}}
\expandafter\def\csname PYGdefault@tok@c1\endcsname{\let\PYGdefault@it=\textit\def\PYGdefault@tc##1{\textcolor[rgb]{0.25,0.50,0.50}{##1}}}
\expandafter\def\csname PYGdefault@tok@cs\endcsname{\let\PYGdefault@it=\textit\def\PYGdefault@tc##1{\textcolor[rgb]{0.25,0.50,0.50}{##1}}}

\makeatother

\makeatletter

\def\Nmc{\Nmc}

\def\1{{\mathbf 1}}

\def\Nmc{N}

\def\[#1{[\![#1]\!]}

\title{Deep Importance Sampling}
\author{Benjamin Virrion\footnote{Natixis and CEREMADE, UMR CNRS, Universit\'e Paris-Dauphine, PSL University.}}
\date{\today}

\begin{document}
\newcommand{\Emmett}[5]{
\draw[#4] (0,1)
\foreach \x in {1,...,#1}
{   -- ++(#2,rand*#3)
}
node[right] {#5};
}

\maketitle

\begin{abstract}
We present a generic path-dependent importance sampling algorithm where the Girsanov induced change of probability on the path space is represented by a sequence of neural networks taking the past of the trajectory as an input. At each learning step, the neural networks' parameters are trained so as to reduce the variance of the Monte Carlo estimator induced by this change of measure. This allows for a generic path dependent change of measure which can be used to reduce the variance of any path-dependent financial payoff. We show in our numerical experiments that for payoffs consisting of either a call, an asymmetric combination of calls and puts, a symmetric combination of calls and puts, a multi coupon autocall or a single coupon autocall, we are able to reduce the variance of the Monte Carlo estimators by factors between 2 and 9. The numerical experiments also show that the method is very robust to changes in the parameter values, which means that in practice, the training can be done offline and only updated on a weekly basis.
\end{abstract}

{\bf Keywords: }Importance Sampling, Neural Networks, Path-Dependence.

\newpage
\tableofcontents
\newpage 

\section{Introduction}

\subsection{Objective}

The objective of this paper is to compute the price of a possibly path-dependent option of payoff $g$, given by:
\begin{equation}
f(x_0) = \mathbb{E}^{\mathbb{Q}}\left[ g(X_{0 \leq t \leq T}^{x_0}) \right]
\end{equation}

Where $\left(X_{0 \leq t \leq T}^{x_0}\right)$ is an Itô diffusion satisfying:

\begin{equation}
\begin{cases}
dX_t^{x_0} = b_t dt + \sigma_t dW_t \\
X_0^{x_0} = x_0
\end{cases}
\end{equation}

Where $\mathbb{Q}$ is the risk neutral measure, $W_t$ is a Brownian process under $\mathbb{Q}$, and $b_t$ and $\sigma_t$ are two locally bounded predictable processes.

In order to reduce the variance of the Monte Carlo, we will compute $f(x)$ by using importance sampling. The change of probability measure used in this importance sampling will be obtained using an adapted process $a^{\theta}\left(t, X_{0 \leq s \leq t}^{x_0}\right)$ and the Girsanov Theorem. \footnote{In full generality, one might want to take as an input for $a^\theta(t, .)$ all the information available at time $t$, that is $\mathcal{F}_t$. For example, for a stochastic volatility diffusion, taking the past of the volatility as an input would be useful. For simplicity's sake, we only take the past of the trajectory of the underlying as an input in this paper.}

\subsection{Litterature}

Using Importance Sampling to reduce the standard deviation of Monte Carlo estimators is a well known and studied practice. Its use has been well described in both non-financial \cite{liu2008monte} and financial \cite{boyle1997monte}, \cite{glasserman2013monte} articles and books on Monte Carlo techniques. Among the many possible ways of using importance sampling, we decide here to separate two broad categories. In the first category, $g$ is a function of a random variable that is not on the path space, whereas in the second category, $g$ is a function of a random process that lives on the path space. 

In the first category, importance sampling using neural networks has been studied in \cite{muller2018neural} and \cite{gu2015neural}. For these papers, as the problem at hand does not inherintly live in the path space, the change of measure is described directly using the densities, and there is no attempt to use the Girsanov theorem to describe the change of measure. 

On the other hand, when importance sampling is done on the path space, describing the change of measure using the Girsanov theorem becomes natural. Such a use of the Girsanov theorem to do importance sampling is well known, and has been been compared to other variance reduction methods in \cite{boyle1997monte} to price out of the money options. Out of this vast litterature, we would like to mention a few articles that stand out in the sense that our method is close to theirs. 

One of the first papers using an adaptive importance sampling scheme in order to reduce variance in a Monte Carlo algorithm for a diffusion process in a financial context is that of \cite{arouna2002variance}. In this paper, the author shows how a Robbins-Monro algorithm allows to find the optimal Girsanov change of measure that reduces the Monte Carlo variance. However, the author does not directly represent the change of measure using a neural network, but restricts himself to using a deterministic drift vector. As our neural-network can take the past of the trajectory as an input, our algorithm is more flexible and allows for a path-dependent change of measure, which is not possible with the approach of \cite{arouna2002variance}. 

In \cite{lemaire2010unconstrained} the authors use importance sampling both for a function of a random variable and a function of a random process. For a function of a random process, the authors use the Girsanov theorem to represent the change of measure. Furthermore, assuming that they have some knowledge on the law of the process, more precisely on $p$ and $\nabla p$ , they show that a Robbins-Monro scheme using this knowledge to update the $\theta$ parameter of a parametrized family of changes of measures converges to a minimizer of this family that minimizes the standard deviation. The method in our paper is extremely close to theirs, except that we do not use a model-dependant analytical formula to obtain $p$ and $\nabla p$, but instead use a neural network and backpropagation to obtain this gradient, which then enables the gradient descent algorithm. So in some sense, our paper is a natural extension of their paper when using a neural network and backpropagation, which enables us to obtain the gradient for diffusion processes where no analytical formulas could be explicited. 

The paper by \cite{muller2018neural} uses importance sampling and a family of changes of measures parametrized by a neural network to learn the change of measure. However, they naturally place themselves in the random variable setting, whereas we place ourselves in the random process setting. This quite naturally brings us to use the neural network as a function of the past of the process,  which we believe is much more adapted to most financial payoffs. 

\subsection{Importance Sampling with a Girsanov Induced Change of Measure}\label{sect:girsanov_importance_sampling}

Let us consider a parametrized family of measurable functions $a^\theta : \left[0, T\right] \times \mathbb{R} \rightarrow \mathbb{R}$, with $\theta \in \Theta$, satisfying:

\begin{equation}
\left| a^\theta \left(t, x\right) \right| \leq C \left(1, + |x| \right), \qquad \textnormal{ for } t \in [0, T],x \in \mathbb{R}
\end{equation}
\begin{equation}
\left| a^\theta (t, x) - a^\theta (t, y) \right| \leq D \left|x - y\right|, \qquad \textnormal{ for } t \in [0, T], x, y \in \mathbb{R}
\end{equation}

for some constants $C$ and $D$.

We can then introduce the process:
\begin{equation}
dW_t^{\theta} = dW_t - a^{\theta}(t, X_{0 \leq s \leq t}^{x_0}) dt
\end{equation}

The diffusion of our underlying process $\left(X_t^{x_0}\right)$ can be rewritten:
\begin{equation}
dX_t^{x_0} = b_t dt + \sigma_t dW_t = b_t dt + \sigma_t dW_t^{\theta} + a^{\theta}(t, X_{0 \leq s \leq t}^{x_0}) \sigma_t dt
\end{equation}

Furthermore, introduce the change of measure process, which is a $\mathbb{Q}$-martingale:
\begin{equation}
\begin{split}
Z_t^{\theta} = \frac{d \mathbb{Q}^{\theta}}{d \mathbb{Q}} \Bigg|_{\mathcal{F}_t}
&= \exp \left( \int_{s = 0}^t a^{\theta}\left(s, X_{0 \leq u \leq s}^{x_0} \right) dW_s
- \frac{1}{2}\int_{s = 0}^t \left(a^{\theta}\left(s, X_{0 \leq u \leq s}^{x_0} \right) \right)^2
ds \right) \\
&= \exp \left( \int_{s = 0}^t a^{\theta}\left(s, X_{0 \leq u \leq s}^{x_0} \right) dW_s^{\theta}
+ \frac{1}{2}\int_{s = 0}^t \left(a^{\theta}\left(s, X_{0 \leq u \leq s}^{x_0} \right) \right)^2
ds \right) \\
\end{split}
\end{equation}

By the Girsanov theorem, $W_t^{\theta}$ is a Brownian Motion under the $\mathbb{Q}^{\theta}$ probability measure.

We then have:

\begin{equation} \label{equation:change_of_measure}
f(x) = \mathbb{E}^{\mathbb{Q}^{\theta}}\left[ g\left( X_{0 \leq t \leq T}^{x}  \right) \frac{1}{Z_T^{\theta}} \right]
\end{equation}

\subsection{Finding the Optimal Change of Measure $a^{\theta}$ Using Neural Networks}

In order to reduce the variance in the Monte-Carlo, our aim is to find the $a^\theta$ that minimizes the variance of $g\left( X_{0 \leq t \leq T}^{x}  \right) \frac{1}{Z_T^{\theta}}$ under $\mathbb{Q}^{\theta}$.  Therefore, let us introduce:
\begin{equation}
\tilde{h}\left( \theta \right) 
:= 
\mathbb{E}^{\mathbb{Q}^\theta}\left[ \left( g\left( X_{0 \leq t \leq T}^{x}  \right) \frac{1}{Z_T^{\theta}} \right)^2 \right]
\end{equation}

Considering that equation (Eq \ref{equation:change_of_measure}) is true for all $\theta$, the square of the mean in the variance term $\mathbb{E}^{\mathbb{Q}^\theta} \left[ g\left( X_{0 \leq t \leq T}^{x}  \right) \frac{1}{Z_T^{\theta}} \right]^2 = f(x)^2$ is the same for all $\theta$. Therefore, we can simply ignore it, and our optimization problem can be rewritten:

\begin{equation}
\tilde{\theta}^* := \underset{\theta \in \Theta}{\textnormal{argmin}} \, \tilde{h}\left( \theta \right)
\end{equation}

To do this, we will use multiple neural networks to represent $a^{\theta}$, and minimize for the parameters of the neural networks using the variance as our loss function.

In practice, we do not want to allow changes of measures that are too extreme. Therefore, we will add the following constraint to the error function:
\begin{equation}
h\left( \theta \right) := \mathbb{E}^{\mathbb{Q}^\theta} \left[ \left( g\left( X_{0 \leq t \leq T}^{x}  \right) \frac{1}{Z_T^{\theta}} \right)^2 \right]  + \lambda \ln \left( 1 + \mathbb{E}^{\mathbb{Q}^\theta} \left[\left( \frac{1}{Z_T^\theta} - C\right)^+ \right] \right)
\end{equation}

and consider the minimizer:
\begin{equation}
\theta^* := \underset{\theta \in \Theta}{\textnormal{argmin}} \, h\left( \theta \right)
\end{equation}

Assuming that $h$ is smooth, when $\Theta$ is compact, such a minimizer exists, albeit not necessarily uniquely. 

\section{Construction of the Neural Networks} \label{sect:construction_neural_networks}

As the number of inputs in $a^{\theta}\left(t, X_{0 \leq s \leq t}^{x_0}\right)$ increases with time, in practice, we need one neural network per time step. If the maturity is long and we want to have a fine mesh for the diffusion process $X_t^{x_0}$, it is possible to introduce a coarser mesh for the $a^{\theta}$, so as not to have too many neural networks to calibrate. We won't do this in this paper so as to keep the notations simple. 

Let us introduce a time grid $[t_0, ..., t_{N^T}] := \left[0, T/N^T, 2T/N^T, ..., T\right]$ with $N^T \in \mathbb{N}^*$ time steps. For $i \in [\![1, N^T-1]\!]$, we introduce the neural network $a^{\theta_i}$ which has $i + 1$ inputs and one output. For $i=0$, as all trajectories start at the same initial point, we introduce instead a neural network $a^{\theta_0}$. 

Mathematically, we have: 

\begin{equation}
\begin{cases}
a^{\theta_i}: \mathbb{R}^{i + 1} \times \Theta^i \rightarrow \mathbb{R},\textnormal{ for } i \in [\![1, N^T - 1]\!] \\
a^{\theta_0}: \Theta^0 \rightarrow \mathbb{R},\textnormal{ for } i = 0 \\
\end{cases}
\end{equation}

For a Bachelier diffusion, the algorithm is then as follows:

\begin{Verbatim}[commandchars=\\\{\}]
\PYG{k+kn}{import} \PYG{n+nn}{numpy} \PYG{k+kn}{as} \PYG{n+nn}{np}
\PYG{k+kn}{import} \PYG{n+nn}{tensorflow} \PYG{k+kn}{as} \PYG{n+nn}{tf}
\PYG{k}{def} \PYG{n+nf}{generate\PYGZus{}trajectories\PYGZus{}z\PYGZus{}and\PYGZus{}a\PYGZus{}list}\PYG{p}{(}\PYG{n+nb+bp}{self}\PYG{p}{):}
    \PYG{c+c1}{\PYGZsh{} Construct neural networks and trajectories}
    \PYG{n}{a\PYGZus{}list} \PYG{o}{=} \PYG{p}{[}\PYG{n+nb+bp}{None} \PYG{k}{for} \PYG{n}{i} \PYG{o+ow}{in} \PYG{n+nb}{range}\PYG{p}{(}\PYG{n+nb+bp}{self}\PYG{o}{.}\PYG{n}{N\PYGZus{}T}\PYG{p}{)]}
    \PYG{n}{trajectories} \PYG{o}{=} \PYG{p}{[}\PYG{n+nb+bp}{None} \PYG{k}{for} \PYG{n}{i} \PYG{o+ow}{in} \PYG{n+nb}{range}\PYG{p}{(}\PYG{n+nb+bp}{self}\PYG{o}{.}\PYG{n}{N\PYGZus{}T} \PYG{o}{+} \PYG{l+m+mi}{1}\PYG{p}{)]}
    \PYG{k}{for} \PYG{n}{n\PYGZus{}time\PYGZus{}step} \PYG{o+ow}{in} \PYG{n+nb}{range}\PYG{p}{(}\PYG{n+nb+bp}{self}\PYG{o}{.}\PYG{n}{N\PYGZus{}T} \PYG{o}{+} \PYG{l+m+mi}{1}\PYG{p}{):}
        \PYG{k}{if} \PYG{n}{n\PYGZus{}time\PYGZus{}step} \PYG{o}{==} \PYG{l+m+mi}{0}\PYG{p}{:}
            \PYG{n}{trajectories}\PYG{p}{[}\PYG{n}{n\PYGZus{}time\PYGZus{}step}\PYG{p}{]} \PYG{o}{=} \PYG{n}{tf}\PYG{o}{.}\PYG{n}{tile}\PYG{p}{(}\PYG{n}{tf}\PYG{o}{.}\PYG{n}{reshape}\PYG{p}{(}\PYG{n+nb+bp}{self}\PYG{o}{.}\PYG{n}{x}\PYG{p}{,} \PYG{p}{[}\PYG{l+m+mi}{1}\PYG{p}{,} \PYG{l+m+mi}{1}\PYG{p}{]),}
                                                \PYG{p}{[}\PYG{n+nb+bp}{self}\PYG{o}{.}\PYG{n}{N\PYGZus{}batch\PYGZus{}size}\PYG{p}{,} \PYG{l+m+mi}{1}\PYG{p}{])}
        \PYG{k}{else}\PYG{p}{:}
            \PYG{n}{a\PYGZus{}list}\PYG{p}{[}\PYG{n}{n\PYGZus{}time\PYGZus{}step} \PYG{o}{\PYGZhy{}} \PYG{l+m+mi}{1}\PYG{p}{]} \PYG{o}{=} \PYG{n+nb+bp}{self}\PYG{o}{.}\PYG{n}{neural\PYGZus{}net}\PYG{p}{(}\PYG{n}{trajectories}\PYG{p}{[:}\PYG{n}{n\PYGZus{}time\PYGZus{}step}\PYG{p}{],}
                                                      \PYG{n+nb+bp}{self}\PYG{o}{.}\PYG{n}{weights\PYGZus{}list}\PYG{p}{[}\PYG{n}{n\PYGZus{}time\PYGZus{}step} \PYG{o}{\PYGZhy{}} \PYG{l+m+mi}{1}\PYG{p}{],}
                                                      \PYG{n+nb+bp}{self}\PYG{o}{.}\PYG{n}{biases\PYGZus{}list}\PYG{p}{[}\PYG{n}{n\PYGZus{}time\PYGZus{}step} \PYG{o}{\PYGZhy{}} \PYG{l+m+mi}{1}\PYG{p}{],}
                                                      \PYG{n}{t\PYGZus{}step}\PYG{o}{=}\PYG{n}{n\PYGZus{}time\PYGZus{}step} \PYG{o}{\PYGZhy{}} \PYG{l+m+mi}{1}\PYG{p}{)}

            \PYG{n}{trajectories}\PYG{p}{[}\PYG{n}{n\PYGZus{}time\PYGZus{}step}\PYG{p}{]} \PYG{o}{=} \PYG{n}{trajectories}\PYG{p}{[}\PYG{n}{n\PYGZus{}time\PYGZus{}step} \PYG{o}{\PYGZhy{}} \PYG{l+m+mi}{1}\PYG{p}{]}
                                        \PYG{o}{+} \PYG{n}{a\PYGZus{}list}\PYG{p}{[}\PYG{n}{n\PYGZus{}time\PYGZus{}step} \PYG{o}{\PYGZhy{}} \PYG{l+m+mi}{1}\PYG{p}{]} \PYG{o}{*} \PYG{n+nb+bp}{self}\PYG{o}{.}\PYG{n}{sigma} \PYG{o}{*} \PYG{n+nb+bp}{self}\PYG{o}{.}\PYG{n}{dt}
                                        \PYG{o}{+} \PYG{n}{gaussian\PYGZus{}term} \PYG{o}{*} \PYG{n+nb+bp}{self}\PYG{o}{.}\PYG{n}{sigma} \PYG{o}{*} \PYG{n+nb+bp}{self}\PYG{o}{.}\PYG{n}{sqrt\PYGZus{}dt}

    \PYG{c+c1}{\PYGZsh{} Construct z}

    \PYG{n}{a\PYGZus{}times\PYGZus{}gaussians} \PYG{o}{=} \PYG{n}{tf}\PYG{o}{.}\PYG{n}{multiply}\PYG{p}{(}\PYG{n}{tf}\PYG{o}{.}\PYG{n}{reduce\PYGZus{}sum}\PYG{p}{(}\PYG{n}{tf}\PYG{o}{.}\PYG{n}{stack}\PYG{p}{(}\PYG{n}{a\PYGZus{}list}\PYG{p}{,} \PYG{n}{axis}\PYG{o}{=}\PYG{l+m+mi}{1}\PYG{p}{),} \PYG{n}{axis}\PYG{o}{=}\PYG{l+m+mi}{2}\PYG{p}{),}
                                    \PYG{n+nb+bp}{self}\PYG{o}{.}\PYG{n}{random\PYGZus{}gaussians} \PYG{o}{*} \PYG{n+nb+bp}{self}\PYG{o}{.}\PYG{n}{sqrt\PYGZus{}dt}\PYG{p}{)}
    \PYG{n}{a\PYGZus{}squared\PYGZus{}list} \PYG{o}{=} \PYG{n}{tf}\PYG{o}{.}\PYG{n}{square}\PYG{p}{(}\PYG{n}{tf}\PYG{o}{.}\PYG{n}{reduce\PYGZus{}sum}\PYG{p}{(}\PYG{n}{tf}\PYG{o}{.}\PYG{n}{stack}\PYG{p}{(}\PYG{n}{a\PYGZus{}list}\PYG{p}{,} \PYG{n}{axis}\PYG{o}{=}\PYG{l+m+mi}{1}\PYG{p}{),} \PYG{n}{axis}\PYG{o}{=}\PYG{l+m+mi}{2}\PYG{p}{))} \PYG{o}{*} \PYG{n+nb+bp}{self}\PYG{o}{.}\PYG{n}{dt}
    \PYG{n}{first\PYGZus{}term\PYGZus{}z} \PYG{o}{=} \PYG{n}{tf}\PYG{o}{.}\PYG{n}{expand\PYGZus{}dims}\PYG{p}{(}\PYG{n}{tf}\PYG{o}{.}\PYG{n}{reduce\PYGZus{}sum}\PYG{p}{(}\PYG{n}{a\PYGZus{}times\PYGZus{}gaussians}\PYG{p}{,} \PYG{n}{axis}\PYG{o}{=}\PYG{l+m+mi}{1}\PYG{p}{),} \PYG{n}{axis}\PYG{o}{=\PYGZhy{}}\PYG{l+m+mi}{1}\PYG{p}{)}
    \PYG{n}{second\PYGZus{}term\PYGZus{}z} \PYG{o}{=} \PYG{l+m+mf}{0.5} \PYG{o}{*} \PYG{n}{tf}\PYG{o}{.}\PYG{n}{expand\PYGZus{}dims}\PYG{p}{(}\PYG{n}{tf}\PYG{o}{.}\PYG{n}{reduce\PYGZus{}sum}\PYG{p}{(}\PYG{n}{a\PYGZus{}squared\PYGZus{}list}\PYG{p}{,} \PYG{n}{axis}\PYG{o}{=}\PYG{l+m+mi}{1}\PYG{p}{),} \PYG{n}{axis}\PYG{o}{=\PYGZhy{}}\PYG{l+m+mi}{1}\PYG{p}{)}
    \PYG{n}{z} \PYG{o}{=} \PYG{n}{tf}\PYG{o}{.}\PYG{n}{exp}\PYG{p}{(}\PYG{n}{first\PYGZus{}term\PYGZus{}z} \PYG{o}{+} \PYG{n}{second\PYGZus{}term\PYGZus{}z}\PYG{p}{)}

    \PYG{k}{return} \PYG{n}{trajectories}\PYG{p}{,} \PYG{n}{z}\PYG{p}{,} \PYG{n}{a\PYGZus{}list}

\PYG{k}{def} \PYG{n+nf}{neural\PYGZus{}net}\PYG{p}{(}\PYG{n+nb+bp}{self}\PYG{p}{,} \PYG{n}{trajectories}\PYG{p}{,} \PYG{n}{weights}\PYG{p}{,} \PYG{n}{biases}\PYG{p}{,} \PYG{n}{t\PYGZus{}step}\PYG{p}{):}
    \PYG{k}{if} \PYG{n}{t\PYGZus{}step} \PYG{o}{==} \PYG{l+m+mi}{0}\PYG{p}{:}
        \PYG{c+c1}{\PYGZsh{} If t\PYGZus{}step is 0, we return the same trainable variable for all trajectories}
        \PYG{n}{variable} \PYG{o}{=} \PYG{n}{tf}\PYG{o}{.}\PYG{n}{Variable}\PYG{p}{(}\PYG{n}{tf}\PYG{o}{.}\PYG{n}{zeros}\PYG{p}{([}\PYG{l+m+mi}{1}\PYG{p}{,} \PYG{l+m+mi}{1}\PYG{p}{],} \PYG{n}{dtype}\PYG{o}{=}\PYG{n}{tf}\PYG{o}{.}\PYG{n}{float64}\PYG{p}{))}
        \PYG{n}{Y} \PYG{o}{=} \PYG{n}{tf}\PYG{o}{.}\PYG{n}{tile}\PYG{p}{(}\PYG{n}{variable}\PYG{p}{,} \PYG{n}{multiples}\PYG{o}{=}\PYG{p}{[}\PYG{n+nb+bp}{self}\PYG{o}{.}\PYG{n}{N\PYGZus{}batch\PYGZus{}size}\PYG{p}{,} \PYG{l+m+mi}{1}\PYG{p}{])}
    \PYG{k}{else}\PYG{p}{:}
        \PYG{c+c1}{\PYGZsh{} If t\PYGZus{}step \PYGZgt{} 0, we return the output of a neural network}
        \PYG{c+c1}{\PYGZsh{} taking t\PYGZus{}step + 1 inputs}
        \PYG{n}{num\PYGZus{}layers} \PYG{o}{=} \PYG{n+nb}{len}\PYG{p}{(}\PYG{n}{weights}\PYG{p}{)} \PYG{o}{+} \PYG{l+m+mi}{1}
        \PYG{c+c1}{\PYGZsh{} Inputs of initial layer are the past values of the trajectories}
        \PYG{c+c1}{\PYGZsh{} We center past of the trajectories by substracting initial value: self.x}
        \PYG{n}{H} \PYG{o}{=} \PYG{n}{tf}\PYG{o}{.}\PYG{n}{reduce\PYGZus{}sum}\PYG{p}{(}\PYG{n}{tf}\PYG{o}{.}\PYG{n}{stack}\PYG{p}{(}\PYG{n}{trajectories}\PYG{p}{,} \PYG{n}{axis}\PYG{o}{=}\PYG{l+m+mi}{1}\PYG{p}{),} \PYG{n}{axis}\PYG{o}{=}\PYG{l+m+mi}{1}\PYG{p}{)} \PYG{o}{\PYGZhy{}} \PYG{n+nb+bp}{self}\PYG{o}{.}\PYG{n}{x}
        \PYG{k}{for} \PYG{n}{l} \PYG{o+ow}{in} \PYG{n+nb}{range}\PYG{p}{(}\PYG{l+m+mi}{0}\PYG{p}{,} \PYG{n}{num\PYGZus{}layers} \PYG{o}{\PYGZhy{}} \PYG{l+m+mi}{2}\PYG{p}{):}
            \PYG{n}{W} \PYG{o}{=} \PYG{n}{weights}\PYG{p}{[}\PYG{n}{l}\PYG{p}{]}
            \PYG{n}{b} \PYG{o}{=} \PYG{n}{biases}\PYG{p}{[}\PYG{n}{l}\PYG{p}{]}
            \PYG{n}{H} \PYG{o}{=} \PYG{n}{tf}\PYG{o}{.}\PYG{n}{nn}\PYG{o}{.}\PYG{n}{relu}\PYG{p}{(}\PYG{n}{tf}\PYG{o}{.}\PYG{n}{add}\PYG{p}{(}\PYG{n}{tf}\PYG{o}{.}\PYG{n}{matmul}\PYG{p}{(}\PYG{n}{H}\PYG{p}{,} \PYG{n}{W}\PYG{p}{),} \PYG{n}{b}\PYG{p}{))}
        \PYG{n}{W} \PYG{o}{=} \PYG{n}{weights}\PYG{p}{[}\PYG{o}{\PYGZhy{}}\PYG{l+m+mi}{1}\PYG{p}{]}
        \PYG{c+c1}{\PYGZsh{} No bias for the neural network output}
        \PYG{n}{Y} \PYG{o}{=} \PYG{n}{tf}\PYG{o}{.}\PYG{n}{matmul}\PYG{p}{(}\PYG{n}{H}\PYG{p}{,} \PYG{n}{W}\PYG{p}{)}
    \PYG{k}{return} \PYG{n}{Y}
\end{Verbatim}

Let us comment on the above algorithm. 

In the generate\_trajectories\_z\_and\_a\_list method, in the loop on the variable n\_time\_step, we do the following. If $\textnormal{n\_time\_step} == 0$, we simply initiate the first value of the trajectories, trajectories[0], with the initial value $self.x$. If $\textnormal{n\_time\_step} > 0$, we do two things. First, we evaluate $a^\theta_{\textnormal{n\_time\_step - 1}}$ on the past of the trajectories. This is done when we call self.neural\_net(trajectories[:n\_time\_step], self.weights\_list[n\_time\_step - 1], self.biases\_list[n\_time\_step - 1], t\_step=n\_time\_step + 1) \footnote{In Python, list[:n] is the sublist of list containing its first n terms. So trajectories[:n\_time\_step] consists in the values of the trajectories up to n\_time\_step, that is, the past of the trajectory}. Second, we construct the next step of the trajectory, with trajectories[n\_time\_step] = trajectories[n\_time\_step - 1] + a\_list[n\_time\_step - 1] * self.sigma * self.dt + gaussian\_term * self.sigma * self.sqrt\_dt. This second step is simply the Euler scheme for the Bachelier process under $\mathbb{Q}^\theta$.

In the function neural\_net, we should notice a few things. First when we call the neural\_net in the method generate\_trajectories\_z\_and\_a\_list, the trajectories input of neural\_net actually consists of the variable trajectories[:n\_time\_step] of the method generate\_trajectories\_z\_and\_a\_list. That is, is consists of all the past of the trajectories up to step n\_time\_step - 1 included. Ignoring the tf.reduce\_sum and tf.stack which are there for reshaping purposes, we then define H as trajectories[:n\_time\_step] - self.x, where self.x is the initial values of the trajectories. That is, we decide to recenter all the trajectories by substracting their initial value. In practice, this helps the algorithm converge better. Finally, for each layer, we multiply H by the weights of the layer weights[l], add the bias terms of the layer biases[l], and apply the tensorflow ReLu function. Finally, for the last layer, we only multiply H by the weights of the final layer weights[-1], but do not add any bias term. We do not put a bias term here because in practice, it hinders the convergence of the algorithm. 

We haven't described in the above code how the self.weights\_list and self.biases\_list list of variables are instansiated. In practice, these are lists of tensorflow trainable variables, with both weights and biases using a xavier initialization. These are the $\theta$ parameters that are trained in the algorithm.

\section{Numerical Implementation}

\subsection{Structure of the Neural Networks}

The neural networks that we use have $(i + 1)$ inputs, 2 intermediate layers with 16 neurons each, and one ouput layer. The intermediate layers have both a weight and a bias term, whereas the output layer only contains a weight term. The activation function that we use is ReLu. 

\begin{figure}[H]
    \includegraphics[width=\linewidth]{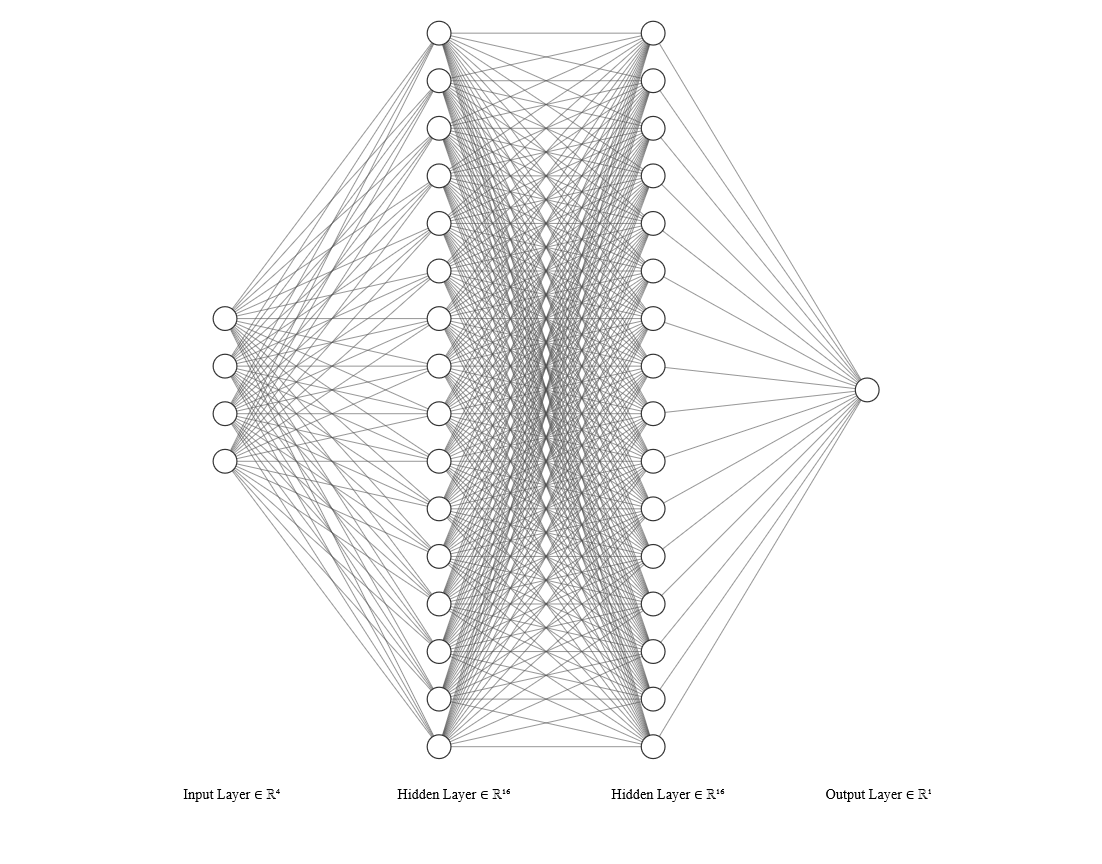}
    \caption{Neural Network Architecture for $a^{\theta_3}$}
    \label{fig:nn_arthitecture}
\end{figure}

\subsection{Discretization of the Processes}

For numerical implementation, one needs to consider discretized versions of $X_t^{x_0}, Z_t^\theta, a^\theta, h(\theta), \theta^*$. The underlying process $X_t^{x_0}$ will be discretized in its numerical version $\overline X_t^{x_0}$ according to two possible Euler schemes in section \ref{sec:diffusion_processes}. Assuming $\overline X_t^{x_0}$ already defined, we now define:

\begin{equation}
\overline a^\theta(t, \overline X^{x}_{0 \leq s \leq t}) := a^{\theta_{\lfloor \frac{t N^T}{T} \rfloor}} \left( \left(\overline X^{x_0}_{t_i} \right)_{i=0}^{\lfloor \frac{t N^T}{T} \rfloor} \right)
\end{equation}

This allows us to define:

\begin{equation}
d \overline W_t^{\theta} := dW_t - \overline a^{\theta}(t, \overline X_{0 \leq s \leq t}^{x_0}) dt
\end{equation}

We now define the martingale:

\begin{equation}
\overline Z_t^\theta := \exp \left( \int_{s=0}^t \overline a^\theta \left(s, \overline X_{0 \leq r \leq s}^{x_0} \right) d \overline W_s^\theta 
+ \frac{1}{2} \int_{s=0}^t \overline a^\theta \left(s, \left( \overline X_{0 \leq r \leq s}^{x_0} \right) \right)^2 ds
\right)
\end{equation}

Which allows us to define $\overline{\mathbb{Q}}^{\theta}$ as a measure such that $\forall t \in [0, T]$:

\begin{equation}
\frac{d \overline{\mathbb{Q}}^{\theta} }{d \mathbb{Q}} \Bigg|_{\mathcal{F}_t} := \overline Z_t^{\theta}
\end{equation}

We can then define:

\begin{equation}
\overline h\left( \theta \right) := \widehat{\mathbb{E}}^{\overline{\mathbb{Q}}^\theta} \left[ \left( g\left( \overline X_{0 \leq t \leq T}^{x}  \right) \frac{1}{\overline Z_T^{\theta}} \right)^2 \right]  + \lambda \ln \left( 1 + \widehat{\mathbb{E}}^{\overline{\mathbb{Q}}^\theta} \left[\left( \frac{1}{\overline Z_T^\theta} - C\right)^+ \right] \right)
\end{equation}

where the operator $\widehat{\mathbb{E}}^\mathbb{A}$ is defined as the empirical average obtained when simulating NBatchSize random variables under the probability $\mathbb{A}$. 

We define $\overline h(\theta)$ as the function to be minimized, and then train our neural networks by doing NBatchChangeProportion learning steps for a given batch of random variables, and fetching NumberOfBatchesForTraining batches of random variables. We thus do in total NBatchChangeProportion x NumberOfBatchesForTraining learning steps. 

We then define $\hat \theta^*$ as the value obtained for $\theta$ after these NBatchChangeProportion x NumberOfBatchesForTraining learning steps. 

The $C$ used for the differente experiments are those of tables \ref{tab:bachelier_call_option_run_params}-\ref{tab:lv_single_coupon_autocall_run_params}. 

For the figures \ref{fig:a_theta_call_bachelier_30_210_a_full}, \ref{fig:a_theta_call_bachelier_30_210_a_local}, \ref{fig:a_theta_calls_and_puts_asym_bachelier_30_210_a_full} \ref{fig:a_theta_calls_and_puts_asym_bachelier_30_210_a_local}, \ref{fig:a_theta_calls_and_puts_asym_bachelier_0_270_a_full} \ref{fig:a_theta_calls_and_puts_asym_bachelier_0_270_a_local}, \ref{fig:a_theta_calls_and_puts_sym_bachelier_30_210_a_full}, \ref{fig:a_theta_calls_and_puts_sym_bachelier_30_210_a_local}, \ref{fig:a_theta_calls_and_puts_sym_bachelier_30_270_a_full}, \ref{fig:a_theta_calls_and_puts_sym_bachelier_30_270_a_local}, \ref{fig:a_theta_autocall_multi_coupons_bachelier_30_210_a_full}, \ref{fig:a_theta_autocall_multi_coupons_bachelier_30_210_a_local}, \ref{fig:a_theta_autocall_multi_coupons_bachelier_30_270_a_full}, \ref{fig:a_theta_autocall_multi_coupons_bachelier_30_270_a_local}, \ref{fig:a_theta_autocall_single_coupon_bachelier_30_210_a_full}, \ref{fig:a_theta_autocall_single_coupon_bachelier_30_210_a_local}, \ref{fig:a_theta_call_lv_30_210_a_full}, \ref{fig:a_theta_call_lv_30_210_a_local}, \ref{fig:a_theta_calls_and_puts_asym_lv_30_210_a_full}, \ref{fig:a_theta_calls_and_puts_asym_lv_30_210_a_local}, \ref{fig:a_theta_calls_and_puts_sym_lv_30_210_a_full}, \ref{fig:a_theta_calls_and_puts_sym_lv_30_210_a_local}, \ref{fig:a_theta_calls_and_puts_sym_lv_30_270_a_full}, \ref{fig:a_theta_calls_and_puts_sym_lv_30_270_a_local}, \ref{fig:a_theta_autocall_multi_coupons_lv_30_210_a_full}, \ref{fig:a_theta_autocall_multi_coupons_lv_30_210_a_local}, \ref{fig:a_theta_autocall_single_coupon_lv_30_210_a_full}, \ref{fig:a_theta_autocall_single_coupon_lv_30_180_a_local}, \ref{fig:a_theta_autocall_single_coupon_lv_30_180_a_full} and \ref{fig:a_theta_autocall_single_coupon_lv_30_180_a_local} showing the surfaces $\overline a^\theta$ and $\tilde a^\theta$, we use the learning rates given by tables \ref{tab:bachelier_call_option_run_params}, \ref{tab:bachelier_asym_call_&_put_options_run_params}, \ref{tab:bachelier_sym_call_&_put_options_run_params}, \ref{tab:bachelier_multi_coupons_autocall_run_params}, \ref{tab:bachelier_single_coupon_autocall_run_params}, \ref{tab:lv_call_option_run_params}, \ref{tab:lv_asym_call_&_put_options_run_params}, \ref{tab:lv_sym_call_&_put_options_run_params}, \ref{tab:lv_multi_coupons_autocall_run_params} and \ref{tab:lv_single_coupon_autocall_run_params}. 
For the graphs \ref{fig:std_adaptative_vs_mc_call_bachelier_a_full}, \ref{fig:std_adaptative_vs_mc_call_bachelier_a_local}, \ref{fig:std_adaptative_vs_mc_calls_and_puts_asym_bachelier_a_full}, \ref{fig:std_adaptative_vs_mc_calls_and_puts_asym_bachelier_a_local}, \ref{fig:std_adaptative_vs_mc_calls_and_puts_sym_bachelier_a_full}, \ref{fig:std_adaptative_vs_mc_calls_and_puts_sym_bachelier_a_local}, \ref{fig:std_adaptative_vs_mc_autocall_multi_coupons_bachelier_a_full}, \ref{fig:std_adaptative_vs_mc_autocall_multi_coupons_bachelier_a_local}, \ref{fig:std_adaptative_vs_mc_autocall_single_coupon_bachelier_a_full}, \ref{fig:std_adaptative_vs_mc_autocall_single_coupon_bachelier_a_local}, \ref{fig:std_adaptative_vs_mc_call_lv_a_full}, \ref{fig:std_adaptative_vs_mc_call_lv_a_local}, \ref{fig:std_adaptative_vs_mc_calls_and_puts_asym_lv_a_full}, \ref{fig:std_adaptative_vs_mc_calls_and_puts_asym_lv_a_local}, \ref{fig:std_adaptative_vs_mc_calls_and_puts_asym_lv_a_full}, \ref{fig:std_adaptative_vs_mc_calls_and_puts_asym_lv_a_local}, \ref{fig:std_adaptative_vs_mc_calls_and_puts_sym_lv_a_full}, \ref{fig:std_adaptative_vs_mc_calls_and_puts_sym_lv_a_local}, \ref{fig:std_adaptative_vs_mc_autocall_multi_coupons_lv_a_full}, \ref{fig:std_adaptative_vs_mc_autocall_multi_coupons_lv_a_local}, \ref{fig:std_adaptative_vs_mc_autocall_single_coupon_lv_a_full} and \ref{fig:std_adaptative_vs_mc_autocall_single_coupon_lv_a_local}, we use the learning rates given by tables \ref{tab:bachelier_call_learning_rates}-\ref{tab:lv_autocall_single_learning_rates}. 

For the values of $\lambda$, we do two things. For the figures \ref{fig:a_theta_call_bachelier_30_210_a_full}, \ref{fig:a_theta_call_bachelier_30_210_a_local}, \ref{fig:a_theta_calls_and_puts_asym_bachelier_30_210_a_full} \ref{fig:a_theta_calls_and_puts_asym_bachelier_30_210_a_local}, \ref{fig:a_theta_calls_and_puts_asym_bachelier_0_270_a_full} \ref{fig:a_theta_calls_and_puts_asym_bachelier_0_270_a_local}, \ref{fig:a_theta_calls_and_puts_sym_bachelier_30_210_a_full}, \ref{fig:a_theta_calls_and_puts_sym_bachelier_30_210_a_local}, \ref{fig:a_theta_calls_and_puts_sym_bachelier_30_270_a_full}, \ref{fig:a_theta_calls_and_puts_sym_bachelier_30_270_a_local}, \ref{fig:a_theta_autocall_multi_coupons_bachelier_30_210_a_full}, \ref{fig:a_theta_autocall_multi_coupons_bachelier_30_210_a_local}, \ref{fig:a_theta_autocall_multi_coupons_bachelier_30_270_a_full}, \ref{fig:a_theta_autocall_multi_coupons_bachelier_30_270_a_local}, \ref{fig:a_theta_autocall_single_coupon_bachelier_30_210_a_full}, \ref{fig:a_theta_autocall_single_coupon_bachelier_30_210_a_local}, \ref{fig:a_theta_call_lv_30_210_a_full}, \ref{fig:a_theta_call_lv_30_210_a_local}, \ref{fig:a_theta_calls_and_puts_asym_lv_30_210_a_full}, \ref{fig:a_theta_calls_and_puts_asym_lv_30_210_a_local}, \ref{fig:a_theta_calls_and_puts_sym_lv_30_210_a_full}, \ref{fig:a_theta_calls_and_puts_sym_lv_30_210_a_local}, \ref{fig:a_theta_calls_and_puts_sym_lv_30_270_a_full}, \ref{fig:a_theta_calls_and_puts_sym_lv_30_270_a_local}, \ref{fig:a_theta_autocall_multi_coupons_lv_30_210_a_full}, \ref{fig:a_theta_autocall_multi_coupons_lv_30_210_a_local}, \ref{fig:a_theta_autocall_single_coupon_lv_30_210_a_full}, \ref{fig:a_theta_autocall_single_coupon_lv_30_180_a_local}, \ref{fig:a_theta_autocall_single_coupon_lv_30_180_a_full} and \ref{fig:a_theta_autocall_single_coupon_lv_30_180_a_local} showing the surfaces $\overline a^\theta$ and $\tilde a^\theta$, we use the values of tables 
\ref{tab:bachelier_call_option_run_params}-\ref{tab:lv_single_coupon_autocall_run_params}. 
However, manually choosing each $\lambda$ for the graphs 
\ref{fig:std_adaptative_vs_mc_call_bachelier_a_full}, \ref{fig:std_adaptative_vs_mc_call_bachelier_a_local}, \ref{fig:std_adaptative_vs_mc_calls_and_puts_asym_bachelier_a_full}, \ref{fig:std_adaptative_vs_mc_calls_and_puts_asym_bachelier_a_local}, \ref{fig:std_adaptative_vs_mc_calls_and_puts_sym_bachelier_a_full}, \ref{fig:std_adaptative_vs_mc_calls_and_puts_sym_bachelier_a_local}, \ref{fig:std_adaptative_vs_mc_autocall_multi_coupons_bachelier_a_full}, \ref{fig:std_adaptative_vs_mc_autocall_multi_coupons_bachelier_a_local}, \ref{fig:std_adaptative_vs_mc_autocall_single_coupon_bachelier_a_full}, \ref{fig:std_adaptative_vs_mc_autocall_single_coupon_bachelier_a_local}, \ref{fig:std_adaptative_vs_mc_call_lv_a_full}, \ref{fig:std_adaptative_vs_mc_call_lv_a_local}, \ref{fig:std_adaptative_vs_mc_calls_and_puts_asym_lv_a_full}, \ref{fig:std_adaptative_vs_mc_calls_and_puts_asym_lv_a_local}, \ref{fig:std_adaptative_vs_mc_calls_and_puts_asym_lv_a_full}, \ref{fig:std_adaptative_vs_mc_calls_and_puts_asym_lv_a_local}, \ref{fig:std_adaptative_vs_mc_calls_and_puts_sym_lv_a_full}, \ref{fig:std_adaptative_vs_mc_calls_and_puts_sym_lv_a_local}, \ref{fig:std_adaptative_vs_mc_autocall_multi_coupons_lv_a_full}, \ref{fig:std_adaptative_vs_mc_autocall_multi_coupons_lv_a_local}, \ref{fig:std_adaptative_vs_mc_autocall_single_coupon_lv_a_full} and \ref{fig:std_adaptative_vs_mc_autocall_single_coupon_lv_a_local}
 would be very cumbersome, as we might need a different $\lambda$ for each point in the graph. Therefore, for these graphs, we instead use a first batch to evaluate the standard deviation $\hat \sigma$ of $g\left(\overline X_{0 \leq t \leq T}^{x_0}\right)$ under $\mathbb{Q}$, and choose $\lambda = \textnormal{BaseForAutomaticLambdaConstraint} \times 10^{- \lfloor \log_{10} \left( \hat \sigma \right) \rfloor}$, where BaseForAutomaticLambdaConstraint is given by table \ref{tab:baseforautomaticlambdaconstraint}. 

\section{Numerical Experiment} \label{sec:numerical_experiment}

\subsection{Experiment Settings}

For the numerical experiments, we will consider two diffusion processes (Bachelier and Local Volatility Diffusion), and 3 types of payoffs (Calls, Calls \& Puts and Autocall). The parameters used for these diffusions and payoffs are those of tables \ref{tab:bachelier_diffusion_params}, \ref{tab:local_volatility_diffusion_params}, \ref{tab:call_option_params}, \ref{tab:asym_call_&_put_options_params}, \ref{tab:sym_call_&_put_options_params}, \ref{tab:multi_coupons_autocall_params} and \ref{tab:single_coupon_autocall_params}.

\subsubsection{Diffusion Processes} \label{sec:diffusion_processes}

\paragraph{Bachelier}

The Bachelier diffusion process is defined by:
\begin{equation}
\begin{cases}
dX_t^{x_0} = \sigma dW_t \qquad \textnormal{ for } t \in [0, T] \\ 
X_0 = x_0
\end{cases}
\end{equation}

In practice, we diffuse the Euler scheme given by:

\begin{equation}
\begin{cases}
\overline X_{\frac{i + 1}{N_t}T}^{x_0} = \overline X_{\frac{i}{N_t}T}^{x_0} + \sigma \left( W_{\frac{i + 1}{N_t}T} - W_{\frac{i}{N_t}T}\right) \qquad \textnormal{ for } i \in [\![0, N_t - 1]\!] \\
\overline X_0 = x_0
\end{cases}
\end{equation}

$x_0, \sigma, T$ and $N_t$ are defined in table (Tab \ref{tab:bachelier_diffusion_params}). 

\paragraph{Local Volatility}

The Local Volatility diffusion process is defined by:
\begin{equation}
\begin{cases}
dX_t^{x_0} = X_t^{x_0} \sigma \left(t, \ln\left(X_t^{x_0}\right)\right) dW_t \\
X_0^{x_0} = x_0
\end{cases}
\end{equation}

In practice, we diffuse the Euler scheme of the log diffusion given by:

\begin{equation}
\begin{cases}
\overline Y_{\frac{i + 1}{N_t}T}^{x_0} = \overline Y_{\frac{i}{N_t}T} + \sigma\left(t, Y_{\frac{i}{N_t}T}^{x_0}\right) \left( W_{\frac{i + 1}{N_t}T} - W_{\frac{i}{N_t}T}\right) \qquad \textnormal{ for } i \in [\![0, N_t - 1]\!] \\
\overline Y_0^{x_0} = \ln(x_0)
\end{cases}
\end{equation}

and define the discrete process as:

\begin{equation}
\overline X_{\frac{i}{N_t}T}^{x_0} = \exp \left(\overline Y_{\frac{i}{N_t}T}^{x_0}\right), \qquad \textnormal{ for } i \in [\![0, N_t]\!]
\end{equation}

To obtain the local volatility $\sigma\left(t, x\right)$, we start from an implied volatility surface given by a  raw SVI parametric, and obtain the corresponding local volatility by using the Dupire formula. 

Taking the definition of \cite{gatheral2014arbitrage}, for a given parameter set $\chi = \{a, b, \rho, m, \sigma\}$, the raw SVI parameterization of total implied variance up to time $t$ reads:
\begin{equation}
\tilde{w}\left(t, \chi \right) = \left(a + b \left\{ \rho \left( k - m \right) + \sqrt{\left(k - m\right)^2 + \sigma^2} \right\} \right)
\end{equation}

where $a \in \mathbb{R}, b \geq 0, |\rho| < 1, m \in \mathbb{R}, \sigma > 0$ and $\chi$ respects the condition $a + b \sigma \sqrt{1 - \rho^2} \geq 0$. $k := \ln \left( \frac{K}{F\left(t, 100\%\right)} \right) =  \ln \left( \frac{K}{X_0} \right)$ is the log-Forward-Strike (we will only consider diffusions with no interest rates, so the forward of the underlying is its spot)

However, as we do not only want one time strand of the volatility surface, but a whole volatility surface, we will very naïvely use the following parameters for the whole volatility surface:

\begin{equation}
w\left(t, k, \chi \right) = t \left(a + b \left\{ \rho \left( k - m \right) + \sqrt{\left(k - m\right)^2 + \sigma^2} \right\} \right)
\end{equation}

As $w\left(t, k, \chi\right) = t \sigma_{imp}^2\left(t, k, \chi\right)$, this gives the following parameterization for the implied volatility $\sigma_{imp} \left(t, k, \chi\right)$:

\begin{equation}
\sigma_{imp}(t, k, \chi) = \sqrt{\left( a + b \left\{ \rho \left( k - m \right) + \sqrt{\left(k - m\right)^2 + \sigma^2} \right\}  \right)}
\end{equation}

By doing so, we consider an implied volatility surface which has the same smile for all maturities. This is not what is observed in practice: the smile tends to smooth out as maturity increases. However, as the main focus of this paper is not the volatility surface, we will still use this simple parameterization in order to have a simple implementation. 

Using the Dupire formula, we can obtain according to the computations in \cite{dasilva2016backtesting} the following local volatility:

\begin{equation}
\sigma^2\left(t, k\right) = \frac{\partial_t w}{1 - \frac{k}{w} \partial_k w + \frac{1}{4} \left( - \frac{1}{4} - \frac{1}{w} + \frac{k^2}{w^2} \right) \left( \partial_k w \right)^2 + \frac{1}{2} \partial^2_{kk} w}
\end{equation}

For $t \in (0, T]$, using the definition for $w$, we obtain:

\begin{equation}
\begin{cases}
\partial_t w = a + b \left\{ \rho \left(k - m\right) + \sqrt{\left( k - m \right)^2 + \sigma^2} \right\} \\
\partial_k w = tb \left\{ \rho + \frac{k - m}{\sqrt{\left( k - m\right)^2 + \sigma^2}} \right\} \\
\partial^2_{kk} w = \frac{tb \sigma^2}{\left(\left( k - m \right)^2 + \sigma^2 \right)^{\frac{1}{3}}} \\
\end{cases}
\end{equation}

For the special case $t=0$, we define the local volatility as the limit for $t \rightarrow 0$ of $\sigma(t, k)$ as previously defined, which gives:

\begin{equation}
\begin{cases}
\sigma^2(0, k) = \frac{\partial_t \omega}{1 - k \frac{\partial_k \omega}{\omega} + \frac{1}{4} \left(- \frac{1}{4} + k^2 \left( \frac{\partial_k \omega}{\omega} \right)^2 \right)} \\
\frac{\partial_k \omega}{\omega} = \frac{b \left\{ \rho + \frac{k - m}{\sqrt{\left( k - m\right)^2 + \sigma^2}} \right\}}{a + b \left\{ \rho \left( k - m \right) + \sqrt{\left(k - m\right)^2 + \sigma^2} \right\}}
\end{cases}
\end{equation}

For a given $\chi$, we can therefore compute both the corresponding implied volatility $\sigma_{imp}\left(t, k\right)$ and local volatility surfaces $\sigma\left(t, k\right)$ numerically. 

\subsubsection{Payoffs}

\paragraph{Call Option}

The Call option payoff of strike $K$ is given by:
\begin{equation}
g(x) = \left(x - K\right)^+
\end{equation}

\begin{figure}[H]
\centering
\begin{tikzpicture}[scale=3.]
\draw[->] (0, 0) -- (2, 0) node[right] {$x$};
\draw[->] (0, 0) -- (0, 1) node[above] {$g(x)$};
\draw[thick] (1.4, -.05) node[below]{$K$} -- (1.4,.05);
\draw[domain=0:2,smooth,variable=\x,blue] plot ({\x},{max(0, \x-1.4)});
\end{tikzpicture}
\label{graph:payoff_call_option}
\end{figure}

\paragraph{Call \& Put Options}

The Call \& Put Options payoff with $N_1$ calls, $N_2$ puts, of strikes $K_1$ and $K_2$ is given by:
\begin{equation}
g(x) = N_1 \left(x - K_1\right)^+ + N_2 \left(K_2 - x \right)^+
\end{equation}

\begin{figure}[H]
\centering
\begin{tikzpicture}[scale=3.]
\draw[->] (0, 0) -- (2, 0) node[right] {$x$};
\draw[->] (0, 0) -- (0, 1) node[above] {$g(x)$};
\draw[thick] (1.2, -.05) node[below]{$K_1$} -- (1.2,.05);
\draw[domain=1:2,smooth,variable=\x,blue] plot ({\x},{max(0, \x-1.2)});
\draw[thick] (0.6, -.05) node[below]{$K_2$} -- (0.6,.05);
\draw[domain=0.53:1,smooth,variable=\x,blue] plot ({\x},{10 * max(0, 0.6 - \x)});
\draw[domain=0.5:0.53,densely dotted,variable=\x,blue] plot ({\x},{10 * max(0, 0.6 - \x)});
\end{tikzpicture}
\label{graph:payoff_call_and_put_options}
\end{figure}

\paragraph{AutoCall}

The AutoCall payoff is a function of the whole trajectory. In practice, we use a smoothed AutoCall payoff, as is used in the finance industry. In our case, the smoothing of the Payoff is necessary for the training of the neural networks, more specifically for the computation of the derivative of our loss function with respect to the neural network parameters, to work properly. For practitioners, we do not see this as a major problem, as it is industry standard to smooth their Payoffs, so as to get reasonable Greeks for the trader's hedge of the product. The smoothing presented here is not the one that practitioners should use in practice. Indeed, as the AutoCall is a non convex payoff, our smoothing can lead to under hedging the product. As the smoothing of the AutoCall payoff is not the topic of this paper, we will do with this very crude smoothing method. 

\begin{equation}
g((X_t^{x_0})_{t=0}^T) = C^{PDI} + \sum_{i=0}^{N^P - 1} C^{P_i}
\end{equation}

with:

\begin{equation}
C^{P_i} = \1^S\left(i, X_{T^A_i}^{x_0} \geq B_{T^A_i}\right) \prod_{\tilde i = 0}^{i - 1} \left( 1 - \1^{S}\left(\tilde i, X_{T^A_{\tilde i}}^{x_0},  B_{T^A_{\tilde i}}\right) \right)
\end{equation}

where

\begin{equation} \label{eq:smooth_indicator}
\1^S(i, x, b) = \frac{(x - b)^+ - (x - b - S_i)^+}{S_i}
\end{equation}

and

\begin{equation}
\begin{split}
C^{PDI} = & - \left( \left( 1 + \frac{1 - K}{S^{PDI}} \right) \left(K + S^{PDI} - X_T^{x_0} \right)^+
-
\frac{1 - K}{S^{PDI}} \left( K - X_T^{x_0} \right)^+ \right) \\
&\prod_{\tilde i = 0}^{N^P - 1} \left( 1 - \1^{S}\left(\tilde i, X_{T^A_{\tilde i}}^{x_0},  B_{T^A_{\tilde i}}\right) \right)
\end{split}
\end{equation}

Under the condition that the product is sill alive at $T_i^A$, the corresponding Phoenix Coupon $C^{P_i}$ payoff is illustrated in figure (Fig. \ref{fig:payoff_autocall_option_coupons}). Similarly, under the condition that the product is still alive at final maturity $T$, we illustrate in figure (Fig. \ref{fig:payoff_autocall_option_pdi}) the Put Down and In $C^{PDI}$ payoff.

\begin{figure}[H]
\begin{minipage}[t]{0.48\linewidth}
\centering
\begin{tikzpicture}[scale=3.]
\draw[->] (0, 0) -- (2, 0) node[right] {$x$};
\draw[->] (0, -1) -- (0, 0.5) node[above] {$C^{PDI}$};

\draw[domain=0:0.6, smooth, variable=\x, blue] plot ({\x}, {-max(0, 1. - \x)});
\draw[domain=0.6:0.7, smooth, variable=\x, blue] plot ({\x}, {-0.4 * (0.7 - \x)/0.1});
\draw[domain=0.7:2., smooth, variable=\x, blue] plot ({\x}, 0.);

\draw[thick] (0.6, -.05) node[below]{$K$} -- (0.6,.05);
\draw[thick, <->] (0.6, 0.1) node[above]{$S^{PDI}$} -- (0.7, 0.1);

\draw[thick] (-0.05, -0.4) node[left]{$\frac{1 - K}{S^{PDI}}$} -- (0.05, - 0.4);
\end{tikzpicture}
\caption{$C^{PDI}$ when product reaches maturity $T_{N^P - 1}^A$}
\label{fig:payoff_autocall_option_pdi}
\end{minipage}
\begin{minipage}[t]{0.04\linewidth}
\hfill
\end{minipage}
\begin{minipage}[t]{0.48\linewidth}
\centering
\begin{tikzpicture}[scale=3.]
\draw[->] (0, 0) -- (2, 0) node[right] {$x$};
\draw[->] (0, -0.5) -- (0, 1) node[above] {$C^{P_i}$};

\draw[domain=0.:1.3, smooth, variable=\x, blue] plot ({\x}, 0.);
\draw[domain=1.3:1.4, smooth, variable=\x, blue] plot ({\x}, {max(0, 10/2. * (\x - 1.3))});
\draw[domain=1.4:2, smooth, variable=\x, blue] plot ({\x}, {1./2.});

\draw[thick] (1.4, -.05) node[below]{$B_{T_{i}^A}$} -- (1.4,.05);
\draw[thick, <->] (1.3, 0.6) node[above]{$S_{i}$} -- (1.4, 0.6);
\end{tikzpicture}
\caption{$C^{P_i}$ when product is alive at time $T_i^A$}
\label{fig:payoff_autocall_option_coupons}
\end{minipage}
\end{figure}

We illustrate in figure (Fig. \ref{fig:autocall_barrier_activations}) how the product is autocalled or not for different trajectories. For trajectory 1, the product is autocalled at the first phoenix barrier date for which the underlying is higher than the barrier value, that is, at time $T_1^A$. The owner of the AutoCall then recieves the corresponding phoenix coupon $C^{P_1}$. For trajectory 2, the product is autocalled slightly later, at time $T_3^A$, and the owner of the AutoCall recieves $C^{P_3}$. Trajectories 3 and 4 never cross a barrier $B_i$, for $0 \leq i \leq 4$. Therefore, for these trajectories, the product reaches maturity. For trajectory 3, at maturity, the underlying is above the put down and in barrier $B^{PDI}$ and below the final phoenix coupon barier $B_4$. Therefore, the owner of the AutoCall doesn't pay or recieve anything. For trajectory 4, at maturity, the underlying is below the put down and in barrier $B^{PDI}$. The owner therefore ``recieves'' $C^{PDI}$ (which is a negative value, so the owner actually pays $|C^{PDI}|$). The above explaination stands for the non-smoothed product. In order to smooth the payoff, we replace the indicator functions by their smoothed versions from equation (Eq. \ref{eq:smooth_indicator}). 

\pgfmathsetseed{1}

\newcommand{\Lathrop}[7]{
\draw[#4] (0,1)
\foreach \x in {1,...,#6}
{   -- ++(#2,rand*#3)
}
coordinate (tempcoord) {};
\pgfmathsetmacro{\remaininglength}{(#1-#6)*#2}
\node[black, right] at (tempcoord) [#4] {#5}
}

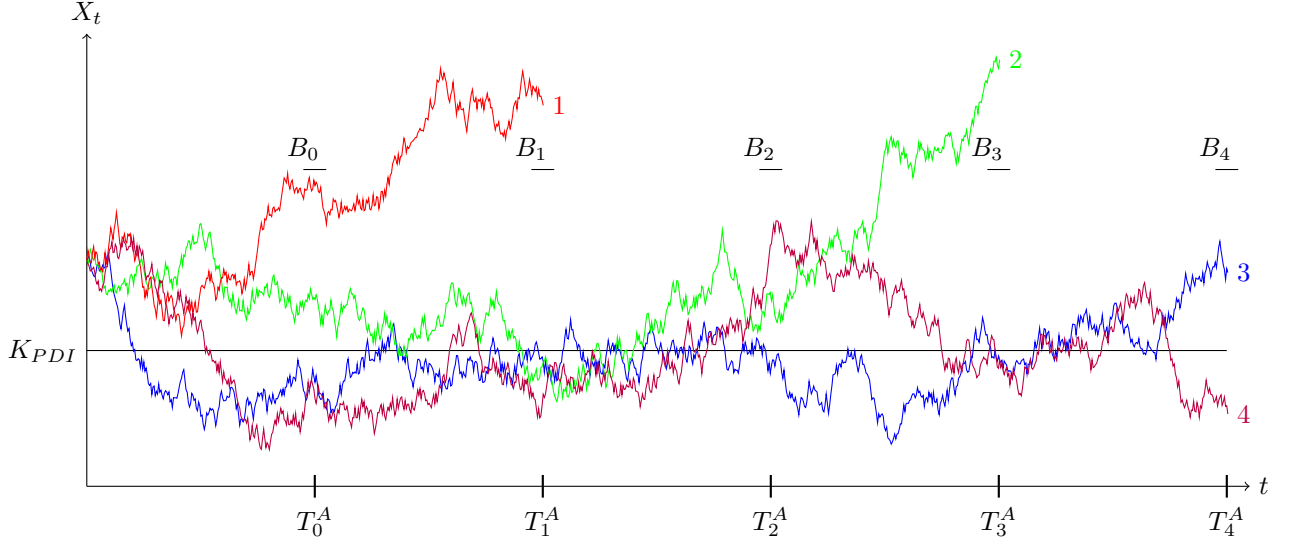
\begin{figure}[H]
\begin{tikzpicture}[scale=3.]
\draw[->] (0, 0) -- (5.1, 0) node[right] {$t$};
\draw[->] (0, 0) -- (0, 2) node[above] {$X_t$};
\draw[thick] (1., -0.05) node[below]{$T_0^A$} -- (1., 0.05);
\draw[thick] (2., -0.05) node[below]{$T_1^A$} -- (2., 0.05);
\draw[thick] (3., -0.05) node[below]{$T_2^A$} -- (3., 0.05);
\draw[thick] (4., -0.05) node[below]{$T_3^A$} -- (4., 0.05);
\draw[thick] (5., -0.05) node[below]{$T_4^A$} -- (5., 0.05);
\draw[-] (0, 0.6) node[left]{$K_{PDI}$} -- (5., 0.6);

\Lathrop{1000}{0.005}{0.05}{blue}{3}{1000}{right};
\Lathrop{1000}{0.005}{0.05}{transparent}{1}{1000}{right};
\Lathrop{1000}{0.005}{0.05}{green}{2}{800}{right};
\Lathrop{1000}{0.005}{0.05}{transparent}{2}{1000}{right};
\Lathrop{1000}{0.005}{0.05}{transparent}{2}{1000}{right};
\Lathrop{1000}{0.005}{0.05}{red}{1}{400}{right};
\Lathrop{1000}{0.005}{0.05}{transparent}{2}{1000}{right};
\Lathrop{1000}{0.005}{0.05}{transparent}{2}{1000}{right};
\Lathrop{1000}{0.005}{0.05}{transparent}{2}{1000}{right};
\Lathrop{1000}{0.005}{0.05}{transparent}{2}{1000}{right};
\Lathrop{1000}{0.005}{0.05}{transparent}{2}{1000}{right};
\Lathrop{1000}{0.005}{0.05}{transparent}{2}{1000}{right};
\Lathrop{1000}{0.005}{0.05}{purple}{4}{1000}{right};

\draw[-] (0.95, 1.4) node[above]{$B_0$} -- (1.05, 1.4);
\draw[-] (1.95, 1.4) node[above]{$B_1$} -- (2.05, 1.4);
\draw[-] (2.95, 1.4) node[above]{$B_2$} -- (3.05, 1.4);
\draw[-] (3.95, 1.4) node[above]{$B_3$} -- (4.05, 1.4);
\draw[-] (4.95, 1.4) node[above]{$B_4$} -- (5.05, 1.4);
\end{tikzpicture}
\caption{AutoCall Barrier Activations for Different Underlying Trajectories}
\label{fig:autocall_barrier_activations}
\end{figure}

\subsection{A Visual Representation of $a^\theta$}

As $a^\theta$ is a function of the pathspace, we cannot represent it visually in a graph. However, we can restrict ourselves to showing its values for some simple trajectories. Let us therefore introduce $\tilde a^\theta(t, x) = a^\theta\left(t, \left(X_s^{x_0}\right)_{s=0}^{t} (\omega(t, x))\right)$, where $\omega(t, x)$ is the set of events such that $\forall s \in [0, t], X_s^{x_0}(\omega(t, x)) = x_0 + \frac{s}{t} \left(x - x_0\right)$. In other words, $\tilde a^\theta(t, x)$ is the evaluation of $a^\theta$ on trajectories that start at $(0, x_0)$ and go in a straight line up to to the point $(t, x)$. Figure (Fig. \ref{fig:a_theta_trajectories}) shows the construction of these trajectories.

\newcommand{\AThetaTrajectories}[5]{
\foreach \tstep in {1,...,#1}
    {
        \foreach \xstep in {0,...,#4}
        {
            \draw[->] (0,1) -- (\tstep * #2, #3 + \xstep * #5);
        }
    }
}

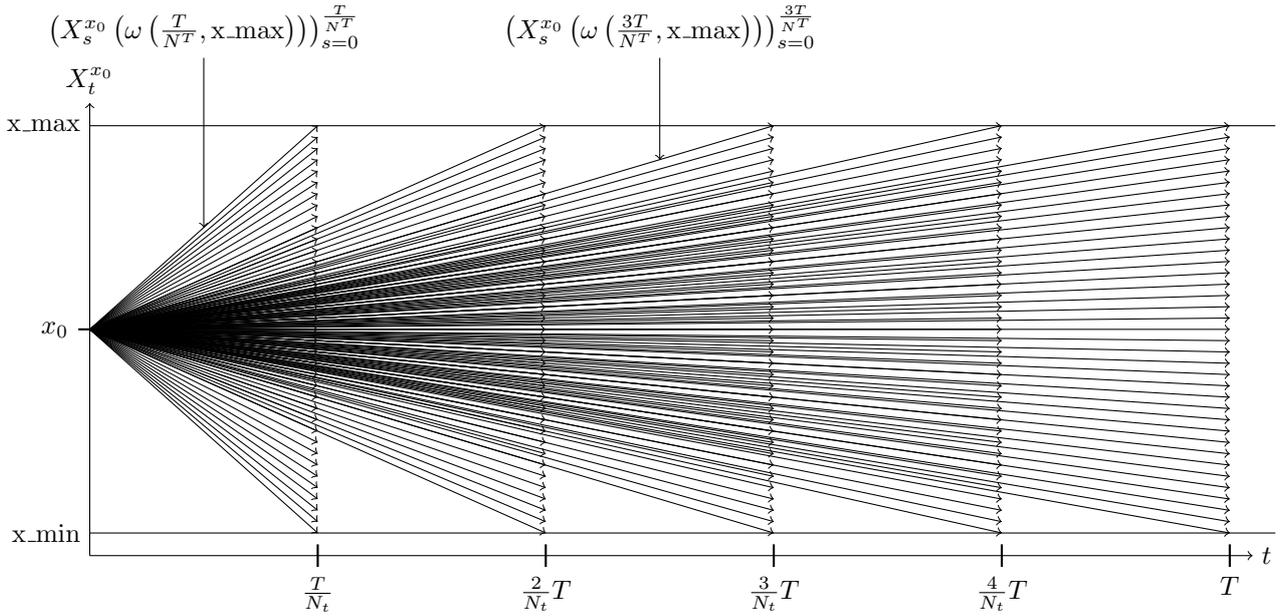
\begin{figure}[H]
\begin{tikzpicture}[scale=3.]
\draw[->] (0, 0) -- (5.1, 0) node[right] {$t$};
\draw[->] (0, 0) -- (0, 2) node[above] {$X_t^{x_0}$};
\draw[thick] (-0.05, 1.) node[left]{$x_0$} -- (0.05, 1);
\draw[thick] (1., -0.05) node[below]{$\frac{T}{N_t}$} -- (1., 0.05);
\draw[thick] (2., -0.05) node[below]{$\frac{2}{N_t}T$} -- (2., 0.05);
\draw[thick] (3., -0.05) node[below]{$\frac{3}{N_t}T$} -- (3., 0.05);
\draw[thick] (4., -0.05) node[below]{$\frac{4}{N_t}T$} -- (4., 0.05);
\draw[thick] (5., -0.05) node[below]{$T$} -- (5., 0.05);
\draw[-] (0, 0.1) node[left]{x\_min} -- (5.2, 0.1);
\draw[-] (0, 1.9) node[left]{x\_max} -- (5.2, 1.9);
\draw[->] (0.5, 2.2) node[above]{$\left( X_s^{x_0} \left( \omega\left( \frac{T}{N^T}, \textnormal{x\_max} \right) \right) \right)_{s=0}^{\frac{T}{N^T}}$} -- (0.5, 1.45);
\draw[->] (2.5, 2.2) node[above]{$\left( X_s^{x_0} \left( \omega\left( \frac{3T}{N^T}, \textnormal{x\_max} \right) \right) \right)_{s=0}^{\frac{3T}{N^T}}$} -- (2.5, 1.75);
\AThetaTrajectories{5}{1.}{0.1}{36}{0.05};
\end{tikzpicture}
\caption{Construction of Trajectories $\left(X_s^{x_0}\left(\omega(t, x)\right)\right)_{s=0}^{t}$}
\label{fig:a_theta_trajectories}
\end{figure}

\subsection{A Simplified Version of the Algorithm with a Local $a$}

In the preceding sections, we have considered a very general change of measure $a^\theta$ taking the whole trajectory of $\left(X^{x_0}_{0 \leq s \leq t}\right)$ as an input. This function being of high dimension, it is difficult to represent it visually. Therefore, one might wonder if a local version of $a^\theta$, taking only the last value of the underlying process at time $t$, $X_t$, as an input, might be sufficient in practice.

Let us therefore consider a function $a^{L, \theta}: [0, T] \times \mathbb{R} \rightarrow \mathbb{R}$. The theory of section (Sect \ref{sect:girsanov_importance_sampling}) still applies. We can then do as in section (Sect. \ref{sect:construction_neural_networks}), and use a list of neural networks to represent $a^{L, \theta}(., .)$ on $\{t_0, ..., t_{N^T - 1}\} \times \mathbb{R}$. Let us introduce the list of neural networks: $a^{L, \theta_i}: \mathbb{R} \rightarrow \mathbb{R}$ for $i \in [\![0, N^T-1]\!]$. We will see in sections (Sect. \ref{sect:results_bachelier}) and (Sect. \ref{sect:results_lv}) that this local version often works nearly as well as the previous full version. Therefore, a practitioner wanting a better interpretability of the method might prefer to restrict himself to this version. We expect this local version to work as well as the full version for European payoffs, but not for fully path-dependent payoffs. 

We define $\overline Z^{L, \theta}$, $\theta^{L, *}$ and $(W_t^{\theta, L})_{0 \leq t \leq T}$ as their counterparts $\overline Z^{\theta}$, $\theta^{*}$ and $W_t^{L}$ except for the fact that we now use the local version $a^{\theta, L}$ instead of $a^\theta$ for the drift in the Girsanov change of measure. 

From now on, we will refer to the algorithm using the full function of the path space $a^{\theta}$ à the ``full'' method, and the one using the local $a^{L, \theta}$ as the ``local'' method.

\subsection{Results for a Bachelier Diffusion} \label{sect:results_bachelier}

In all the following graphs, the full lines and dotted lines represent on the left axis the standard deviations obtained when pricing with a plain Monte Carlo and a Deep Importance Sampling Monte Carlo for different values of the spot price $x_0$ \footnote{For each value of the spot price $x_0$, we train a separate set of neural networks. This will not be the case in the section \ref{sec:robust}, where the neural networks will be trained only with $x_0=1$, and we will see the results obtained via a plain and a Deep Importance Sampling Monte Carlo when changing the different parameters, without retraining the neural networks}. The dashed lines (right axis) show the ratios of these standard deviations. The graphs on the left use the "full" version of the Deep Importance Sampling algorithm, whereas graphs on the right use the "local" version of the Deep Importance Sampling algorithm.

\subsubsection{Call}

\begin{figure}[H]
  \begin{minipage}[t]{0.48\linewidth}
    \includegraphics[width=\linewidth]{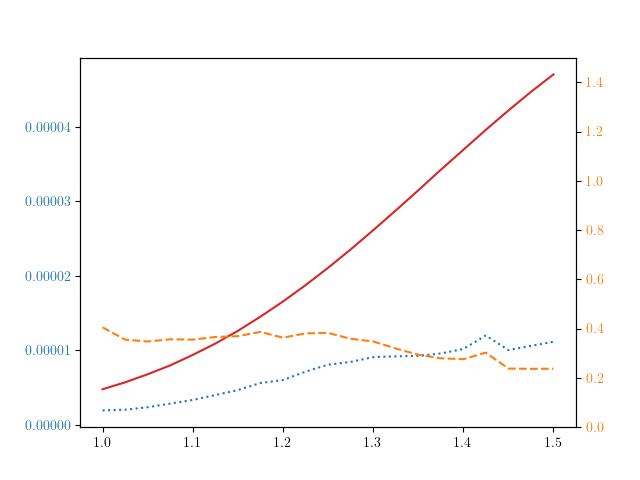}
    \caption{Standard Deviation vs $x_0$ for Full Method}
    \label{fig:std_adaptative_vs_mc_call_bachelier_a_full}
  \end{minipage}
  \begin{minipage}[t]{0.04\linewidth}
  \hfill
  \end{minipage}
  \begin{minipage}[t]{0.48\linewidth}
    \includegraphics[width=\linewidth]{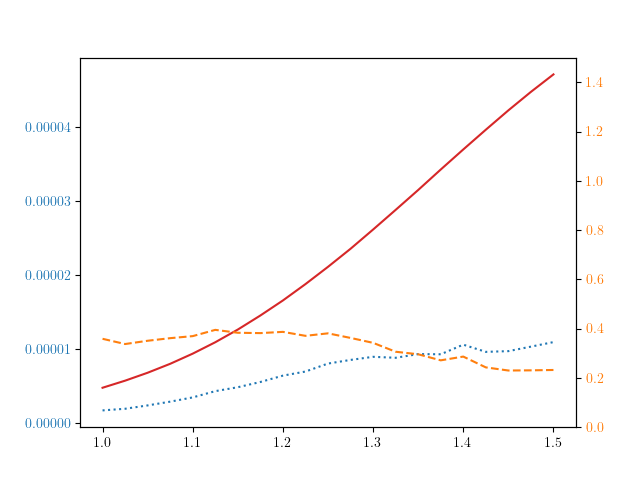}
    \caption{Standard Deviation vs $x_0$ for Local Method}
    \label{fig:std_adaptative_vs_mc_call_bachelier_a_local}
  \end{minipage}
\end{figure}

We see in figures (Fig. \ref{fig:std_adaptative_vs_mc_call_bachelier_a_full}) and (Fig. \ref{fig:std_adaptative_vs_mc_call_bachelier_a_local}) that the Monte Carlo obtained via our adaptative importance sampling has a lower standard deviation than a plain Monte Carlo. 

\begin{figure}[H]
\begin{minipage}[t]{0.48\linewidth}
\includegraphics[width=\linewidth]{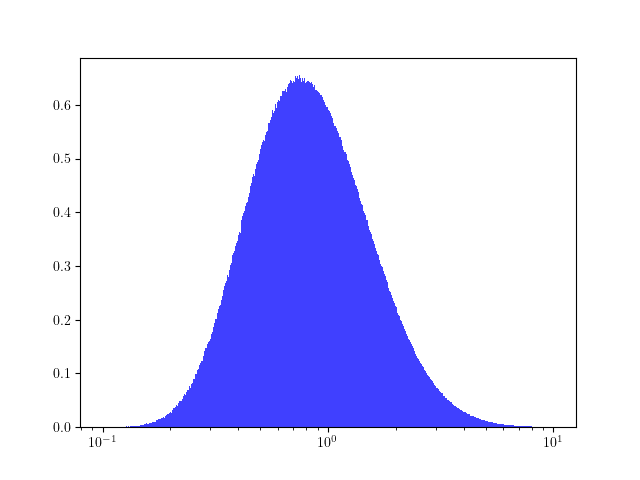}
\caption{$\overline Z_T^{\hat \theta^*}$ Distribution Under $\overline{\mathbb{Q}}^{\hat \theta^*}$}
\label{fig:weight_log_scale_distribution_call_bachelier_a_full}
\end{minipage}
\begin{minipage}[t]{0.04\linewidth}
\hfill
\end{minipage}
\begin{minipage}[t]{0.48\linewidth}
\includegraphics[width=\linewidth]{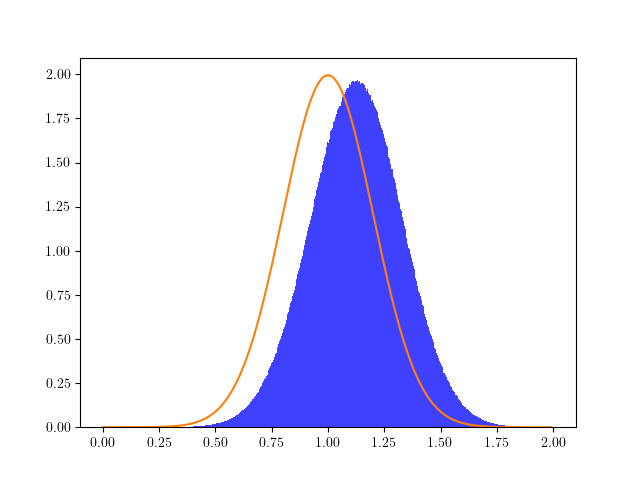}
\caption{$\overline X_T$ Distribution Under $\overline{\mathbb{Q}}^{\hat \theta^*}$ for Full Method}
\label{fig:trajectories_distribution_call_bachelier_a_full}
\end{minipage}
\begin{minipage}[t]{0.48\linewidth}
\includegraphics[width=\linewidth]{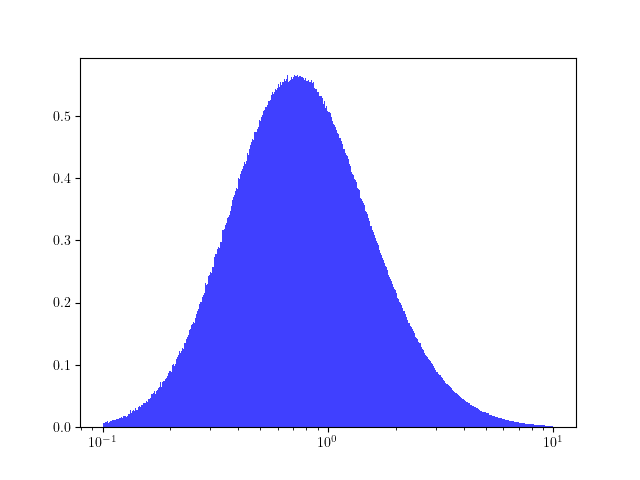}
\caption{$\overline Z_T^{\hat \theta^*}$ Distribution Under $\overline{\mathbb{Q}}^{L, \hat \theta^{L, *}}$}
\label{fig:weight_log_scale_distribution_call_bachelier_a_local}
\end{minipage}
\begin{minipage}[t]{0.04\linewidth}
\hfill
\end{minipage}
\begin{minipage}[t]{0.48\linewidth}
\includegraphics[width=\linewidth]{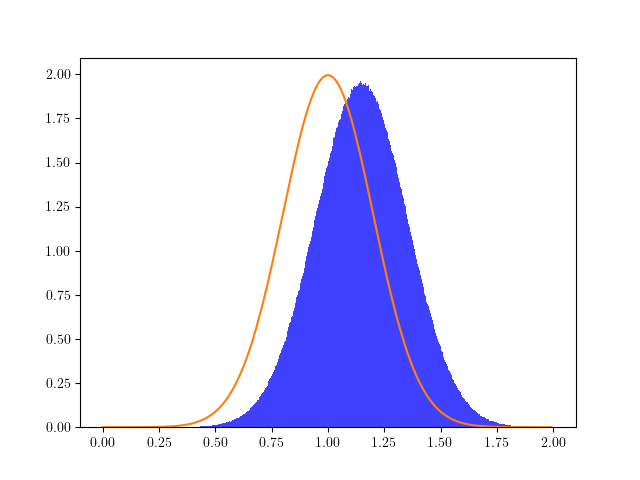}
\caption{$\overline X_T$ Distribution Under $\mathbb{Q}^{\hat \theta^*}$ for Local Method}
\label{fig:trajectories_distribution_call_bachelier_a_local}
\end{minipage}
\end{figure}

To get an idea of how the trajectories are modified, we show in figures (Fig. \ref{fig:weight_log_scale_distribution_call_bachelier_a_full}) and (Fig. \ref{fig:weight_log_scale_distribution_call_bachelier_a_local}) the distributions of the weights $\overline Z_T^{\hat \theta^*}$ and $\overline Z_T^{L, \hat \theta^{L, *}}$ in log scale. Figures (Fig. \ref{fig:trajectories_distribution_call_bachelier_a_full}) and (Fig. \ref{fig:trajectories_distribution_call_bachelier_a_local}) show the distributions of $\overline X_T$ under $\mathbb{Q}^{\hat \theta^*}$ and $\mathbb{Q}^{\hat \theta^{L, *}}$ as histograms, and their theoretical distributions under $\mathbb{Q}$ as a solid line. We can see that the trajectories are modified so as to get closer to the call's strike at $K = 1.4$.

\begin{figure}[H]
\begin{minipage}[t]{0.48\linewidth}
\includegraphics[width=\linewidth]{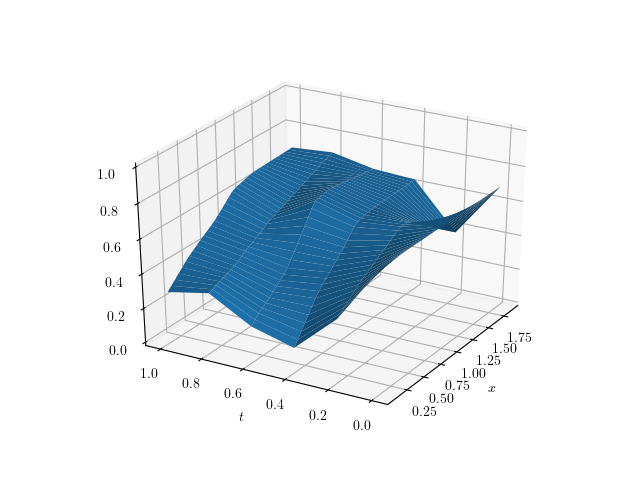}
\caption{$\tilde a^{\hat \theta^*} (t, x)$}
\label{fig:a_theta_call_bachelier_30_210_a_full}
\end{minipage}
\begin{minipage}[t]{0.04\linewidth}
\hfill
\end{minipage}
\begin{minipage}[t]{0.48\linewidth}
\includegraphics[width=\linewidth]{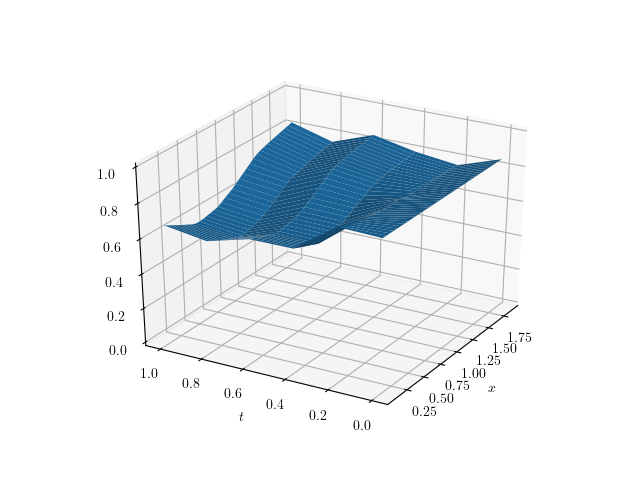}
\caption{$\overline a^{L, \hat \theta^{L, *}} (t, x)$}
\label{fig:a_theta_call_bachelier_30_210_a_local}
\end{minipage}
\end{figure}

In figures (Fig. \ref{fig:a_theta_call_bachelier_30_210_a_full}) and (Fig. \ref{fig:a_theta_call_bachelier_30_210_a_local}), we show the surfaces $\tilde a^{\hat \theta^*}(t,x)$ and $\overline a^{L, \hat \theta^{L, *}}(t,x)$. For the call option, $\tilde a^\theta$ and $\overline a^{L, \theta}$ are always positive. This agrees with the intuition that we need to apply a positive drift in order to have more trajectories reach the strike region of the product. 

\subsubsection{Asymmetric Calls and Puts}

For this experiment, we price $N_1$ calls of strike $K$ and $N_2$ puts of strike $K_2$. The parameters used are those of table (Tab. \ref{tab:asym_call_&_put_options_params}), where $N_1, N_2, K$ and $K_2$ have been chosen so that the $N_1$ calls and $N_2$ puts have roughly the same price. 

\begin{figure}[H]
  \begin{minipage}[t]{0.48\linewidth}
    \includegraphics[width=\linewidth]{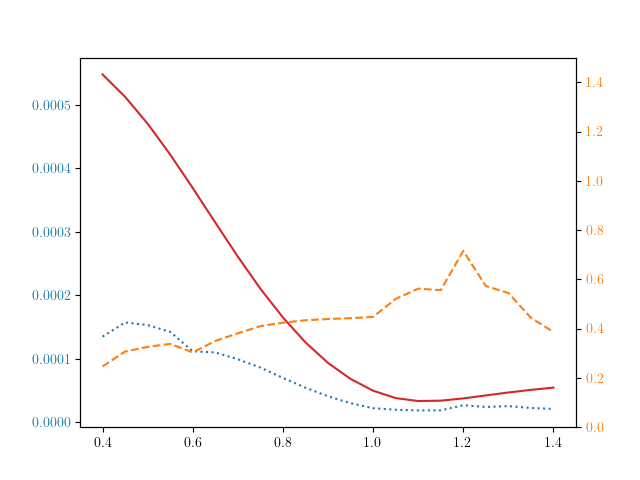}
    \caption{Standard Deviation vs $x_0$ for Full Method}
    \label{fig:std_adaptative_vs_mc_calls_and_puts_asym_bachelier_a_full}
  \end{minipage}
  \begin{minipage}[t]{0.04\linewidth}
  \hfill
  \end{minipage}
  \begin{minipage}[t]{0.48\linewidth}
    \includegraphics[width=\linewidth]{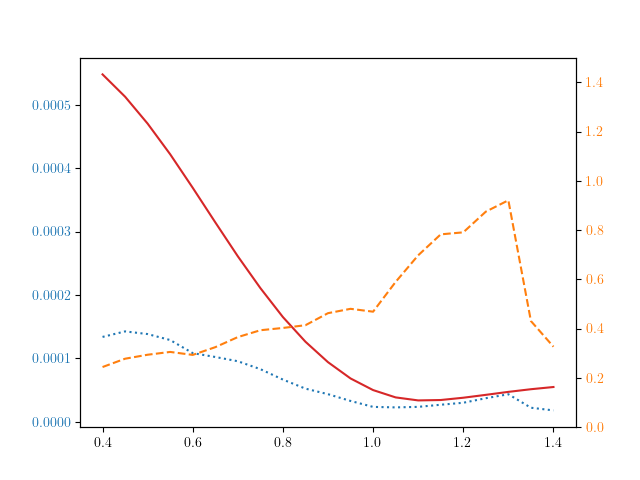}
    \caption{Standard Deviation vs $x_0$ for Local Method}
    \label{fig:std_adaptative_vs_mc_calls_and_puts_asym_bachelier_a_local}
  \end{minipage}
\end{figure}

Again, we see in figures (Fig. \ref{fig:std_adaptative_vs_mc_calls_and_puts_asym_bachelier_a_full}) and (Fig. \ref{fig:std_adaptative_vs_mc_calls_and_puts_asym_bachelier_a_local}) that the Monte Carlo obtained using our adaptative importance sampling has a lower standard deviation than a plain Monte Carlo. 

\begin{figure}[H]
\begin{minipage}[t]{0.48\linewidth}
\includegraphics[width=\linewidth]{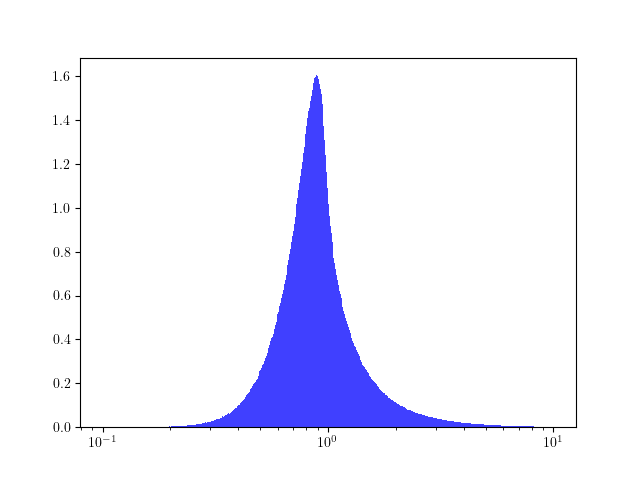}
\caption{$\overline Z_T^{\hat \theta^{*}}$ Distribution Under $\overline{\mathbb{Q}}^{\hat \theta^*}$}
\label{fig:weight_log_scale_distribution_calls_and_puts_asym_bachelier_a_full}
\end{minipage}
\begin{minipage}[t]{0.04\linewidth}
\hfill
\end{minipage}
\begin{minipage}[t]{0.48\linewidth}
\includegraphics[width=\linewidth]{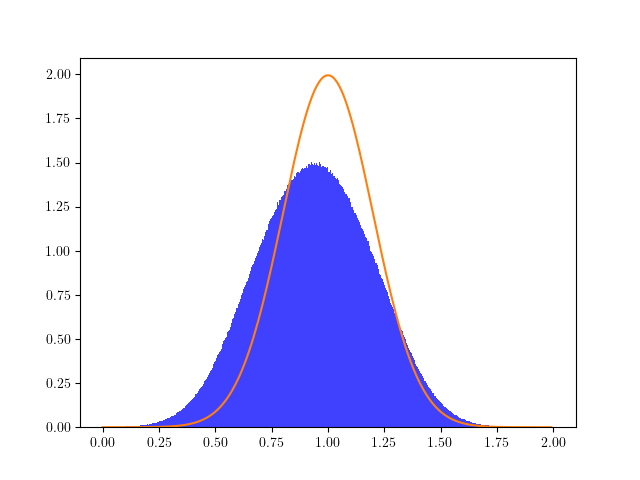}
\caption{$\overline X_T$ Distribution Under $\overline{\mathbb{Q}}^{\hat \theta^*}$}
\label{fig:trajectories_distribution_calls_and_puts_asym_bachelier_a_full}
\end{minipage}
\begin{minipage}[t]{0.48\linewidth}
\includegraphics[width=\linewidth]{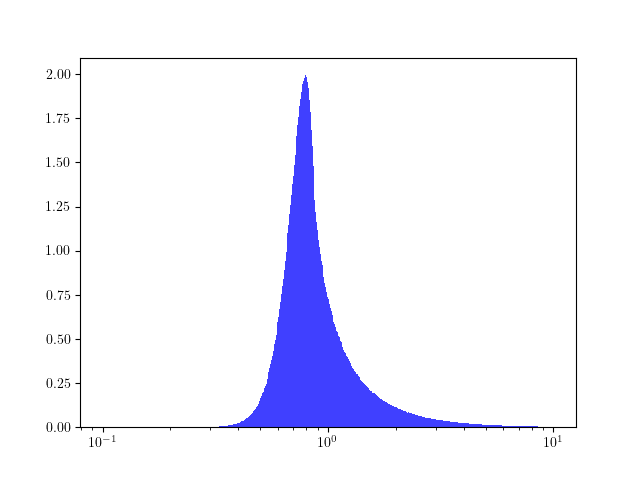}
\caption{$\overline Z_T^{L, \hat \theta^{L, *}}$ Distribution Under $\overline{\mathbb{Q}}^{L, \hat \theta^{L, *}}$}
\label{fig:weight_log_scale_distribution_calls_and_puts_asym_bachelier_a_local}
\end{minipage}
\begin{minipage}[t]{0.04\linewidth}
\hfill
\end{minipage}
\begin{minipage}[t]{0.48\linewidth}
\includegraphics[width=\linewidth]{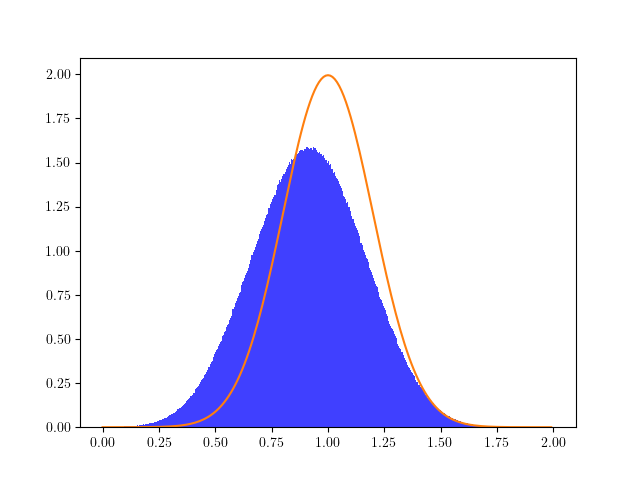}
\caption{$\overline X_T$ Distribution Under $\overline{\mathbb{Q}}^{L, \hat \theta^{L, *}}$}
\label{fig:trajectories_distribution_calls_and_puts_asym_bachelier_a_local}
\end{minipage}
\end{figure}

To get an idea of how the trajectories are modified, we again show in figures (Fig. \ref{fig:weight_log_scale_distribution_calls_and_puts_asym_bachelier_a_full}) and (Fig. \ref{fig:weight_log_scale_distribution_calls_and_puts_asym_bachelier_a_local}) the distributions of the weights $\overline Z_T^{\hat \theta^*}$ and $\overline Z_T^{L, \hat \theta^{L, *}}$ under $\overline{\mathbb{Q}}^{\hat \theta^*}$ and $\overline{\mathbb{Q}}^{L, \hat \theta^{L, *}}$. Figures (Fig. \ref{fig:trajectories_distribution_calls_and_puts_asym_bachelier_a_full}) and (Fig. \ref{fig:trajectories_distribution_calls_and_puts_asym_bachelier_a_local}) show the distributions of $\overline X_T$ under $\overline{\mathbb{Q}}^{\hat \theta^*}$ and $\overline{\mathbb{Q}}^{L, \hat \theta^{L, *}}$. In figures (Fig. \ref{fig:trajectories_distribution_calls_and_puts_asym_bachelier_a_full}) and (Fig. \ref{fig:trajectories_distribution_calls_and_puts_asym_bachelier_a_full}), we can see that the mode of the distributions are lower than 1, which shows that many trajectories are now sent downwards.

\begin{figure}[H]
\begin{minipage}[t]{0.48\linewidth}
\includegraphics[width=\linewidth]{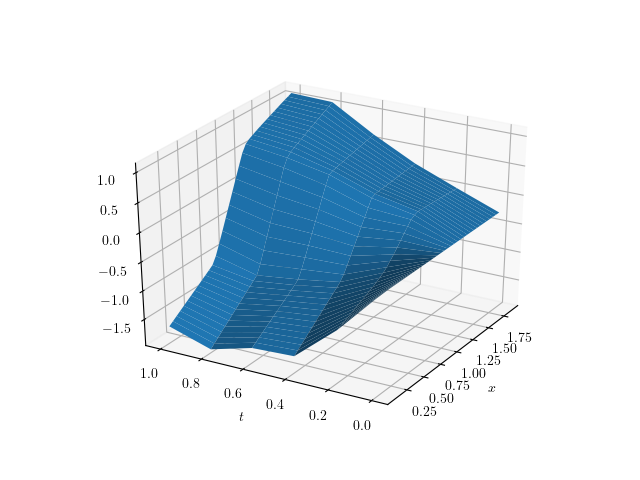}
\caption{$\tilde a^{\hat \theta^*} (t, x)$}
\label{fig:a_theta_calls_and_puts_asym_bachelier_30_210_a_full}
\end{minipage}
\begin{minipage}[t]{0.04\linewidth}
\hfill
\end{minipage}
\begin{minipage}[t]{0.48\linewidth}
\includegraphics[width=\linewidth]{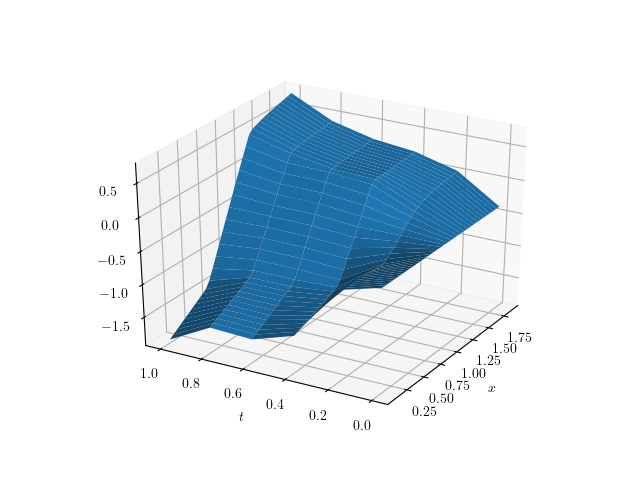}
\caption{$\overline a^{L, \hat \theta^{L, *}} (t, x)$}
\label{fig:a_theta_calls_and_puts_asym_bachelier_30_210_a_local}
\end{minipage}
\begin{minipage}[t]{0.48\linewidth}
\includegraphics[width=\linewidth]{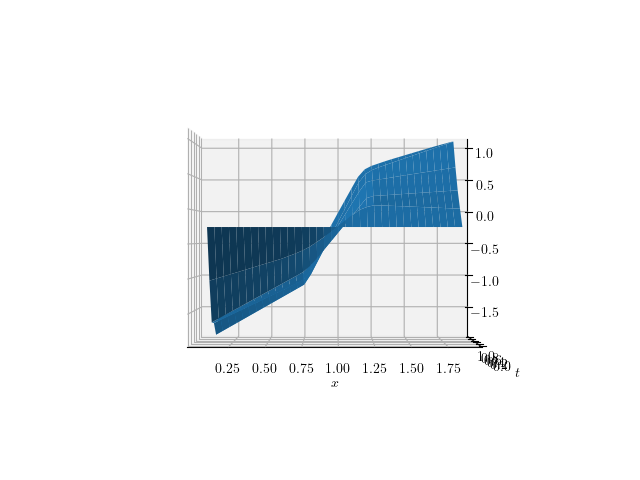}
\caption{$\tilde a^{\hat \theta^*} (t, x)$}
\label{fig:a_theta_calls_and_puts_asym_bachelier_0_270_a_full}
\end{minipage}
\begin{minipage}[t]{0.04\linewidth}
\hfill
\end{minipage}
\begin{minipage}[t]{0.48\linewidth}
\includegraphics[width=\linewidth]{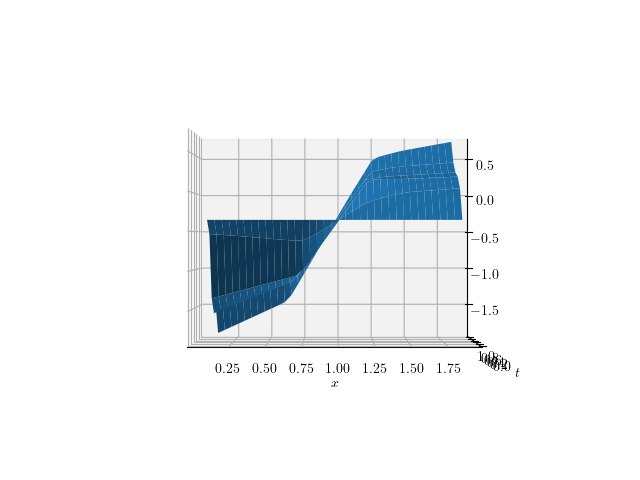}
\caption{$\overline a^{L, \hat \theta^{L, *}} (t, x)$}
\label{fig:a_theta_calls_and_puts_asym_bachelier_0_270_a_local}
\end{minipage}
\end{figure}

We also show in figures (Fig. \ref{fig:a_theta_calls_and_puts_asym_bachelier_30_210_a_full}), (Fig. \ref{fig:a_theta_calls_and_puts_asym_bachelier_0_270_a_full}) and (Fig. \ref{fig:a_theta_calls_and_puts_asym_bachelier_30_210_a_local}), (Fig. \ref{fig:a_theta_calls_and_puts_asym_bachelier_0_270_a_local}) the surfaces $\tilde a^{\hat \theta^*}(t,x)$ and $\overline a^{L, \hat \theta^{L, *}}$ from different viewpoints. Notice that now, $\tilde a^{\theta^*}$ and $\overline a^{L, \theta^{L, *}}$ are now positive roughly speaking when $x > 1$. , and negative when $x < 1$. This is expected, as the payoff now has two strikes, one for the call options, and one for the put options, so it needs to separate the trajectories in two groups. One group goes up to get closer to the call strike $K=1.2$. Another group goes down to join the more extreme put strike $K_2 = 0.6$. Our option consists of 1 call of strike $K = 1.2$, and 10 puts of strike $0.6$. The ratio of $10$ has been chosen such that the price of the call is roughly the same as the price of the $10$ puts. However, we can notice that the surface is asymmetric: the part of the surface that goes down roughly reaches the values around $2$, whereas the positive part of the surface reaches around $1$. This is because the trajectories that go down need to reach a more extreme strike of 0.6, whereas trajectories that go up only need to reach a strike of 1.2.

\subsubsection{Symmetric Calls and Puts}

For this experiment, we again price $N_1$ calls of strike $K_1$ and $N_2$ puts of strike $K_2$. However, the parameters used are now those of table (Tab. \ref{tab:sym_call_&_put_options_params}), and have been chosen so that the situation is perfectly symmetric: $K_1 = 0.6$, $K_2=1.4$, $N_1 = 10$ and $N_2 = 10$. As the diffusion is of Bachelier type, its distribution is also symmetric with respect to $x=1$. We therefore expect $a^\theta$ to have a symmetry with respect to $x=1$. 

\begin{figure}[H]
  \begin{minipage}[t]{0.48\linewidth}
    \includegraphics[width=\linewidth]{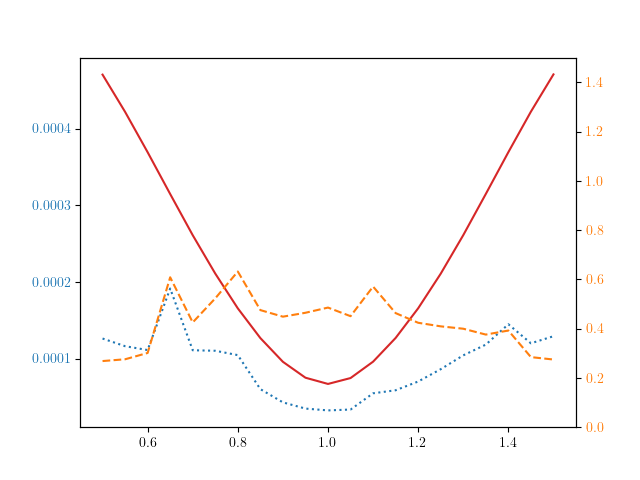}
    \caption{Standard Deviation vs $x_0$ for Full Method}
    \label{fig:std_adaptative_vs_mc_calls_and_puts_sym_bachelier_a_full}
  \end{minipage}
  \begin{minipage}[t]{0.04\linewidth}
  \hfill
  \end{minipage}
  \begin{minipage}[t]{0.48\linewidth}
    \includegraphics[width=\linewidth]{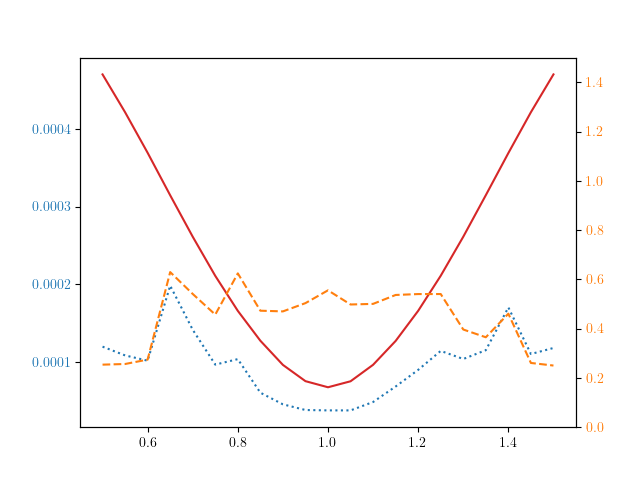}
    \caption{Standard Deviation vs $x_0$ for Local Method}
    \label{fig:std_adaptative_vs_mc_calls_and_puts_sym_bachelier_a_local}
  \end{minipage}
\end{figure}

Again, we see in figures (Fig. \ref{fig:std_adaptative_vs_mc_calls_and_puts_sym_bachelier_a_full}) and (Fig. \ref{fig:std_adaptative_vs_mc_calls_and_puts_sym_bachelier_a_local}) that the Monte Carlo obtained using our adaptative importance sampling has a lower standard deviation than a plain Monte Carlo. 

\begin{figure}[H]
\begin{minipage}[t]{0.48\linewidth}
\includegraphics[width=\linewidth]{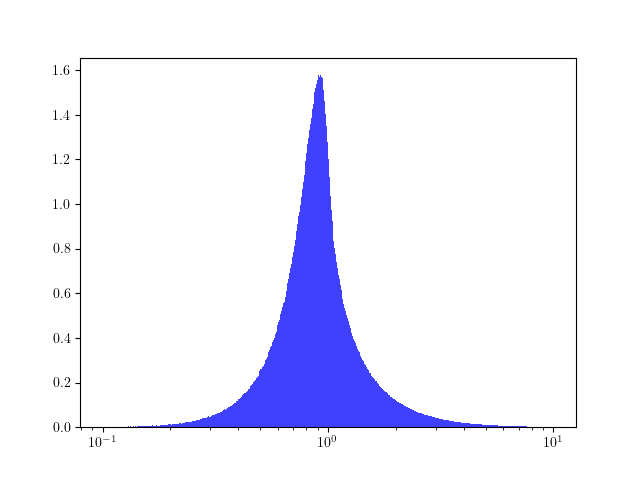}
\caption{$\overline Z_T^{\hat \theta^*}$ Distribution Under $\overline{\mathbb{Q}}^{\hat \theta^*}$}
\label{fig:weight_log_scale_distribution_calls_and_puts_sym_bachelier_a_full}
\end{minipage}
\begin{minipage}[t]{0.04\linewidth}
\hfill
\end{minipage}
\begin{minipage}[t]{0.48\linewidth}
\includegraphics[width=\linewidth]{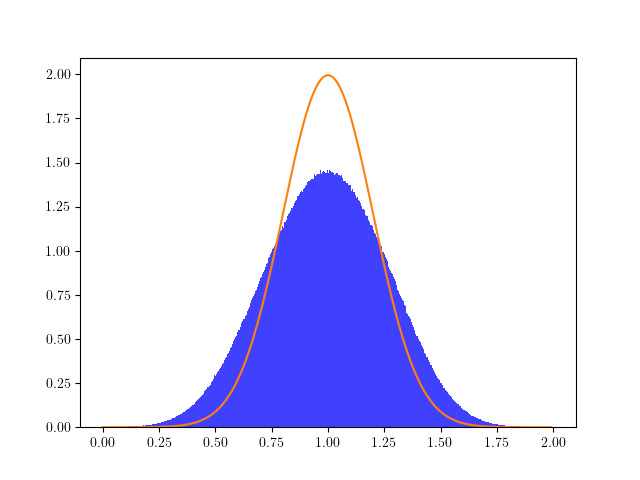}
\caption{$\overline X_T$ Distribution Under $\overline{\mathbb{Q}}^{\hat \theta^*}$}
\label{fig:trajectories_distribution_calls_and_puts_sym_bachelier_a_full}
\end{minipage}
\begin{minipage}[t]{0.48\linewidth}
\includegraphics[width=\linewidth]{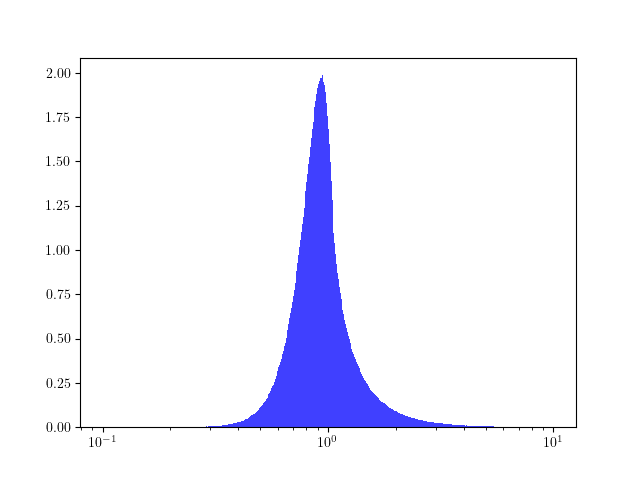}
\caption{$\overline Z_T^{L, \hat \theta^{L, *}}$ Distribution Under $\overline{\mathbb{Q}}^{L, \hat \theta^{L, *}}$}
\label{fig:weight_log_scale_distribution_calls_and_puts_sym_bachelier_a_local}
\end{minipage}
\begin{minipage}[t]{0.04\linewidth}
\hfill
\end{minipage}
\begin{minipage}[t]{0.48\linewidth}
\includegraphics[width=\linewidth]{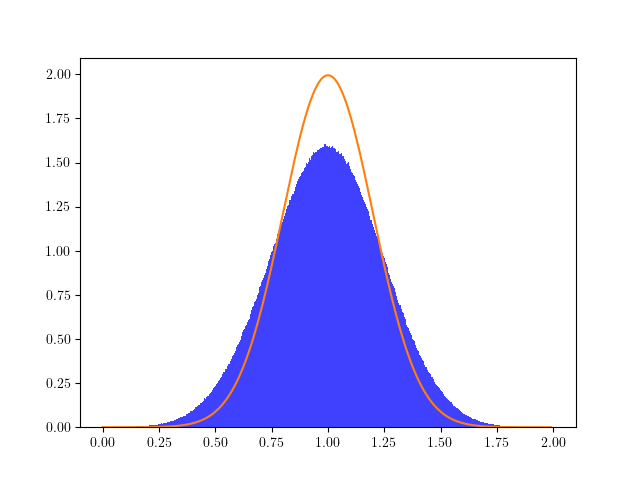}
\caption{$\overline X_T$ Distribution Under $\overline{\mathbb{Q}}^{L, \hat \theta^{L, *}}$}
\label{fig:trajectories_distribution_calls_and_puts_sym_bachelier_a_local}
\end{minipage}
\end{figure}

As previously, we show in figures (Fig. \ref{fig:weight_log_scale_distribution_calls_and_puts_sym_bachelier_a_full}) and (Fig. \ref{fig:weight_log_scale_distribution_calls_and_puts_sym_bachelier_a_local}) the distribution of the weights $\overline Z_T^{\hat \theta^*}$ and $\overline Z_T^{L, \hat \theta^{L, *}}$ in log scales and in figures (Fig. \ref{fig:trajectories_distribution_calls_and_puts_sym_bachelier_a_full}) and (Fig. \ref{fig:trajectories_distribution_calls_and_puts_sym_bachelier_a_local}) the distribution of $\overline X_T$ under $\mathbb{Q}^{\hat \theta^*}$ and $\mathbb{Q}^{L, \hat \theta^{L, *}}$. The mode of the distributions are now at 1, which is expected from the symmetries of both the product and the Bachelier process law with respect to $x=1$.

\begin{figure}[H]
\begin{minipage}[t]{0.48\linewidth}
\includegraphics[width=\linewidth]{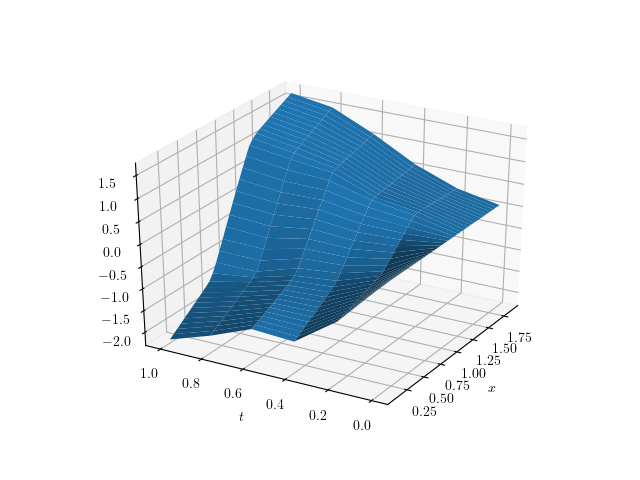}
\caption{$\tilde a^{\hat \theta^*} (t, x)$}
\label{fig:a_theta_calls_and_puts_sym_bachelier_30_210_a_full}
\end{minipage}
\begin{minipage}[t]{0.04\linewidth}
\hfill
\end{minipage}
\begin{minipage}[t]{0.48\linewidth}
\includegraphics[width=\linewidth]{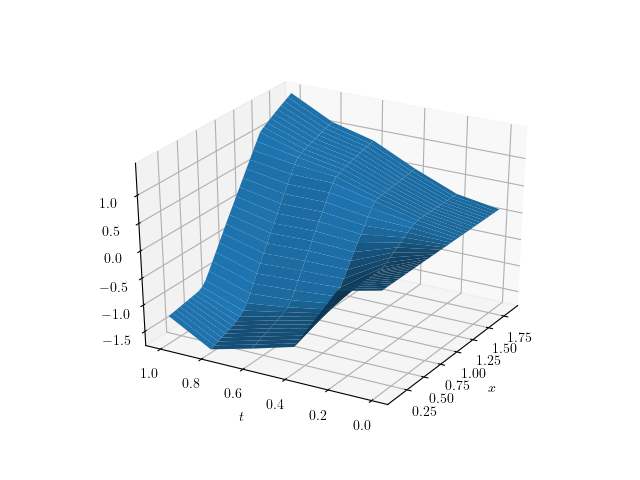}
\caption{$\overline a^{L, \hat \theta^{L, *}} (t, x)$}
\label{fig:a_theta_calls_and_puts_sym_bachelier_30_210_a_local}
\end{minipage}
\begin{minipage}[t]{0.48\linewidth}
\centering
\includegraphics[width=\linewidth]{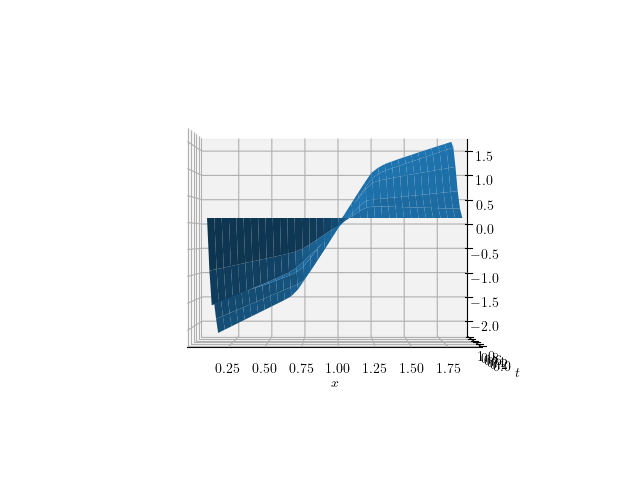}
\caption{$\tilde a^{\hat \theta^*} (t, x)$}
\label{fig:a_theta_calls_and_puts_sym_bachelier_30_270_a_full}
\end{minipage}
\begin{minipage}[t]{0.04\linewidth}
\end{minipage}
\begin{minipage}[t]{0.48\linewidth}
\centering
\includegraphics[width=\linewidth]{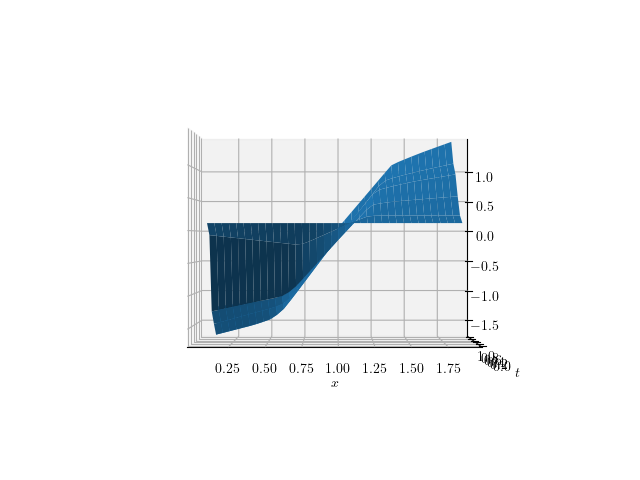}
\caption{$\overline a^{L, \hat \theta^{L, *}} (t, x)$}
\label{fig:a_theta_calls_and_puts_sym_bachelier_30_270_a_local}
\end{minipage}
\end{figure}

As previously, we show in figures (Fig. \ref{fig:a_theta_calls_and_puts_sym_bachelier_30_210_a_full}), (Fig. \ref{fig:a_theta_calls_and_puts_sym_bachelier_30_210_a_local}) and (Fig. \ref{fig:a_theta_calls_and_puts_sym_bachelier_30_270_a_full}) (Fig. \ref{fig:a_theta_calls_and_puts_sym_bachelier_30_270_a_local}) the surfaces $\tilde a^{\hat \theta^*}(t,x)$ and $\overline a^{L, \hat \theta^{L, *}}$ from different viewpoints. Results are similar to those obtained with the asymmetric calls and puts, except that we can now notice in figures (Fig. \ref{fig:a_theta_calls_and_puts_sym_bachelier_30_270_a_full}) and (Fig. \ref{fig:a_theta_calls_and_puts_sym_bachelier_30_270_a_local}), that $\tilde a^{\hat \theta^*}$ and $\overline a^{L, \hat \theta^{L, *}}$ present a symmetry with respect to $x = 1$, which is expected from the symmetries of the product and the Bachelier process law. 

\subsubsection{Multi Coupons AutoCall}

\begin{figure}[H]
  \begin{minipage}[t]{0.48\linewidth}
    \includegraphics[width=\linewidth]{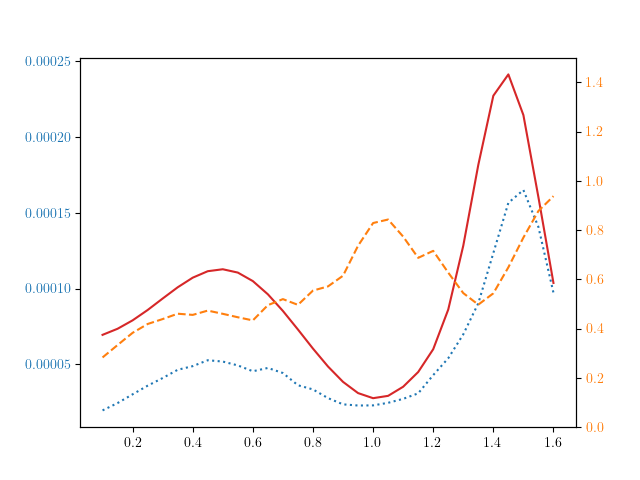}
    \caption{Standard Deviation vs $x_0$ for Full Method}
    \label{fig:std_adaptative_vs_mc_autocall_multi_coupons_bachelier_a_full}
  \end{minipage}
  \begin{minipage}[t]{0.04\linewidth}
  \hfill
  \end{minipage}
  \begin{minipage}[t]{0.48\linewidth}
    \includegraphics[width=\linewidth]{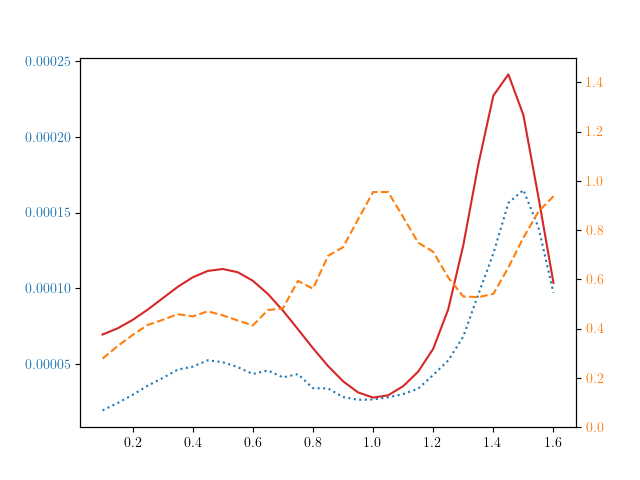}
    \caption{Standard Deviation vs $x_0$ for Local Method}
    \label{fig:std_adaptative_vs_mc_autocall_multi_coupons_bachelier_a_local}
  \end{minipage}
\end{figure}

For this experiment, we price an AutoCall that has multiple coupons. The precise characteristics are those of table (Tab. \ref{tab:multi_coupons_autocall_params}). Again, we see in figures (Fig. \ref{fig:std_adaptative_vs_mc_autocall_multi_coupons_bachelier_a_full}) and (Fig. \ref{fig:std_adaptative_vs_mc_autocall_multi_coupons_bachelier_a_local}) that the Monte Carlo obtained using our adaptative importance sampling has a lower standard deviation than a plain Monte Carlo. 

\begin{figure}[H]
\begin{minipage}[t]{0.48\linewidth}
\includegraphics[width=\linewidth]{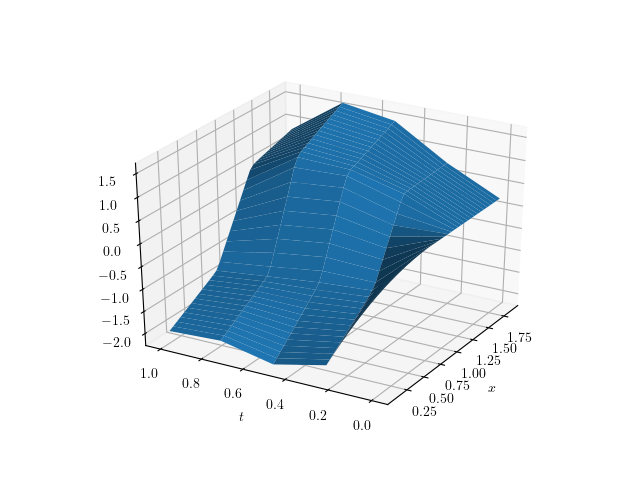}
\caption{$\tilde a^\theta (t, x)$}
\label{fig:a_theta_autocall_multi_coupons_bachelier_30_210_a_full}
\end{minipage}
\begin{minipage}[t]{0.04\linewidth}
\hfill
\end{minipage}
\begin{minipage}[t]{0.48\linewidth}
\includegraphics[width=\linewidth]{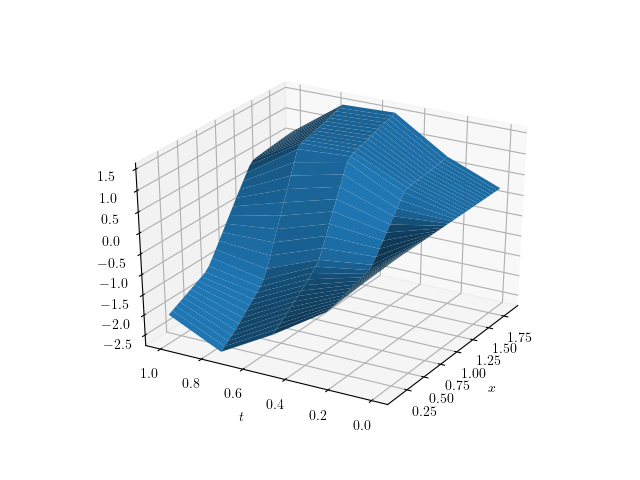}
\caption{$\tilde a^\theta (t, x)$}
\label{fig:a_theta_autocall_multi_coupons_bachelier_30_210_a_local}
\end{minipage}
\begin{minipage}[t]{0.48\linewidth}
\includegraphics[width=\linewidth]{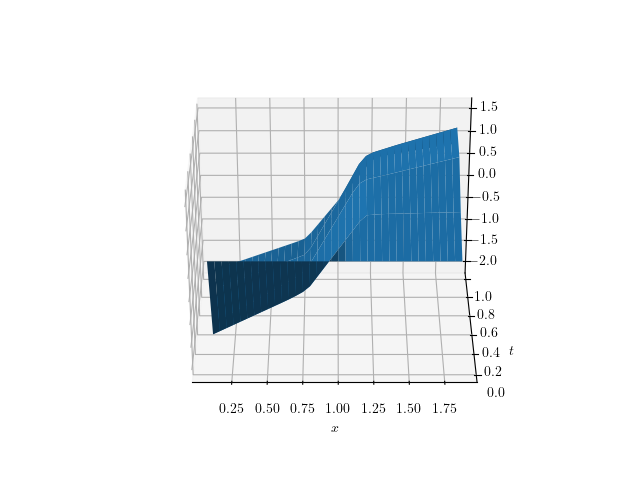}
\caption{$\tilde a^\theta (t, x)$}
\label{fig:a_theta_autocall_multi_coupons_bachelier_30_270_a_full}
\end{minipage}
\begin{minipage}[t]{0.04\linewidth}
\hfill
\end{minipage}
\begin{minipage}[t]{0.48\linewidth}
\includegraphics[width=\linewidth]{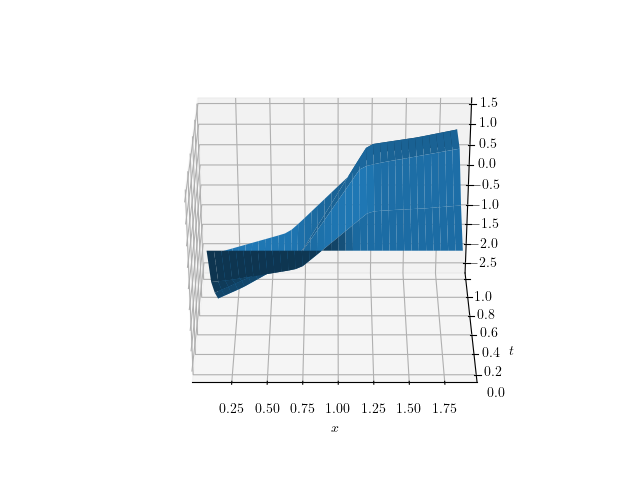}
\caption{$\tilde a^\theta (t, x)$}
\label{fig:a_theta_autocall_multi_coupons_bachelier_30_270_a_local}
\end{minipage}
\end{figure}

In figures (Fig. \ref{fig:a_theta_autocall_multi_coupons_bachelier_30_210_a_full}), (Fig. \ref{fig:a_theta_autocall_multi_coupons_bachelier_30_270_a_full}) and (Fig. \ref{fig:a_theta_autocall_multi_coupons_bachelier_30_210_a_local}), (Fig. \ref{fig:a_theta_autocall_multi_coupons_bachelier_30_270_a_local}), we show the surfaces $\tilde a^{\hat \theta^*}(t,x)$ and $\overline a^{L, \hat \theta^{L, *}}$ from different viewpoints. We can see that that the values are rougly speaking positive when $x > 1$, and negative for $x < 1$. This is again expected from the fact that some trajectories need to get closer to the barrier strikes region $B_.^A = 1.5$, while others need to get close to the put down and in strike region $K^{PDI} = 0.5$. 

\subsubsection{Single Coupon AutoCall}

\begin{figure}[H]
  \begin{minipage}[t]{0.48\linewidth}
    \includegraphics[width=\linewidth]{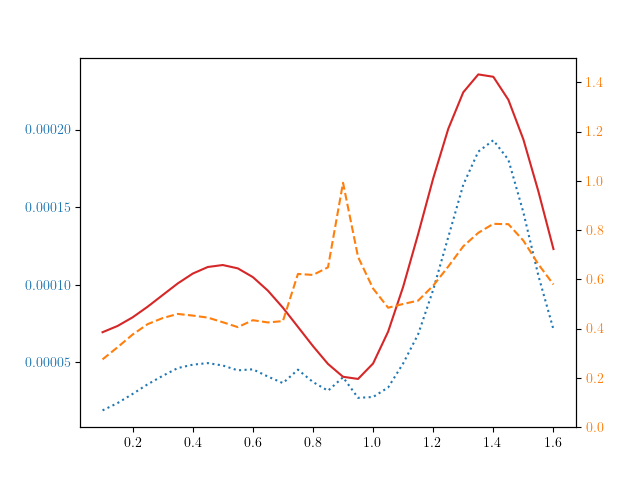}
    \caption{Standard Deviation vs $x_0$ for Full Method}
    \label{fig:std_adaptative_vs_mc_autocall_single_coupon_bachelier_a_full}
  \end{minipage}
  \begin{minipage}[t]{0.04\linewidth}
  \hfill
  \end{minipage}
  \begin{minipage}[t]{0.48\linewidth}
    \includegraphics[width=\linewidth]{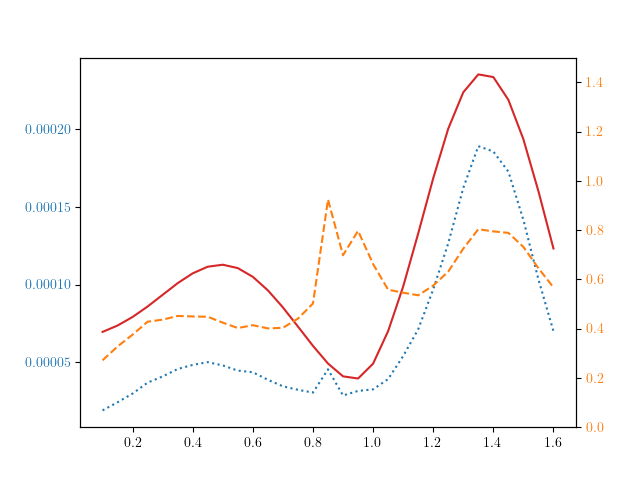}
    \caption{Standard Deviation vs $x_0$ for Local Method}
    \label{fig:std_adaptative_vs_mc_autocall_single_coupon_bachelier_a_local}
  \end{minipage}
\end{figure}

For this experiment, we price an AutoCall that has a single coupon (but still multiple AutoCall barriers). The precise characteristics are those of table (Tab. \ref{tab:single_coupon_autocall_params}). Again, we see in figures (Fig. \ref{fig:std_adaptative_vs_mc_autocall_single_coupon_bachelier_a_full}) and (Fig. \ref{fig:std_adaptative_vs_mc_autocall_single_coupon_bachelier_a_local}) that the Monte Carlo obtained using our adaptative importance sampling has a lower standard deviation than a plain Monte Carlo. 

\begin{figure}[H]
\begin{minipage}[t]{0.48\linewidth}
\includegraphics[width=\linewidth]{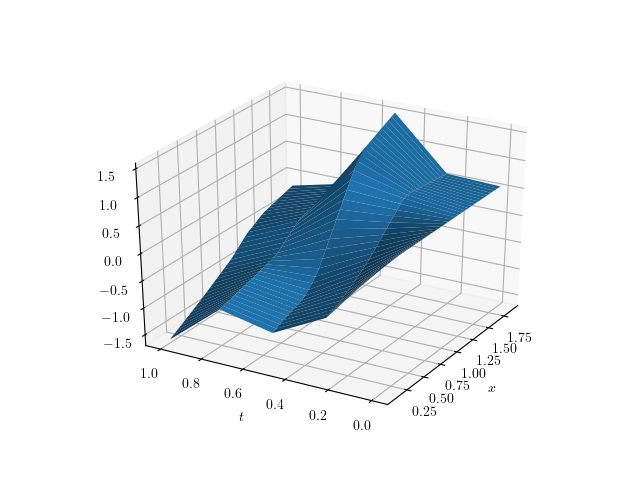}
\caption{$\tilde a^{\hat \theta^*} (t, x)$}
\label{fig:a_theta_autocall_single_coupon_bachelier_30_210_a_full}
\end{minipage}
\begin{minipage}[t]{0.04\linewidth}
\hfill
\end{minipage}
\begin{minipage}[t]{0.48\linewidth}
\includegraphics[width=\linewidth]{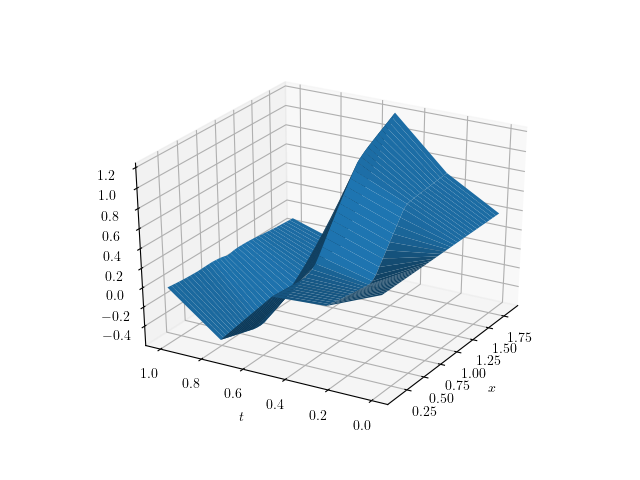}
\caption{$\overline a^{L, \hat \theta^{L, *}} (t, x)$}
\label{fig:a_theta_autocall_single_coupon_bachelier_30_210_a_local}
\end{minipage}
\end{figure}

In figures (Fig. \ref{fig:a_theta_autocall_single_coupon_bachelier_30_210_a_full}) and (Fig. \ref{fig:a_theta_autocall_single_coupon_bachelier_30_210_a_local}), we again show the surfaces $\tilde a^{\hat \theta^*}(t,x)$ and $\overline a^{L, \hat \theta^{L, *}}$ from different viewpoints. Compared to the multiple coupons AutoCall, we can see one striking difference: for $x > 1$, the surface increases from time $t=0$ up to the coupon date $t=T_3^A$, then suddenly drop to values close to 0, and slightly negative. This is expected, as once the coupon date has passed, there is no reason to deviate trajectories towards the $x > 1$. Indeed, once the coupon date has passed, the only thing left to price is the put down and in, so trajectories now need to get to values closer to the put down and in strike $K^{PDI} = 0.5$. 

\subsection{Results for a Local Volatility Diffusion} \label{sect:results_lv}

Results for the local volatility diffusion, which is the most used diffusion by practitioners, are very similar to those obtained with the Bachelier diffusion. We will therefore be more brief in our comments. 

\subsubsection{Call}

\begin{figure}[H]
  \begin{minipage}[t]{0.48\linewidth}
    \includegraphics[width=\linewidth]{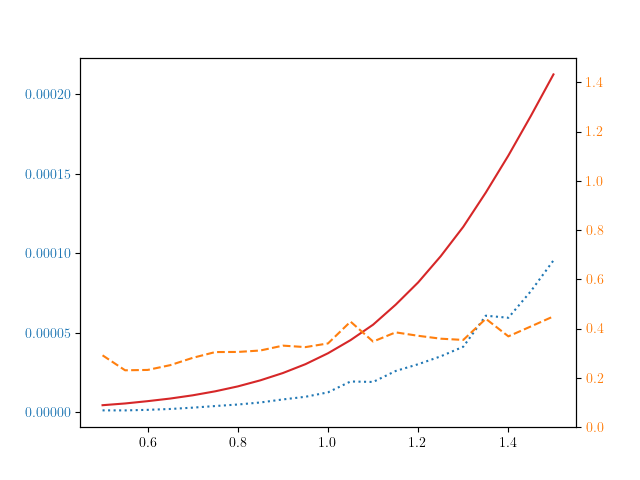}
    \caption{Standard Deviation vs $x_0$ for Full Method}
    \label{fig:std_adaptative_vs_mc_call_lv_a_full}
  \end{minipage}
  \begin{minipage}[t]{0.04\linewidth}
  \hfill
  \end{minipage}
  \begin{minipage}[t]{0.48\linewidth}
    \includegraphics[width=\linewidth]{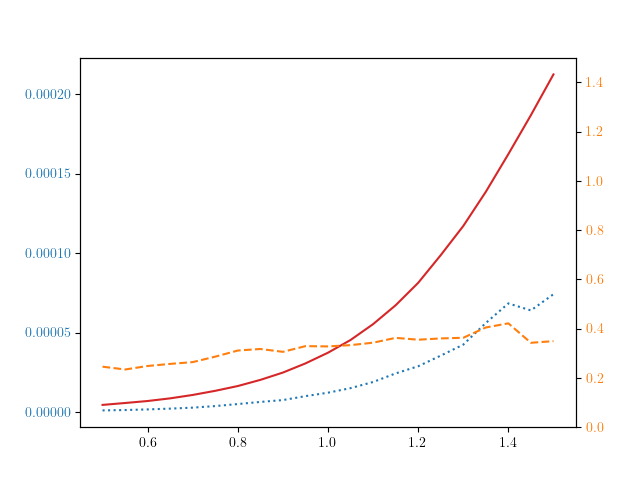}
    \caption{Standard Deviation vs $x_0$ for Local Method}
    \label{fig:std_adaptative_vs_mc_call_lv_a_local}
  \end{minipage}
\end{figure}

As in the previous section, we see in figures (Fig. \ref{fig:std_adaptative_vs_mc_call_lv_a_full}) and (Fig. \ref{fig:std_adaptative_vs_mc_call_lv_a_local}) that the Monte Carlo obtained via our adaptative importance sampling has a lower standard deviation than a plain Monte Carlo. 

\begin{figure}[H]
\begin{minipage}[t]{0.48\linewidth}
\includegraphics[width=\linewidth]{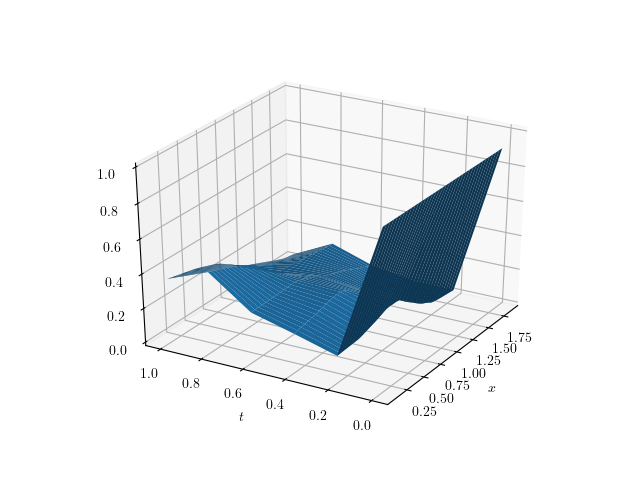}
\caption{$\tilde a^\theta (t, x)$}
\label{fig:a_theta_call_lv_30_210_a_full}
\end{minipage}
\begin{minipage}[t]{0.04\linewidth}
\hfill
\end{minipage}
\begin{minipage}[t]{0.48\linewidth}
\includegraphics[width=\linewidth]{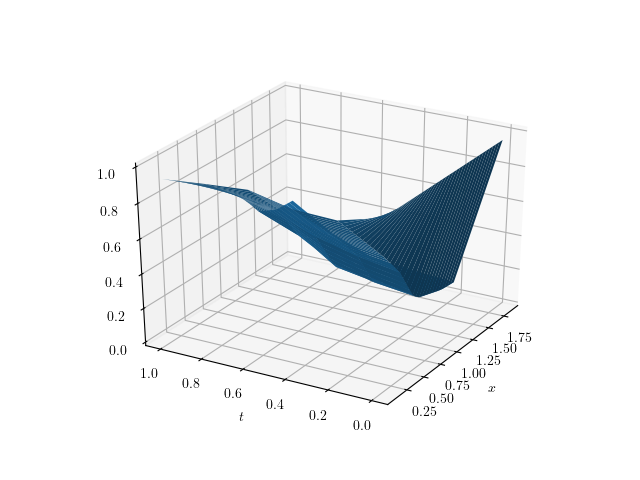}
\caption{$\tilde a^\theta (t, x)$}
\label{fig:a_theta_call_lv_30_210_a_local}
\end{minipage}
\end{figure}

In figures (Fig. \ref{fig:a_theta_call_lv_30_210_a_full}) and (Fig. \ref{fig:a_theta_call_lv_30_210_a_local}), we show the surfaces $\tilde a^{\hat \theta^*}(t,x)$ and $\overline a^{L, \hat \theta^{L, *}}$. As in the Bachelier diffusion case, for the call option, $\tilde a^{\hat \theta^*}$ and $\overline a^{L, \hat \theta^{L, *}}$ are always positive, as expected.

\subsubsection{Asymmetric Calls \& Puts}

\begin{figure}[H]
  \begin{minipage}[t]{0.48\linewidth}
    \includegraphics[width=\linewidth]{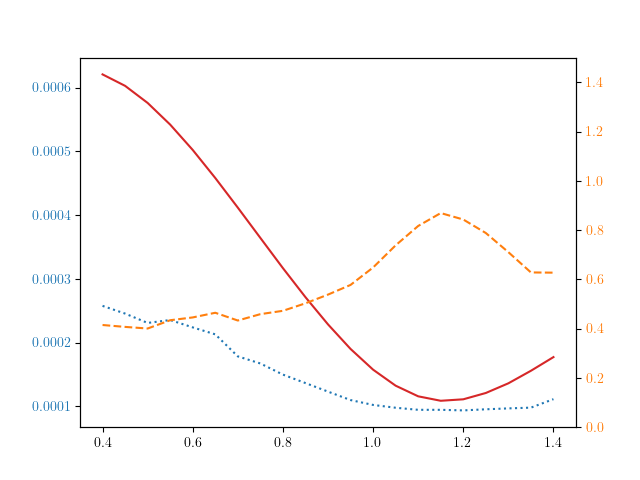}
    \caption{Standard Deviation vs $x_0$ for Full Method}
    \label{fig:std_adaptative_vs_mc_calls_and_puts_asym_lv_a_full}
  \end{minipage}
  \begin{minipage}[t]{0.04\linewidth}
  \hfill
  \end{minipage}
  \begin{minipage}[t]{0.48\linewidth}
    \includegraphics[width=\linewidth]{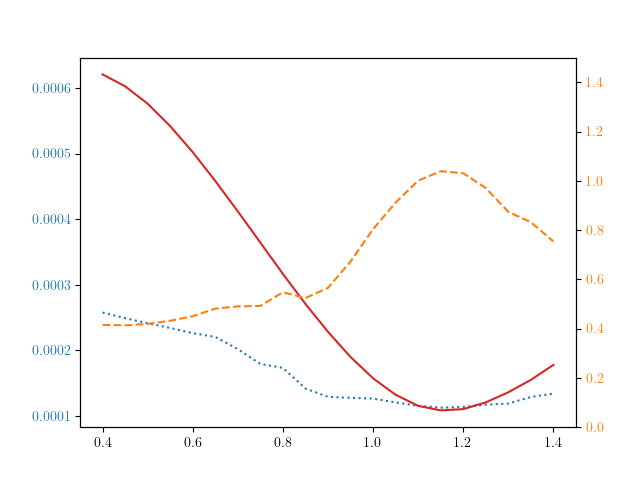}
    \caption{Standard Deviation vs $x_0$ for Local Method}
    \label{fig:std_adaptative_vs_mc_calls_and_puts_asym_lv_a_local}
  \end{minipage}
\end{figure}

For this experiment, we again price $N_1$ calls of strike $K_1$ and $N_2$ puts of strike $K_2$ with the parameters of table (Tab. \ref{tab:asym_call_&_put_options_params}).
As for the Bachelier diffusion case, we see in figures (Fig. \ref{fig:std_adaptative_vs_mc_calls_and_puts_asym_lv_a_full}) and (Fig. \ref{fig:std_adaptative_vs_mc_calls_and_puts_asym_lv_a_local}) that the Monte Carlo obtained using our adaptative importance sampling has a lower standard deviation than a plain Monte Carlo. 

\begin{figure}[H]
\begin{minipage}[t]{0.48\linewidth}
\includegraphics[width=\linewidth]{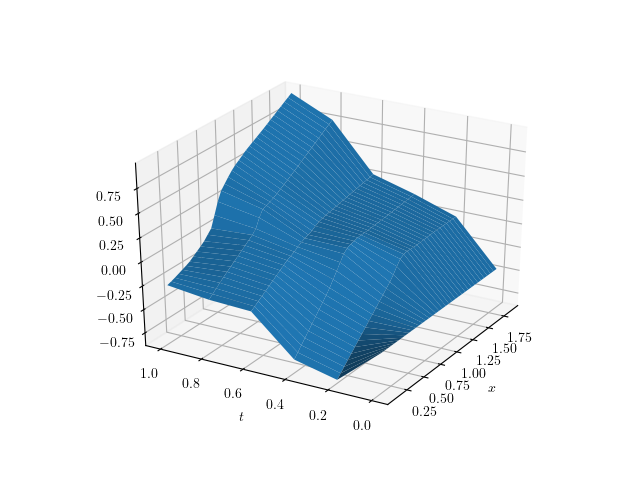}
\caption{$\tilde a^{\hat \theta^*} (t, x)$}
\label{fig:a_theta_calls_and_puts_asym_lv_30_210_a_full}
\end{minipage}
\begin{minipage}[t]{0.04\linewidth}
\hfill
\end{minipage}
\begin{minipage}[t]{0.48\linewidth}
\centering
\includegraphics[width=\linewidth]{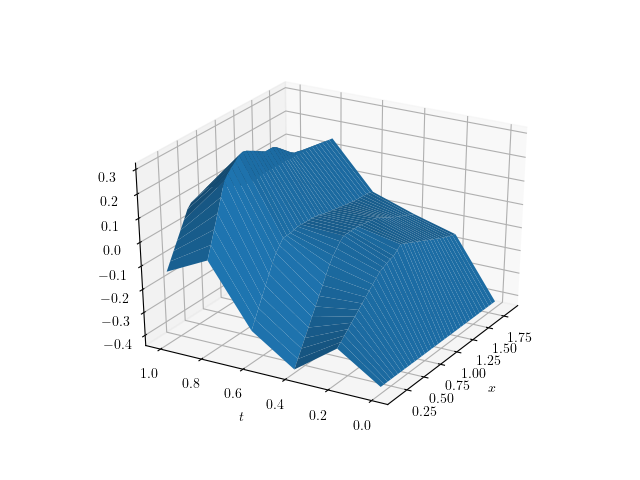}
\caption{$\overline a^{L, \hat \theta^{L, *}} (t, x)$}
\label{fig:a_theta_calls_and_puts_asym_lv_30_210_a_local}
\end{minipage}
\end{figure}

As previously, we show in figures (Fig. \ref{fig:a_theta_calls_and_puts_asym_lv_30_210_a_full}) and (Fig. \ref{fig:a_theta_calls_and_puts_asym_lv_30_210_a_local}) the surfaces $\tilde a^{\hat \theta^*}(t,x)$ and $\overline a^{L, \hat \theta^{L, *}}(t,x)$. 

\subsubsection{Symmetric Calls \& Puts}

For this experiment, we again price $N_1$ calls of strike $K_1$ and $N_2$ puts of strike $K_2$ with the parameters of table (Tab. \ref{tab:sym_call_&_put_options_params}). 

\begin{figure}[H]
  \begin{minipage}[t]{0.48\linewidth}
    \includegraphics[width=\linewidth]{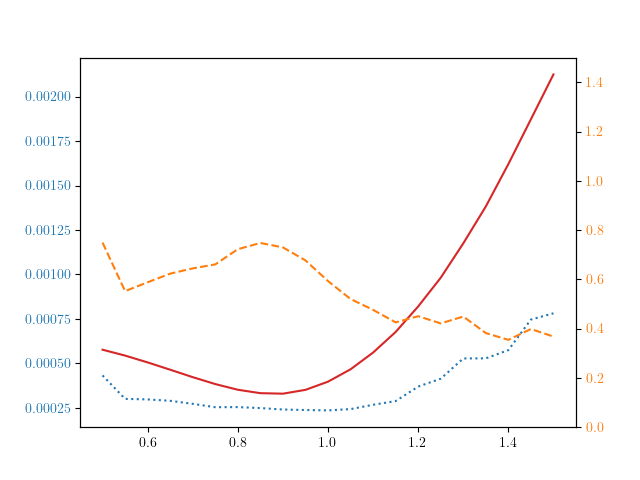}
    \caption{Standard Deviation vs $x_0$ for Full Method}
    \label{fig:std_adaptative_vs_mc_calls_and_puts_sym_lv_a_full}
  \end{minipage}
  \begin{minipage}[t]{0.04\linewidth}
  \hfill
  \end{minipage}
  \begin{minipage}[t]{0.48\linewidth}
    \includegraphics[width=\linewidth]{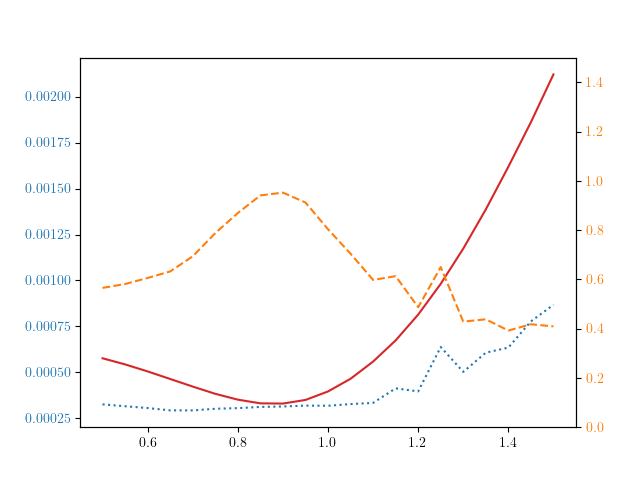}
    \caption{Standard Deviation vs $x_0$ for Local Method}
    \label{fig:std_adaptative_vs_mc_calls_and_puts_sym_lv_a_local}
  \end{minipage}
\end{figure}

Again, we see in figures (Fig. \ref{fig:std_adaptative_vs_mc_calls_and_puts_sym_lv_a_full}) and (Fig. \ref{fig:std_adaptative_vs_mc_calls_and_puts_sym_lv_a_local}) that the Monte Carlo obtained using our adaptative importance sampling has a lower standard deviation than that of a plain Monte Carlo. 

\begin{figure}[H]
\begin{minipage}[t]{0.48\linewidth}
\includegraphics[width=\linewidth]{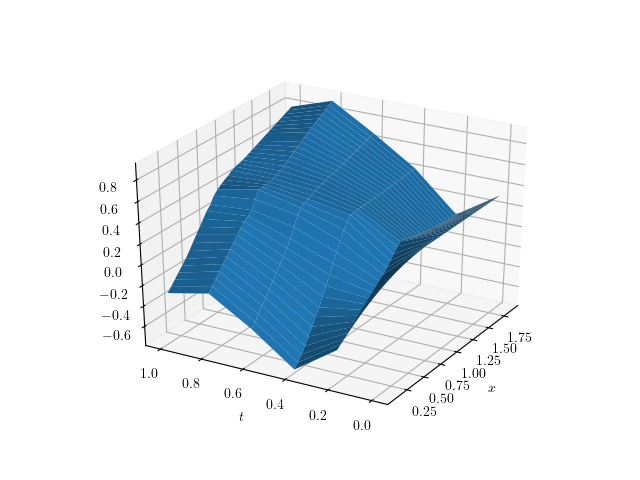}
\caption{$\tilde a^{\hat \theta^*} (t, x)$}
\label{fig:a_theta_calls_and_puts_sym_lv_30_210_a_full}
\end{minipage}
\begin{minipage}[t]{0.04\linewidth}
\hfill
\end{minipage}
\begin{minipage}[t]{0.48\linewidth}
\includegraphics[width=\linewidth]{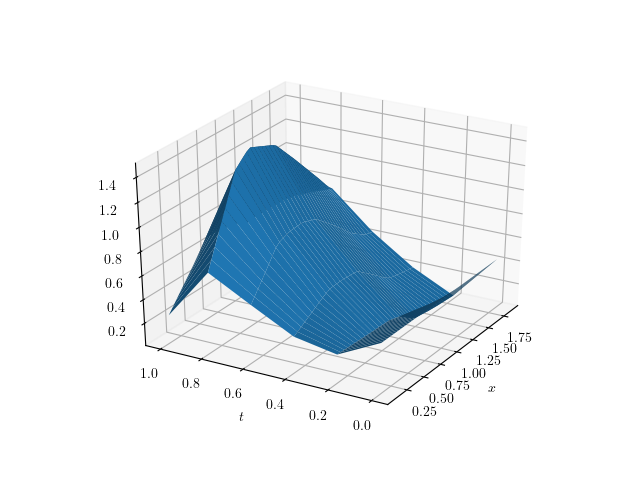}
\caption{$\overline a^{L, \hat \theta^ {L, *}} (t, x)$}
\label{fig:a_theta_calls_and_puts_sym_lv_30_210_a_local}
\end{minipage}
\begin{minipage}[t]{0.48\linewidth}
\includegraphics[width=\linewidth]{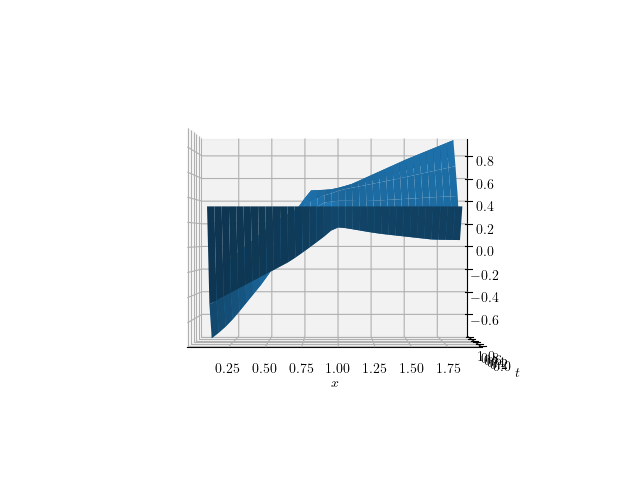}
\caption{$\tilde a^\theta (t, x)$}
\label{fig:a_theta_calls_and_puts_sym_lv_30_270_a_full}
\end{minipage}
\begin{minipage}[t]{0.04\linewidth}
\hfill
\end{minipage}
\begin{minipage}[t]{0.48\linewidth}
\includegraphics[width=\linewidth]{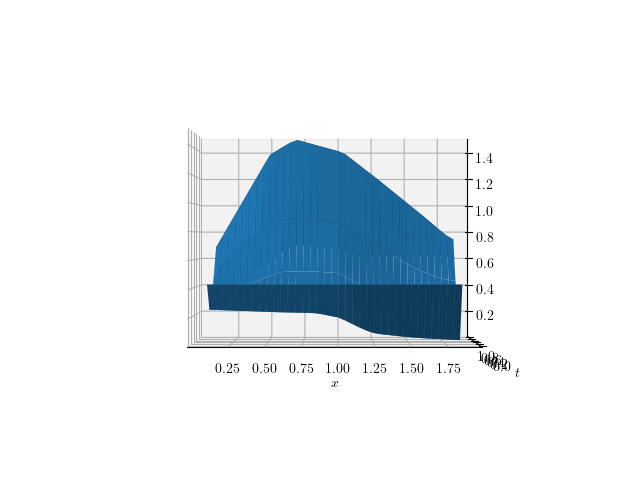}
\caption{$\overline a^{L, \hat \theta^{L, *}} (t, x)$}
\label{fig:a_theta_calls_and_puts_sym_lv_30_270_a_local}
\end{minipage}
\end{figure}

As previously, we show in figures (Fig. \ref{fig:a_theta_calls_and_puts_sym_lv_30_210_a_full}), (Fig. \ref{fig:a_theta_calls_and_puts_sym_lv_30_270_a_full}) and (Fig. \ref{fig:a_theta_calls_and_puts_sym_lv_30_210_a_local}), (Fig. \ref{fig:a_theta_calls_and_puts_sym_lv_30_270_a_local}) the surfaces $\tilde a^{\hat \theta^*}(t,x)$ and $\overline a^{L, \hat \theta^{L, *}}$ from different viewpoints. We see in figures (Fig. \ref{fig:a_theta_calls_and_puts_sym_lv_30_270_a_full}) and (Fig. \ref{fig:a_theta_calls_and_puts_sym_lv_30_270_a_local}), that contrary to the Bachelier diffusion case, $\tilde a^{\hat \theta^*}$ and $\overline a^{L, \hat \theta^{L, *}}$ do not present a symmetry with respect to $x = 1$. This is expected, as the local volatility diffusion process does not present the same symmetry with respect to $x = 1$ that the Bachelier diffusion process presents.

\subsubsection{Multi Coupons AutoCall}

\begin{figure}[H]
\begin{minipage}[t]{0.48\linewidth}
\centering
\includegraphics[width=\linewidth]{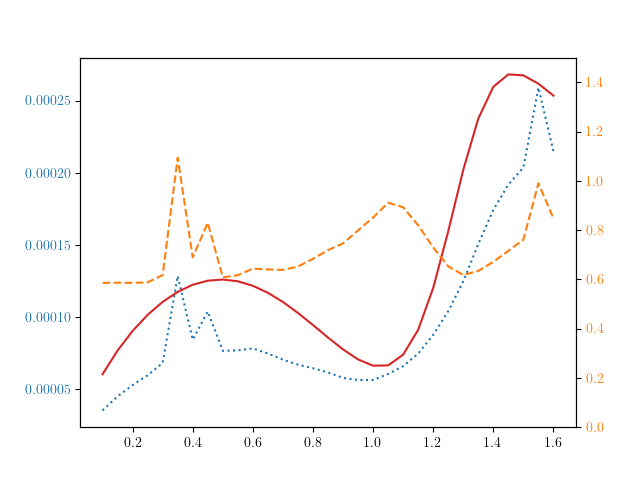}
\caption{Standard Deviation vs $x_0$ for Full Method}
\label{fig:std_adaptative_vs_mc_autocall_multi_coupons_lv_a_full}
\end{minipage}
\begin{minipage}[t]{0.04\linewidth}
\hfill
\end{minipage}
\begin{minipage}[t]{0.48\linewidth}
\centering
\includegraphics[width=\linewidth]{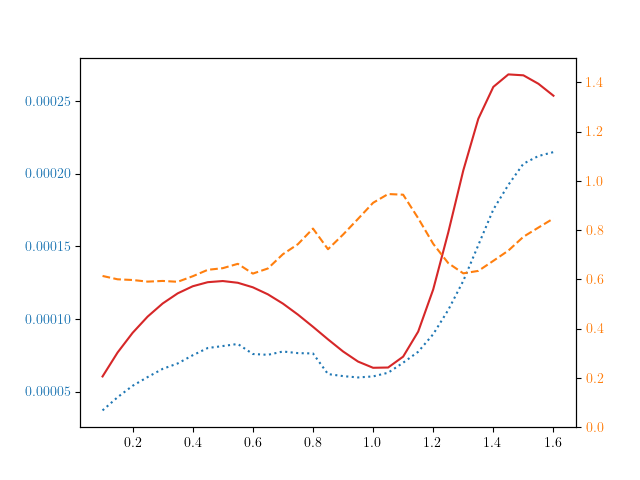}
\caption{Standard Deviation vs $x_0$ for Local Method}
\label{fig:std_adaptative_vs_mc_autocall_multi_coupons_lv_a_local}
\end{minipage}
\end{figure}

For this experiment, we again price an AutoCall that has multiple coupons with the characteristics of table (Tab. \ref{tab:multi_coupons_autocall_params}). Again, we see in figures (Fig. \ref{fig:std_adaptative_vs_mc_autocall_multi_coupons_lv_a_full}) and (Fig. \ref{fig:std_adaptative_vs_mc_autocall_multi_coupons_lv_a_local}) that the Monte Carlo obtained using our adaptative importance sampling has a lower standard deviation than a plain Monte Carlo. 

\begin{figure}[H]
\begin{minipage}[t]{0.48\linewidth}
\centering
\includegraphics[width=\linewidth]{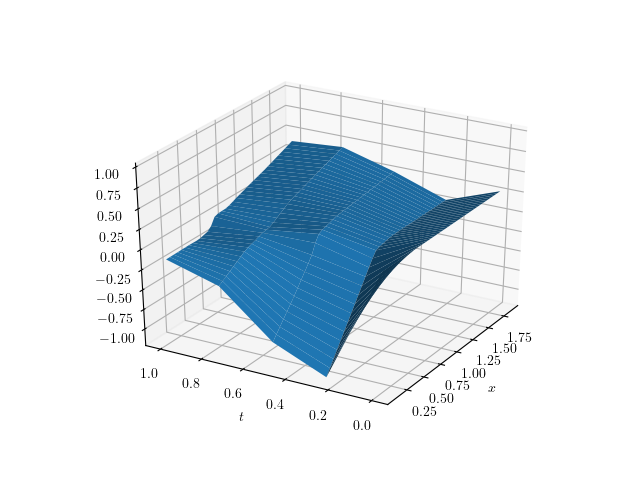}
\caption{$\tilde a^{\hat \theta^*} (t, x)$}
\label{fig:a_theta_autocall_multi_coupons_lv_30_210_a_full}
\end{minipage}
\begin{minipage}[t]{0.04\linewidth}
\hfill
\end{minipage}
\begin{minipage}[t]{0.48\linewidth}
\centering
\includegraphics[width=\linewidth]{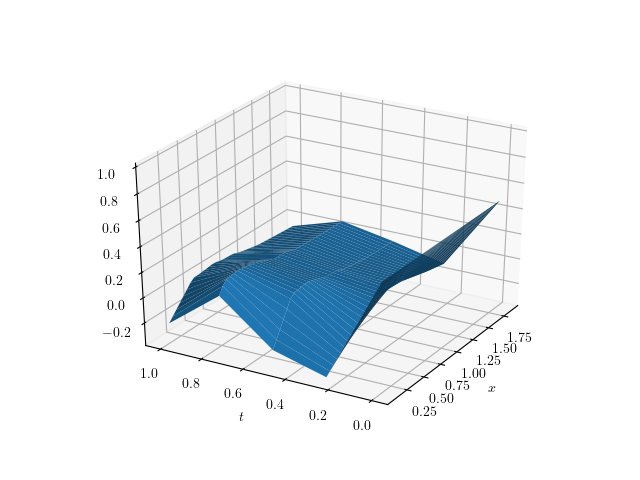}
\caption{$\overline a^{L, \hat \theta^{L, *}} (t, x)$}
\label{fig:a_theta_autocall_multi_coupons_lv_30_210_a_local}
\end{minipage}
\end{figure}

In figures (Fig. \ref{fig:a_theta_autocall_multi_coupons_lv_30_210_a_full}) and (Fig. \ref{fig:a_theta_autocall_multi_coupons_lv_30_210_a_local}), we again show the surfaces $\tilde a^{\hat \theta^*}(t,x)$ and $\overline a^{L, \hat \theta^{L, *}}(t,x)$. As with the Bachelier diffusion case, we can see that that the values are rougly speaking positive when $x > 1$, and negative for $x < 1$. This is again expected from the fact that some trajectories need to get closer to the barrier strikes region $B_.^A = 1.5$, while others need to get close to the put down and in strike region $K^{PDI} = 0.5$. 

\subsubsection{Single Coupon AutoCall}

\begin{figure}[H]
  \begin{minipage}[t]{0.48\linewidth}
    \includegraphics[width=\linewidth]{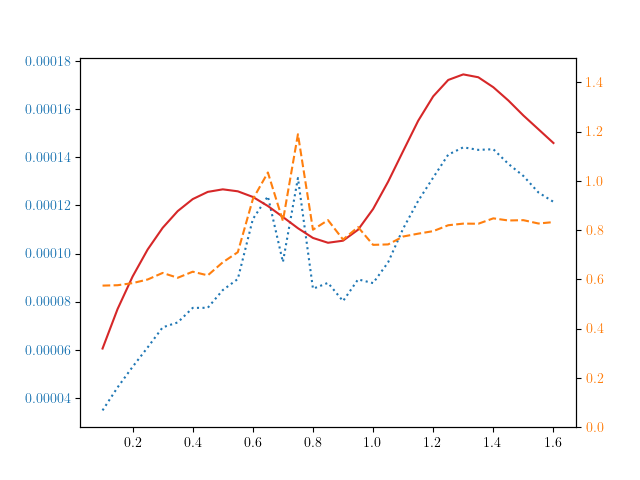}
    \caption{Standard Deviation vs $x_0$ for Full Method}
    \label{fig:std_adaptative_vs_mc_autocall_single_coupon_lv_a_full}
  \end{minipage}
  \begin{minipage}[t]{0.04\linewidth}
  \hfill
  \end{minipage}
  \begin{minipage}[t]{0.48\linewidth}
    \includegraphics[width=\linewidth]{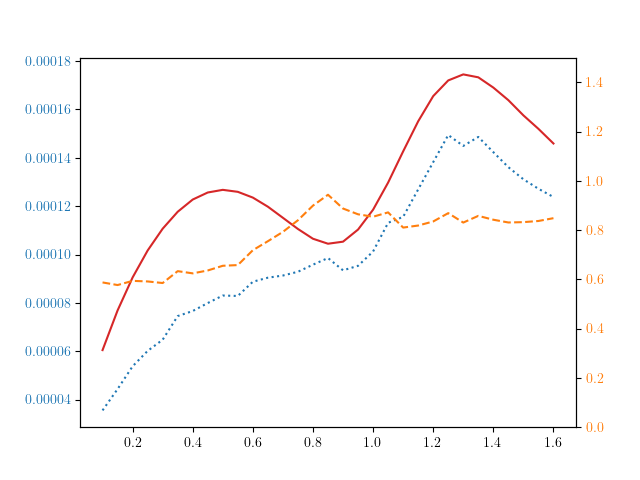}
    \caption{Standard Deviation vs $x_0$ for Local Method}
    \label{fig:std_adaptative_vs_mc_autocall_single_coupon_lv_a_local}
  \end{minipage}
\end{figure}

For this experiment, we again price an AutoCall that has a single coupon (but still multiple AutoCall barriers) with the characteristics of table (Tab. \ref{tab:single_coupon_autocall_params}). Again, we see in figures (Fig. \ref{fig:std_adaptative_vs_mc_autocall_single_coupon_lv_a_full}) and (Fig. \ref{fig:std_adaptative_vs_mc_autocall_single_coupon_lv_a_local}) that the Monte Carlo obtained using our adaptative importance sampling has a lower standard deviation than a plain Monte Carlo. 

\begin{figure}[H]
\begin{minipage}[t]{0.48\linewidth}
\includegraphics[width=\linewidth]{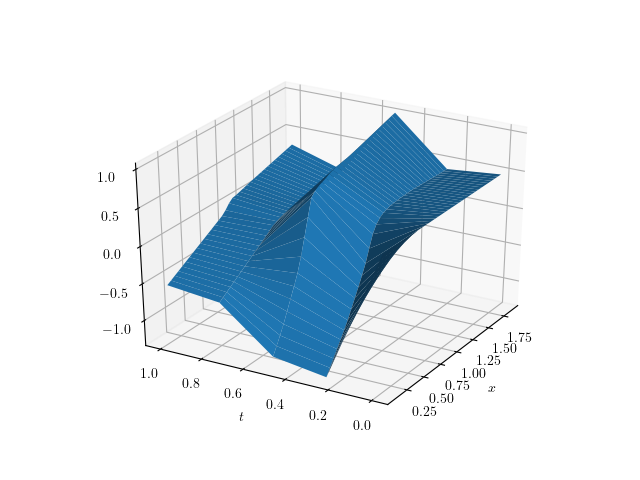}
\caption{$\tilde a^\theta (t, x)$}
\label{fig:a_theta_autocall_single_coupon_lv_30_210_a_full}
\end{minipage}
\begin{minipage}[t]{0.04\linewidth}
\hfill
\end{minipage}
\begin{minipage}[t]{0.48\linewidth}
\includegraphics[width=\linewidth]{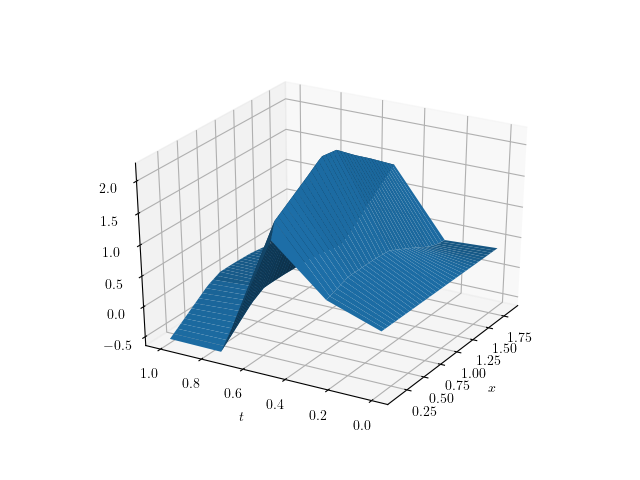}
\caption{$\tilde a^\theta (t, x)$}
\label{fig:a_theta_autocall_single_coupon_lv_30_210_a_local}
\end{minipage}
\end{figure}

\begin{figure}[H]
\begin{minipage}[t]{0.48\linewidth}
\includegraphics[width=\linewidth]{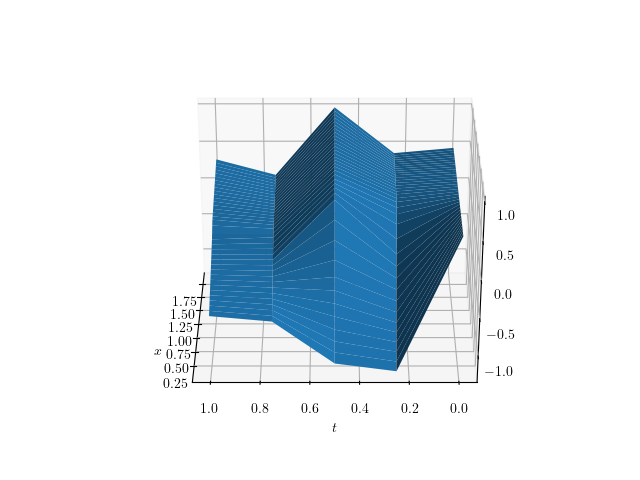}
\caption{$\tilde a^\theta (t, x)$}
\label{fig:a_theta_autocall_single_coupon_lv_30_180_a_full}
\end{minipage}
\begin{minipage}[t]{0.04\linewidth}
\hfill
\end{minipage}
\begin{minipage}[t]{0.48\linewidth}
\includegraphics[width=\linewidth]{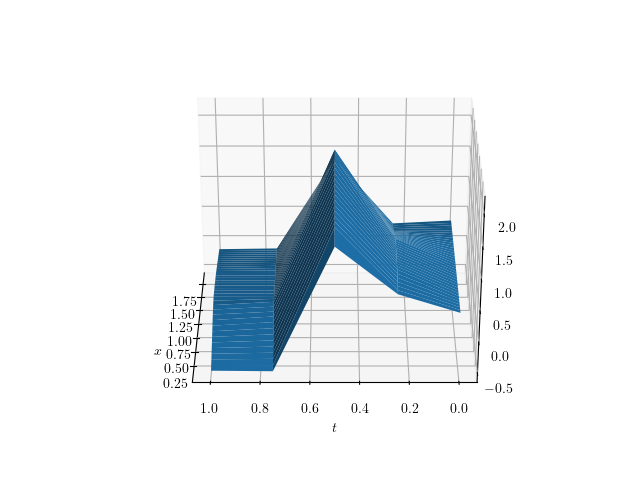}
\caption{$\tilde a^\theta (t, x)$}
\label{fig:a_theta_autocall_single_coupon_lv_30_180_a_local}
\end{minipage}
\end{figure}

In figures (Fig. \ref{fig:a_theta_autocall_single_coupon_lv_30_210_a_full}), (Fig. \ref{fig:a_theta_autocall_single_coupon_lv_30_180_a_full}) and (Fig. \ref{fig:a_theta_autocall_single_coupon_lv_30_210_a_local}), (Fig. \ref{fig:a_theta_autocall_single_coupon_lv_30_180_a_local}), we show the surfaces $\tilde a^{\hat \theta^*}(t,x)$ and $\overline a^{L, \hat \theta^{L, *}}$ from different viewpoints. As with the Bachelier diffusion case, compared to the multiple coupons AutoCall, we can see one striking difference: for $x > 1$, the surface increases from time $t=0$ up to the coupon date $t=T_3^A$, then suddenly drop to values close to 0 and slightly negative. This is expected, as once the coupon date has passed, there is no reason to deviate trajectories towards the $x > 1$. As with the Bachelier diffusion case, once the coupon date has passed, the only thing left to price is the put down and in, so trajectories now need to get to values closer to the put down and in strike $K^{PDI} = 0.5$. 


\section{Robustness of the Algorithm} \label{sec:robust}

In the previous sections, we have considered a user that trains the neural network prior to each pricing. In most cases, such a training can take only a small percentage of the time taken for the pricing and thus only add a small overhead. Indeed, our experience is that training the neural networks applying a number of training steps of only $10\%$ of the number of trajectories used in the final pricing is often enough to obtain good results. Furthermore, our algorithm could easily be improved upon by keeping the values obtained during the training step and use them in the the final Monte Carlo estimator, instead of throwing them away. 

However, an alternative approach is to only train the neural network once for many pricings. For example, a bank might want to train the neural networks only once per week and use the same neural network to price its book for the whole week. To test the feasibility of such a methodology, we study here how well the algorithm performs when we change each of the model parameters: $x_0$ and $\sigma$ for the Bachelier diffusion, $x$, $\sigma$, $a$, $b$, $m$ and $\rho$ for the local volatility diffusion. 

In order to do this, for each payoff and each diffusion type, we train the neural network with the parameters of tables (Tab. \ref{tab:bachelier_call_option_run_params} - Tab. \ref{tab:lv_single_coupon_autocall_run_params}). Once the neural networks are trained, we then vary each parameter from $-50\%$ to $+50\%$ of its original value, and plot the algorithm's and a plain Monte Carlo's standard deviations. We see that in practice, the algorithm is very robust for each tested payoff and diffusion type. Indeed, for all parameters except the spot price $x_0$, the algorithm systematically outperforms the plain Monte Carlo, even though it was trained with the different parameters than those used for the pricing. For the spot price, it usually outperforms the plain Monte Carlo in the range $-90\%$ to $110\%$. For most assets, especially indices, variation of $10\%$ in the underlying is quite rare in a week, so we conclude that a bank could definitely only train the neural networks on a regular basis of around a week. 

In order to limit the number of graphs, We only show the results obtained when using the full version of $a^\theta$. Results for the local version $a^{L, \theta}$ are similar.

\subsection{Bachelier Diffusion}

\subsubsection{Call}

\begin{figure}[H]
\begin{minipage}[t]{0.48\linewidth}
\includegraphics[width=\linewidth]{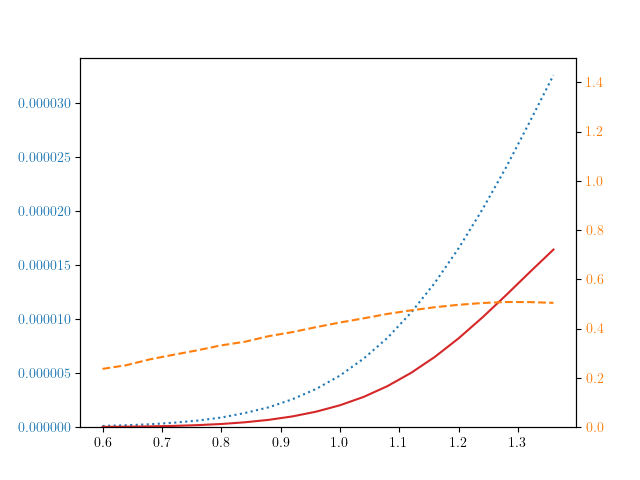}
\caption{Standard Deviation against $x_0$, Adaptative vs MC - Bachelier Diffusion - Call}
\label{fig:robust_X_bachelier_call_a_full}
\end{minipage}
\begin{minipage}[t]{0.04\linewidth}
\hfill
\end{minipage}
\begin{minipage}[t]{0.48\linewidth}
\includegraphics[width=\linewidth]{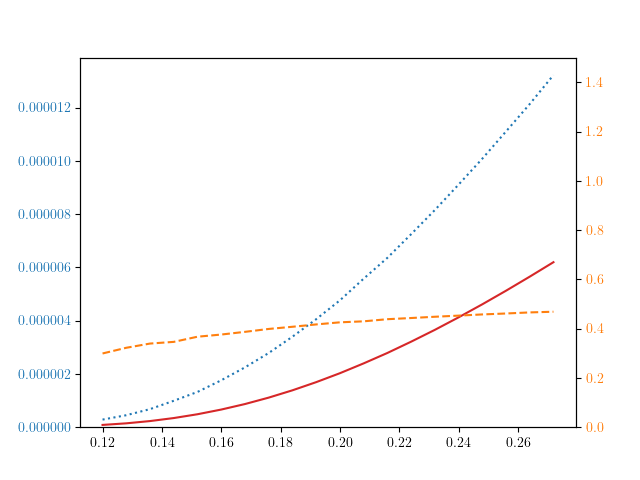}
\caption{Standard Deviation vs $\sigma$, Adaptative vs MC - Bachelier Diffusion - Call}
\label{fig:robust_Sigma_bachelier_call_a_full}
\end{minipage}
\end{figure}

\subsubsection{Asymmetric Calls and Puts}

\begin{figure}[H]
\begin{minipage}[t]{0.48\linewidth}
\includegraphics[width=\linewidth]{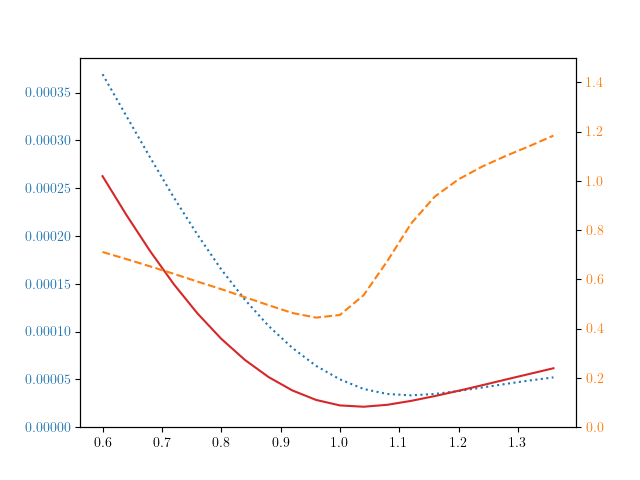}
\caption{Standard Deviation against $x_0$, Adaptative vs MC - Bachelier Diffusion - Asymmetric Calls and Puts}
\label{fig:robust_X_bachelier_calls_and_puts_asymmetric_a_full}
\end{minipage}
\begin{minipage}[t]{0.04\linewidth}
\hfill
\end{minipage}
\begin{minipage}[t]{0.48\linewidth}
\includegraphics[width=\linewidth]{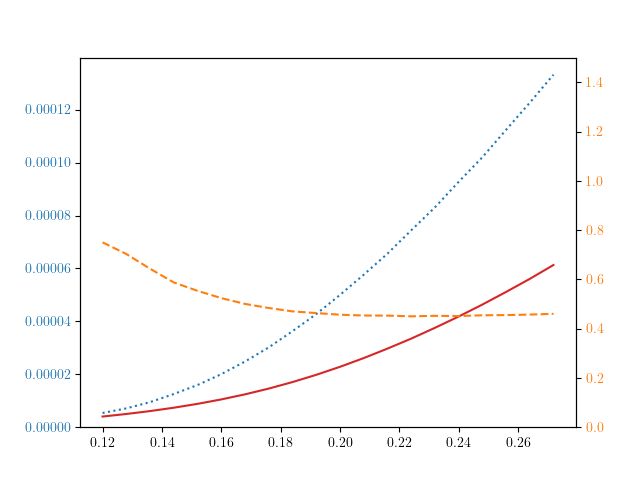}
\caption{Standard Deviation vs $\sigma$, Adaptative vs MC - Bachelier Diffusion - Asymmetric Calls and Puts}
\label{fig:robust_Sigma_bachelier_calls_and_puts_asymmetric_a_full}
\end{minipage}
\end{figure}

\subsubsection{Symmetric Calls and Puts}

\begin{figure}[H]
\begin{minipage}[t]{0.48\linewidth}
\includegraphics[width=\linewidth]{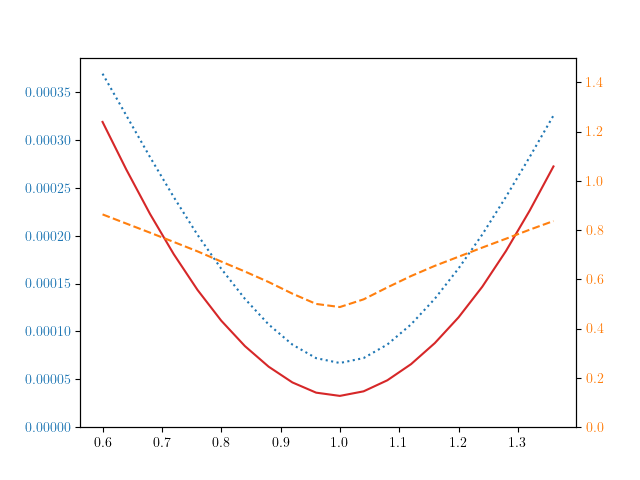}
\caption{Standard Deviation against $x_0$, Adaptative vs MC - Bachelier Diffusion - Symmetric Calls and Puts}
\label{fig:robust_X_bachelier_calls_and_puts_symmetric_a_full}
\end{minipage}
\begin{minipage}[t]{0.04\linewidth}
\hfill
\end{minipage}
\begin{minipage}[t]{0.48\linewidth}
\includegraphics[width=\linewidth]{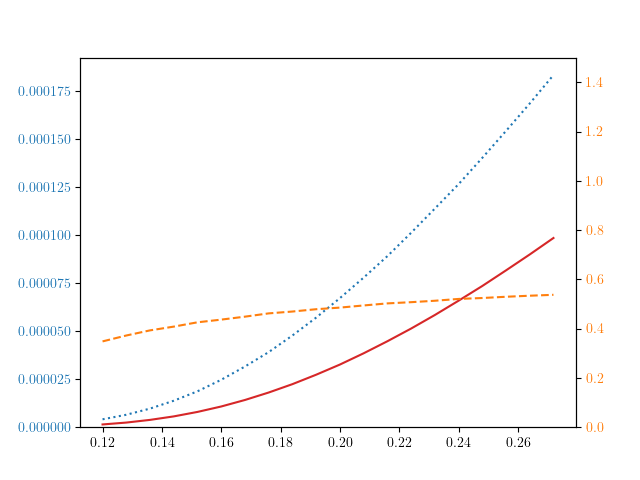}
\caption{Standard Deviation vs $\sigma$, Adaptative vs MC - Bachelier Diffusion - Symmetric Calls and Puts}
\label{fig:robust_Sigma_bachelier_calls_and_puts_symmetric_a_full}
\end{minipage}
\end{figure}

\subsubsection{Multi Coupons AutoCall}

\begin{figure}[H]
\begin{minipage}[t]{0.48\linewidth}
\includegraphics[width=\linewidth]{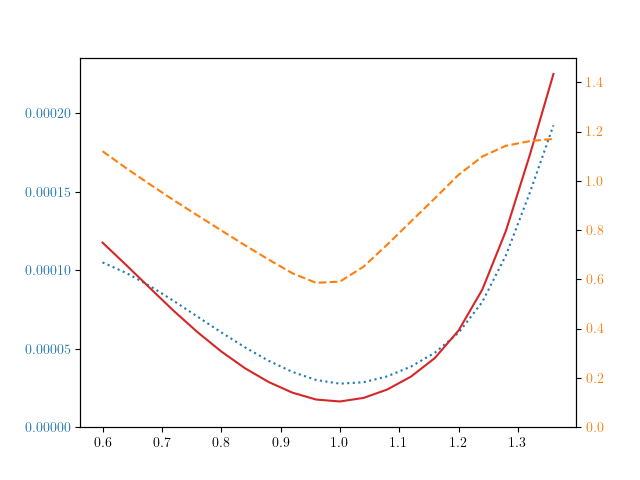}
\caption{Standard Deviation against $x_0$, Adaptative vs MC - Bachelier Diffusion - Multi Coupons AutoCall}
\label{fig:robust_X_bachelier_autocall_multi_coupons_a_full}
\end{minipage}
\begin{minipage}[t]{0.04\linewidth}
\hfill
\end{minipage}
\begin{minipage}[t]{0.48\linewidth}
\includegraphics[width=\linewidth]{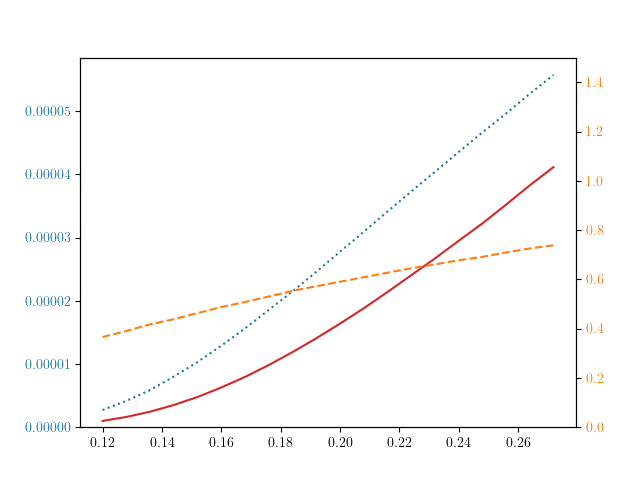}
\caption{Standard Deviation vs $\sigma$, Adaptative vs MC - Bachelier Diffusion - Multi Coupons AutoCall}
\label{fig:robust_Sigma_bachelier_autocall_multi_coupons_a_full}
\end{minipage}
\end{figure}

\subsubsection{Single Coupon AutoCall}

\begin{figure}[H]
\begin{minipage}[t]{0.48\linewidth}
\includegraphics[width=\linewidth]{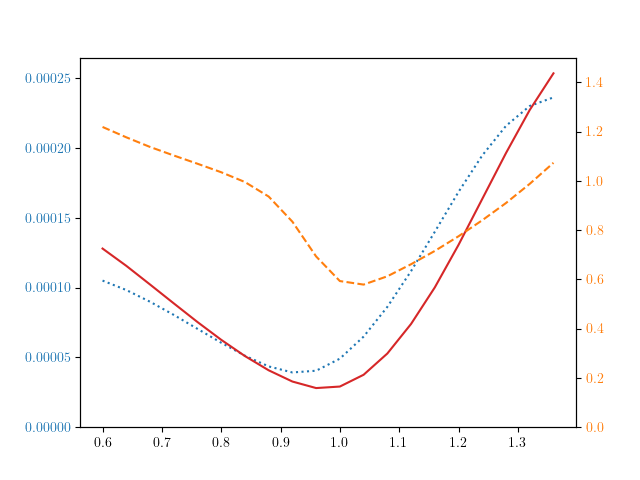}
\caption{Standard Deviation against $x_0$, Adaptative vs MC - Bachelier Diffusion - Single Coupon AutoCall}
\label{fig:robust_X_bachelier_autocall_single_coupon_a_full}
\end{minipage}
\begin{minipage}[t]{0.04\linewidth}
\hfill
\end{minipage}
\begin{minipage}[t]{0.48\linewidth}
\includegraphics[width=\linewidth]{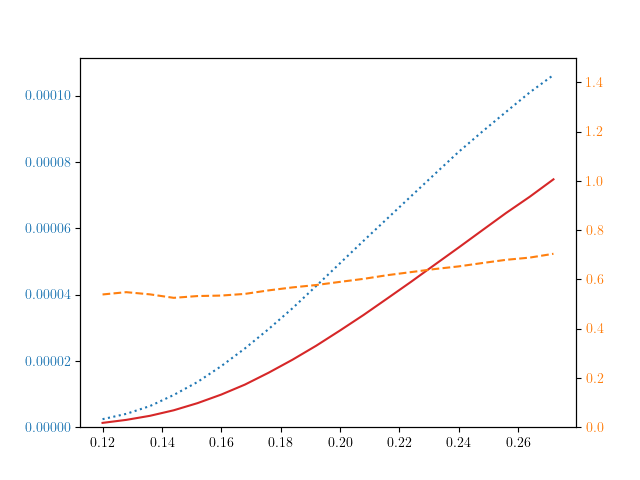}
\caption{Standard Deviation vs $\sigma$, Adaptative vs MC - Bachelier Diffusion - Single Coupon AutoCall}
\label{fig:robust_Sigma_bachelier_autocall_single_coupon_a_full}
\end{minipage}
\end{figure}

\subsection{LV Diffusion}

\subsubsection{Call}

\begin{figure}[H]
\begin{minipage}[t]{0.48\linewidth}
\includegraphics[width=\linewidth]{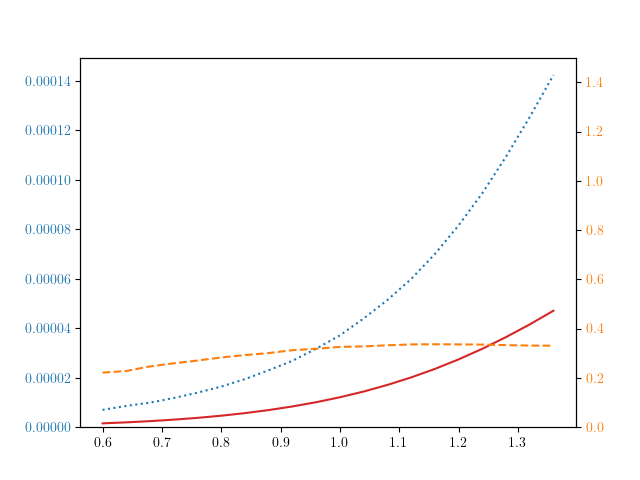}
\caption{Standard Deviation against $x_0$, Adaptative vs MC - LV Diffusion - Call}
\label{fig:robust_X_lv_call_a_full}
\end{minipage}
\begin{minipage}[t]{0.04\linewidth}
\hfill
\end{minipage}
\begin{minipage}[t]{0.48\linewidth}
\includegraphics[width=\linewidth]{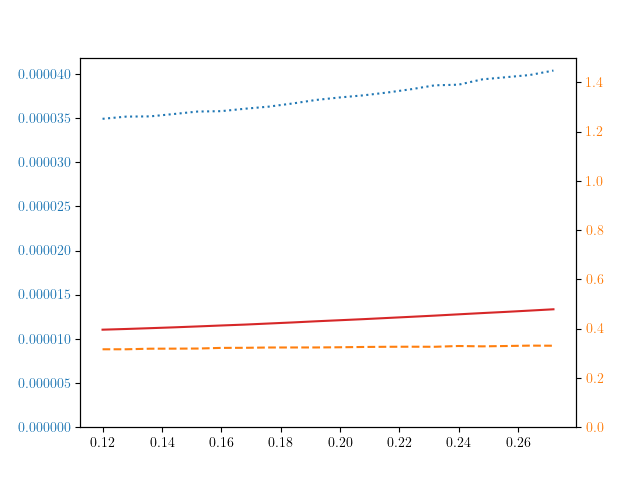}
\caption{Standard Deviation against $\sigma$, Adaptative vs MC - LV Diffusion - Call}
\label{fig:robust_Sigma_lv_call_a_full}
\end{minipage}
\end{figure}

\begin{figure}[H]
\begin{minipage}[t]{0.48\linewidth}
\includegraphics[width=\linewidth]{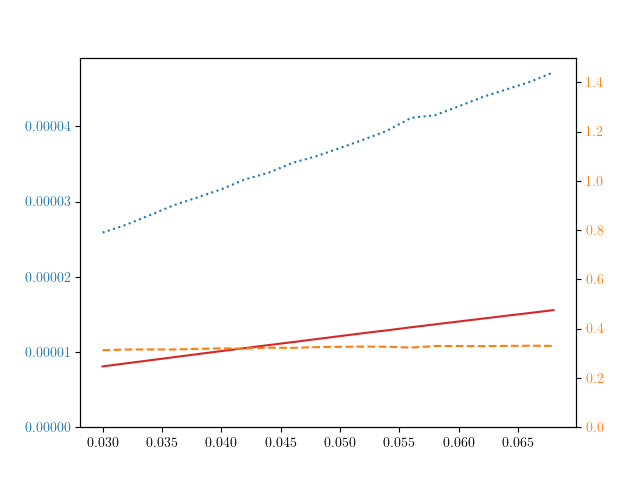}
\caption{Standard Deviation against $a$, Adaptative vs MC - LV Diffusion - Call}
\label{fig:robust_A_lv_call_a_full}
\end{minipage}
\begin{minipage}[t]{0.04\linewidth}
\hfill
\end{minipage}
\begin{minipage}[t]{0.48\linewidth}
\includegraphics[width=\linewidth]{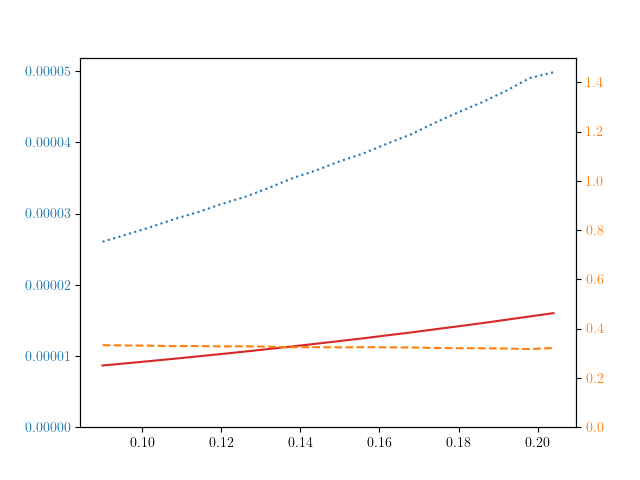}
\caption{Standard Deviation against $b$, Adaptative vs MC - LV Diffusion - Call}
\label{fig:robust_B_lv_call_a_full}
\end{minipage}
\end{figure}

\begin{figure}[H]
\begin{minipage}[t]{0.48\linewidth}
\includegraphics[width=\linewidth]{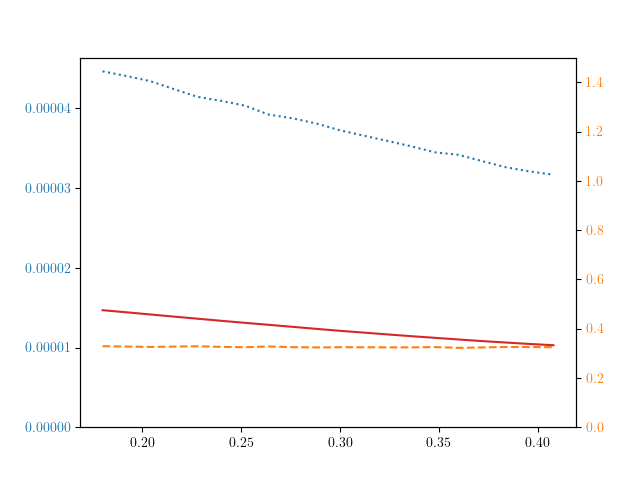}
\caption{Standard Deviation against $m$, Adaptative vs MC - LV Diffusion - Call}
\label{fig:robust_M_lv_call_a_full}
\end{minipage}
\begin{minipage}[t]{0.04\linewidth}
\hfill
\end{minipage}
\begin{minipage}[t]{0.48\linewidth}
\includegraphics[width=\linewidth]{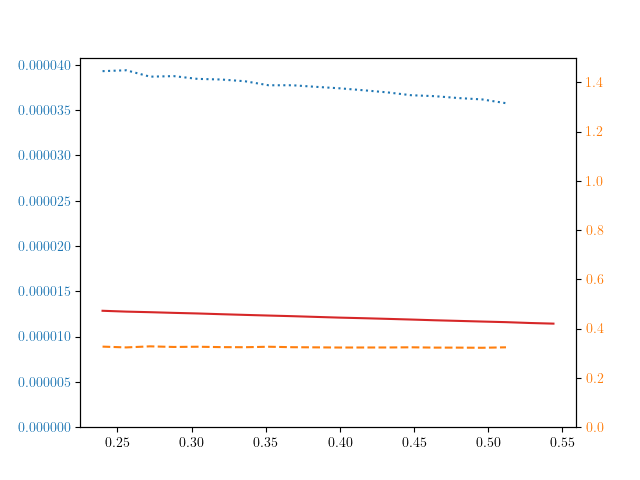}
\caption{Standard Deviation against $\rho$, Adaptative vs MC - LV Diffusion - Call}
\label{fig:robust_Rho_lv_call_a_full}
\end{minipage}
\end{figure}

\subsubsection{Asymmetric Calls and Puts}

\begin{figure}[H]
\begin{minipage}[t]{0.48\linewidth}
\includegraphics[width=\linewidth]{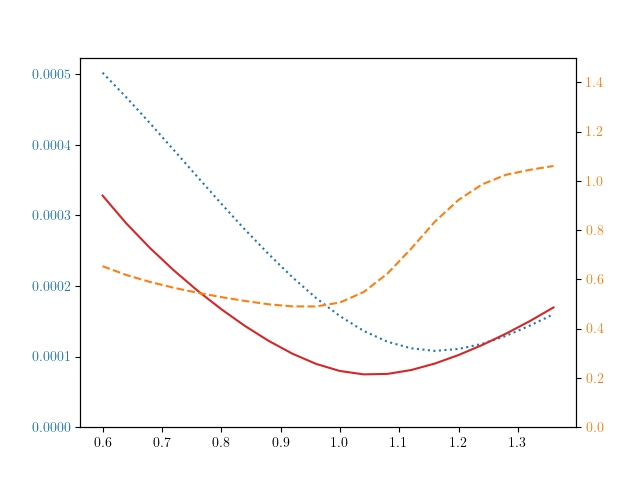}
\caption{Standard Deviation against $x_0$, Adaptative vs MC - LV Diffusion - Asymmetric Calls and Puts}
\label{fig:robust_X_lv_calls_and_puts_asymmetric_a_full}
\end{minipage}
\begin{minipage}[t]{0.04\linewidth}
\hfill
\end{minipage}
\begin{minipage}[t]{0.48\linewidth}
\includegraphics[width=\linewidth]{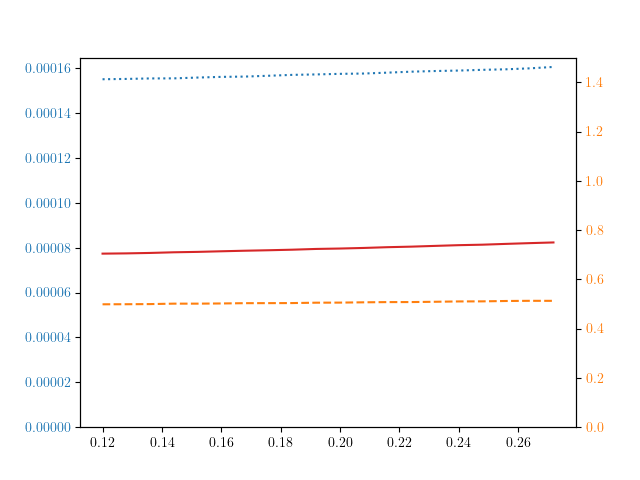}
\caption{Standard Deviation vs $\sigma$, Adaptative vs MC - LV Diffusion - Asymmetric Calls and Puts}
\label{fig:robust_Sigma_lv_calls_and_puts_asymmetric_a_full}
\end{minipage}
\end{figure}

\begin{figure}[H]
\begin{minipage}[t]{0.48\linewidth}
\includegraphics[width=\linewidth]{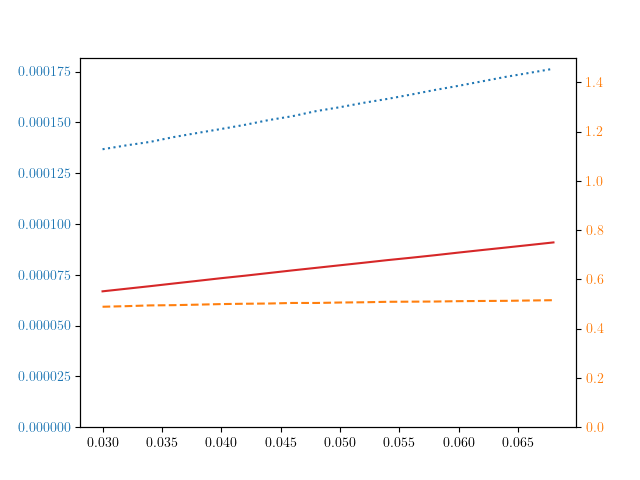}
\caption{Standard Deviation against $a$, Adaptative vs MC - LV Diffusion - Asymmetric Calls and Puts}
\label{fig:robust_A_lv_calls_and_puts_asymmetric_a_full}
\end{minipage}
\begin{minipage}[t]{0.04\linewidth}
\hfill
\end{minipage}
\begin{minipage}[t]{0.48\linewidth}
\includegraphics[width=\linewidth]{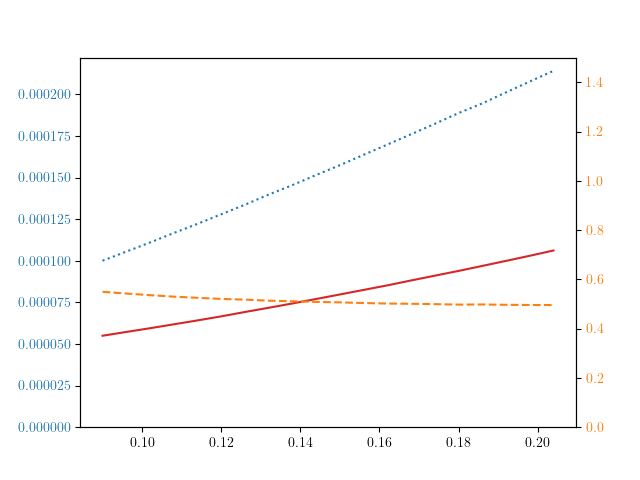}
\caption{Standard Deviation against $b$, Adaptative vs MC - LV Diffusion - Asymmetric Calls and Puts}
\label{fig:robust_B_lv_calls_and_puts_asymmetric_a_full}
\end{minipage}
\end{figure}

\begin{figure}[H]
\begin{minipage}[t]{0.48\linewidth}
\includegraphics[width=\linewidth]{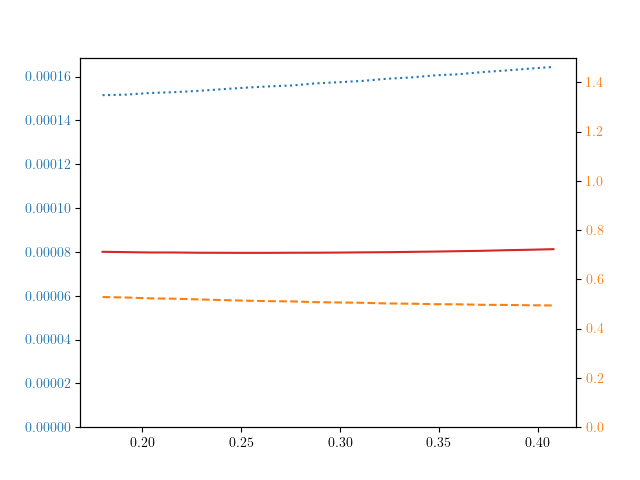}
\caption{Standard Deviation against $m$, Adaptative vs MC - LV Diffusion - Asymmetric Calls and Puts}
\label{fig:robust_M_lv_calls_and_puts_asymmetric_a_full}
\end{minipage}
\begin{minipage}[t]{0.04\linewidth}
\hfill
\end{minipage}
\begin{minipage}[t]{0.48\linewidth}
\includegraphics[width=\linewidth]{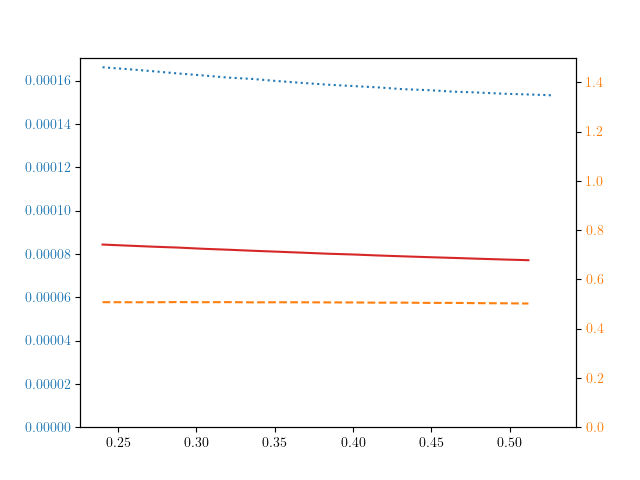}
\caption{Standard Deviation against $\rho$, Adaptative vs MC - LV Diffusion - Asymmetric Calls and Puts}
\label{fig:robust_Rho_lv_calls_and_puts_asymmetric_a_full}
\end{minipage}
\end{figure}

\subsubsection{Symmetric Calls and Puts}

\begin{figure}[H]
\begin{minipage}[t]{0.48\linewidth}
\includegraphics[width=\linewidth]{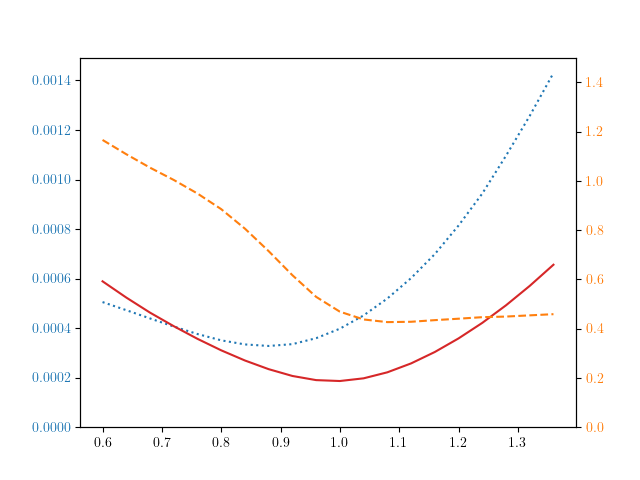}
\caption{Standard Deviation against $x_0$, Adaptative vs MC - LV Diffusion - Symmetric Calls and Puts}
\label{fig:robust_X_lv_calls_and_puts_symmetric_a_full}
\end{minipage}
\begin{minipage}[t]{0.04\linewidth}
\hfill
\end{minipage}
\begin{minipage}[t]{0.48\linewidth}
\includegraphics[width=\linewidth]{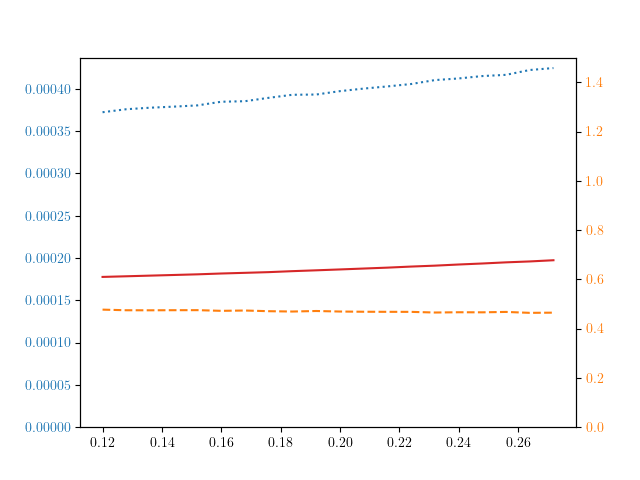}
\caption{Standard Deviation against $\sigma$, Adaptative vs MC - LV Diffusion - Symmetric Calls and Puts}
\label{fig:robust_Sigma_lv_calls_and_puts_symmetric_a_full}
\end{minipage}
\end{figure}

\begin{figure}[H]
\begin{minipage}[t]{0.48\linewidth}
\includegraphics[width=\linewidth]{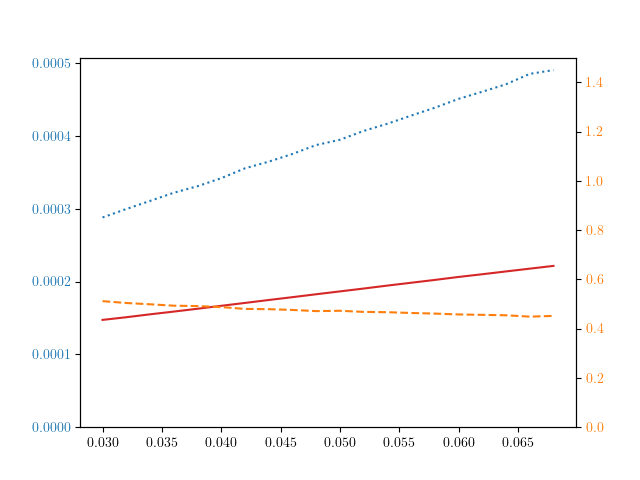}
\caption{Standard Deviation against $a$, Adaptative vs MC - LV Diffusion - Symmetric Calls and Puts}
\label{fig:robust_A_lv_calls_and_puts_symmetric_a_full}
\end{minipage}
\begin{minipage}[t]{0.04\linewidth}
\hfill
\end{minipage}
\begin{minipage}[t]{0.48\linewidth}
\includegraphics[width=\linewidth]{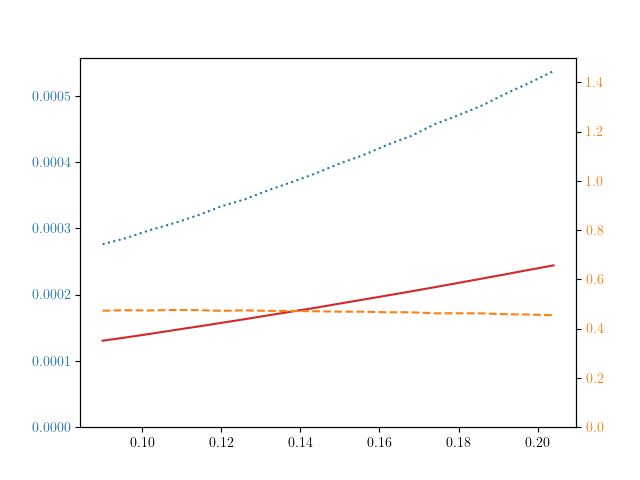}
\caption{Standard Deviation against $b$, Adaptative vs MC - LV Diffusion - Symmetric Calls and Puts}
\label{fig:robust_B_lv_calls_and_puts_symmetric_a_full}
\end{minipage}
\end{figure}

\begin{figure}[H]
\begin{minipage}[t]{0.48\linewidth}
\includegraphics[width=\linewidth]{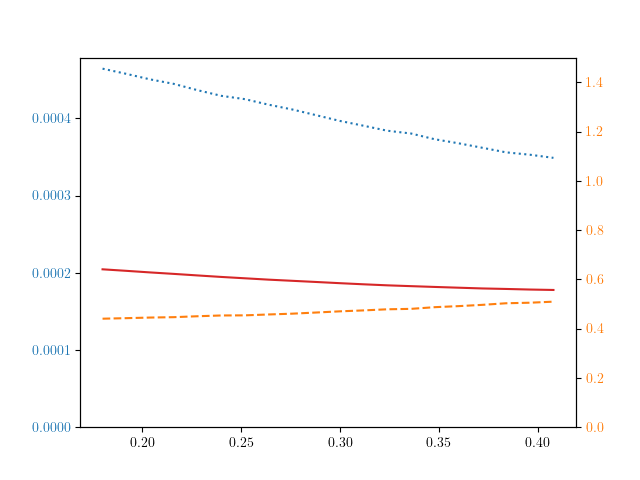}
\caption{Standard Deviation against $m$, Adaptative vs MC - LV Diffusion - Symmetric Calls and Puts}
\label{fig:robust_M_lv_calls_and_puts_symmetric_a_full}
\end{minipage}
\begin{minipage}[t]{0.04\linewidth}
\hfill
\end{minipage}
\begin{minipage}[t]{0.48\linewidth}
\includegraphics[width=\linewidth]{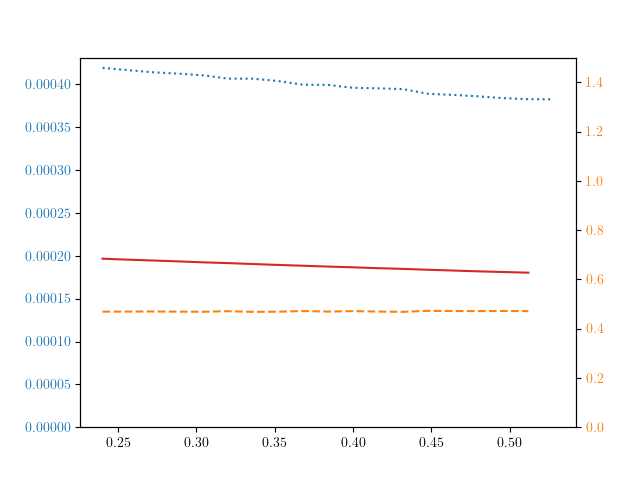}
\caption{Standard Deviation against $\rho$, Adaptative vs MC - LV Diffusion - Symmetric Calls and Puts}
\label{fig:robust_Rho_lv_calls_and_puts_symmetric_a_full}
\end{minipage}
\end{figure}

\subsubsection{Multi Coupons AutoCall}

\begin{figure}[H]
\begin{minipage}[t]{0.48\linewidth}
\includegraphics[width=\linewidth]{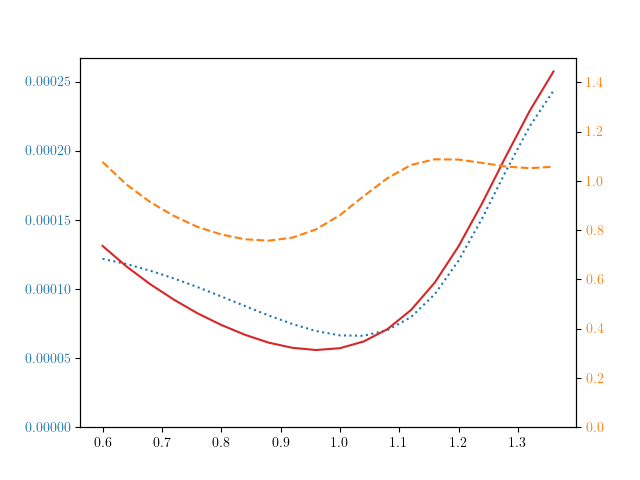}
\caption{Standard Deviation against $x_0$, Adaptative vs MC - LV Diffusion - Multi Coupons AutoCall}
\label{fig:robust_X_lv_autocall_multi_coupons_a_full}
\end{minipage}
\begin{minipage}[t]{0.04\linewidth}
\hfill
\end{minipage}
\begin{minipage}[t]{0.48\linewidth}
\includegraphics[width=\linewidth]{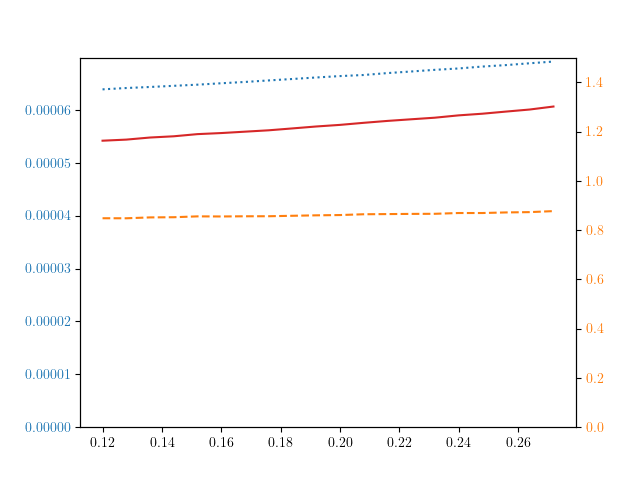}
\caption{Standard Deviation vs $\sigma$, Adaptative vs MC - LV Diffusion - Multi Coupons AutoCall}
\label{fig:robust_Sigma_lv_autocall_multi_coupons_a_full}
\end{minipage}
\end{figure}

\begin{figure}[H]
\begin{minipage}[t]{0.48\linewidth}
\includegraphics[width=\linewidth]{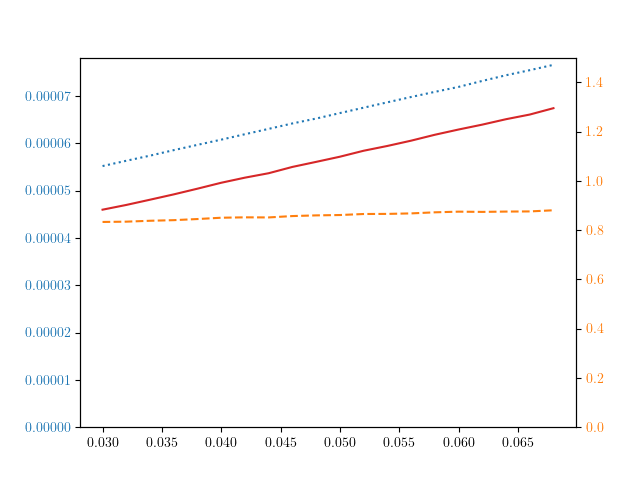}
\caption{Standard Deviation against $a$, Adaptative vs MC - LV Diffusion - Multi Coupons AutoCall}
\label{fig:robust_A_lv_autocall_multi_coupons_a_full}
\end{minipage}
\begin{minipage}[t]{0.04\linewidth}
\hfill
\end{minipage}
\begin{minipage}[t]{0.48\linewidth}
\includegraphics[width=\linewidth]{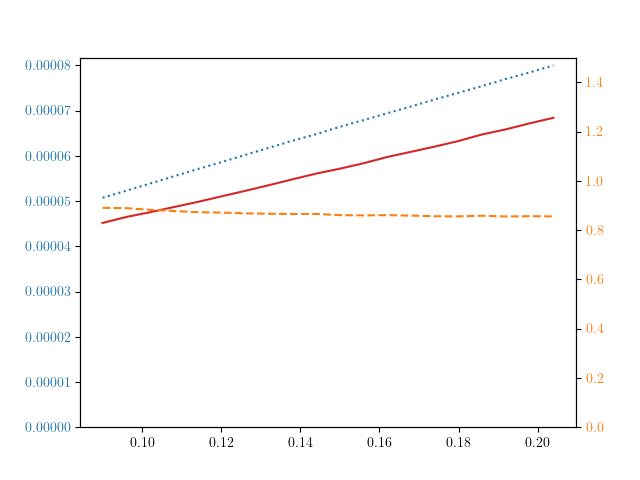}
\caption{Standard Deviation against $b$, Adaptative vs MC - LV Diffusion - Multi Coupons AutoCall}
\label{fig:robust_B_lv_autocall_multi_coupons_a_full}
\end{minipage}
\end{figure}

\begin{figure}[H]
\begin{minipage}[t]{0.48\linewidth}
\includegraphics[width=\linewidth]{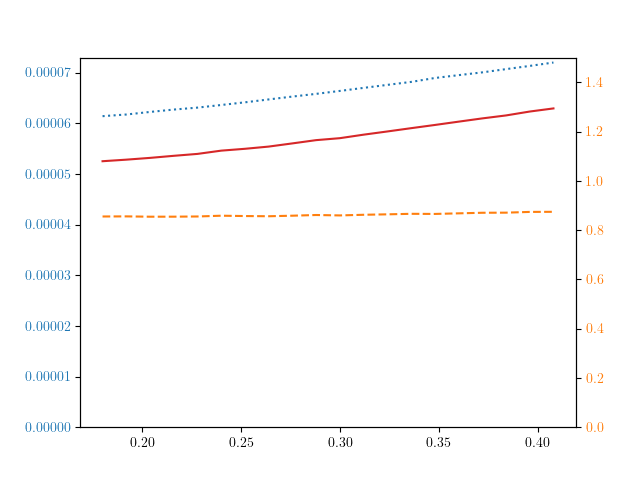}
\caption{Standard Deviation against $m$, Adaptative vs MC - LV Diffusion - Multi Coupons AutoCall}
\label{fig:robust_M_lv_autocall_multi_coupons_a_full}
\end{minipage}
\begin{minipage}[t]{0.04\linewidth}
\hfill
\end{minipage}
\begin{minipage}[t]{0.48\linewidth}
\includegraphics[width=\linewidth]{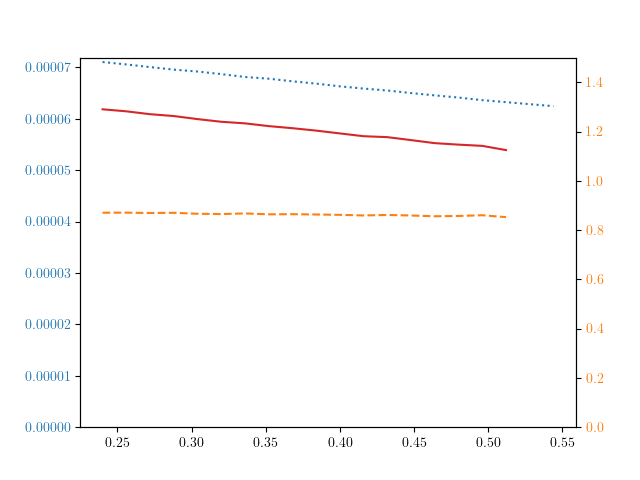}
\caption{Standard Deviation against $\rho$, Adaptative vs MC - LV Diffusion - Multi Coupons AutoCall}
\label{fig:robust_Rho_lv_autocall_multi_coupons_a_full}
\end{minipage}
\end{figure}

\subsubsection{Single Coupon AutoCall}

\begin{figure}[H]
\begin{minipage}[t]{0.48\linewidth}
\includegraphics[width=\linewidth]{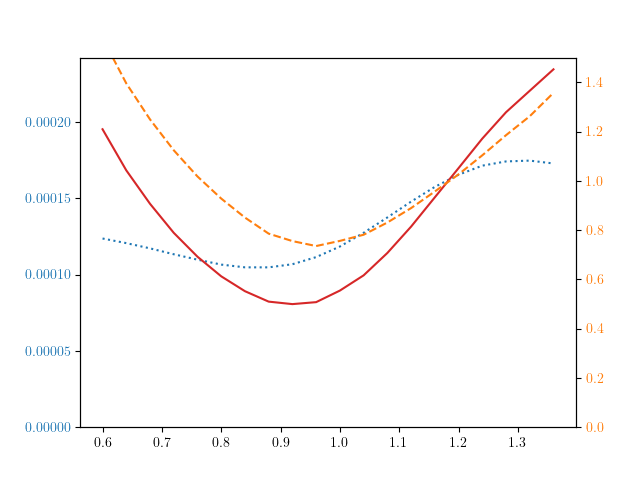}
\caption{Standard Deviation against $x_0$, Adaptative vs MC - LV Diffusion - Single Coupon AutoCall}
\label{fig:robust_X_lv_autocall_single_coupon_a_full}
\end{minipage}
\begin{minipage}[t]{0.04\linewidth}
\hfill
\end{minipage}
\begin{minipage}[t]{0.48\linewidth}
\includegraphics[width=\linewidth]{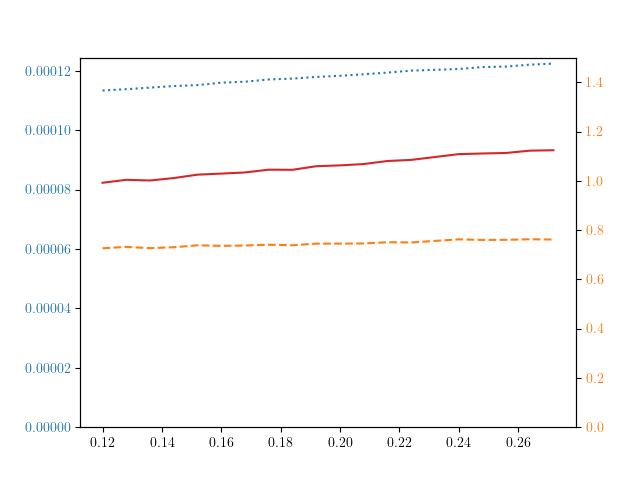}
\caption{Standard Deviation vs $\sigma$, Adaptative vs MC - LV Diffusion - Single Coupon AutoCall}
\label{fig:robust_Sigma_lv_autocall_single_coupon_a_full}
\end{minipage}
\end{figure}

\begin{figure}[H]
\begin{minipage}[t]{0.48\linewidth}
\includegraphics[width=\linewidth]{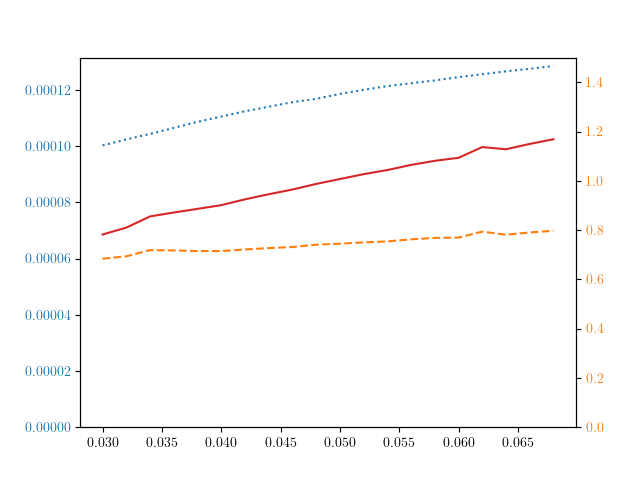}
\caption{Standard Deviation against $a$, Adaptative vs MC - LV Diffusion - Single Coupon AutoCall}
\label{fig:robust_A_lv_autocall_single_coupon_a_full}
\end{minipage}
\begin{minipage}[t]{0.04\linewidth}
\hfill
\end{minipage}
\begin{minipage}[t]{0.48\linewidth}
\includegraphics[width=\linewidth]{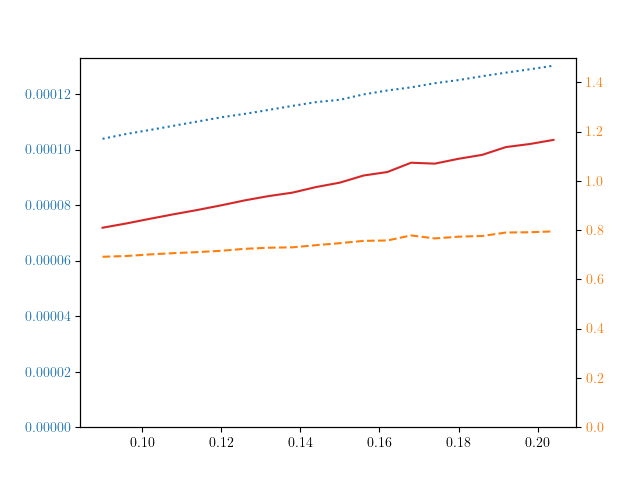}
\caption{Standard Deviation against $b$, Adaptative vs MC - LV Diffusion - Single Coupon AutoCall}
\label{fig:robust_B_lv_autocall_single_coupon_a_full}
\end{minipage}
\end{figure}

\begin{figure}[H]
\begin{minipage}[t]{0.48\linewidth}
\includegraphics[width=\linewidth]{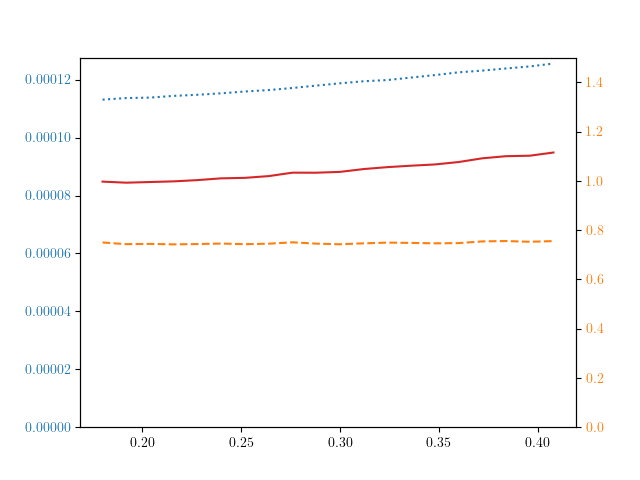}
\caption{Standard Deviation against $m$, Adaptative vs MC - LV Diffusion - Single Coupon AutoCall}
\label{fig:robust_M_lv_autocall_single_coupon_a_full}
\end{minipage}
\begin{minipage}[t]{0.04\linewidth}
\hfill
\end{minipage}
\begin{minipage}[t]{0.48\linewidth}
\includegraphics[width=\linewidth]{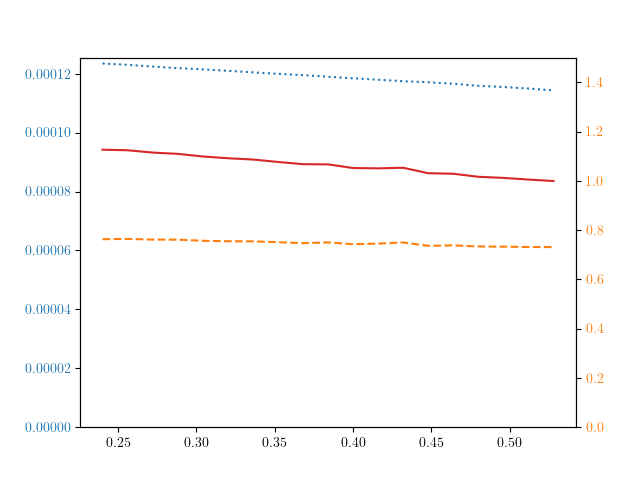}
\caption{Standard Deviation against $\rho$, Adaptative vs MC - LV Diffusion - Single Coupon AutoCall}
\label{fig:robust_Rho_lv_autocall_single_coupon_a_full}
\end{minipage}
\end{figure}

\section{Code on Github Repository}

The code used to generate the above graphs is available on the github repository \cite{github_repo}. The code runs with Python 3.6.5, matplotlib 3.1.1, pandas 0.25.2, scikit-learn 0.21.3, numpy 1.16.4 and tensorflow 1.13.1. 

\section{Conclusion}

We have presented here a generic algorithm that finds the path-dependent change of measure one needs to apply to any particular payoff to reduce the variance of the corresponding Monte Carlo estimator. We have shown in our numerical results of section \ref{sec:numerical_experiment} that this enables for  a wide range of payoffs a reduction of variance of a factor between 2 and 9. In section \ref{sec:robust}, we show that even if one uses the Deep Importance Sampling algorithm with parameters with different values than those used during the training, these values can vary by large amounts (10\% for the spot price, 40\% for the volatility parameters) and the algorithm still performs better than a plain Monte Carlo. 

However, if one were to implement this algorithm, one could still improve it in multiple ways. For an on-the-fly version of the Deep Importance Sampling algorithm, contrary to what we did in the paper, one might want to keep the values obtained during the training, so as to use them in the final Monte Carlo estimator, and thus not throw them away. Furthermore, if one is worried about the Deep Importance Sampling algorithm performing less well than a plain Monte Carlo, one can estimate on the fly the standard deviations obtained both via the Deep Importance Sampling algorithm and a plain Monte Carlo, and automatically switch back to the plain Monte Carlo if results are unsatisfactory. 
 For a user wishing to only train the neural networks on a regular basis (such as once per week), training the neural networks on multiple initial spot prices $x_0$ should make the Deep Importance Sampling algorithm more robust to changes in the spot price. 

Although we have not explored these applications, this method can also naturally be extended to payoffs with multiple underlyings as well as diffusion models with more than one driving brownian motion. One might then want to add all relevant factors in the input of $a^\theta$. The Deep Importance Sampling algorithm should also be useful for rare event simulation, where one might expect even larger gains in variance reduction than in the examples presented in this paper.

\newpage
\appendix
\section{Parameter Tables}

\begin{table}[H]
\parbox[b]{.45\linewidth}{
\centering
\begin{tabular}{ |c|c|}
\hline
Parameter & Value  \\
\hline
$x_0$ & 1.0 \\
$\sigma$ & 0.2 \\
$T$ & 1. \\
$N^{T}$ & 6 \\
\hline
\end{tabular}
\caption{Bachelier Diffusion}
\label{tab:bachelier_diffusion_params}
}
\hfill
\parbox[b]{.45\linewidth}{
\centering
\begin{tabular}{ |c|c|}
\hline
Parameter & Value  \\
\hline
$x_0$ & 1.0 \\
$T$ & 1.0 \\
$N^{T}$ & 6 \\
$a$ & 0.05 \\
$b$ & 0.15 \\
$\rho$ & 0.40 \\
$m$ & 0.30 \\
$\sigma$ & 0.45 \\
\hline
\end{tabular}
\caption{Local Volatility Diffusion}
\label{tab:local_volatility_diffusion_params}
}
\end{table}

\begin{table}[H]
\parbox[b]{.30\linewidth}{
\centering
\begin{tabular}{ |c|c|}
\hline
Parameter & Value  \\
\hline
$K$ & 1.4 \\
\hline
\end{tabular}
\caption{Call Option}
\label{tab:call_option_params}
}
\hfill
\parbox[b]{.30\linewidth}{
\centering
\begin{tabular}{ |c|c|}
\hline
Parameter & Value  \\
\hline
$N_1$ & 1 \\
$K_1$ & 1.2\\
$N_2$ & 10 \\
$K_2$ & 0.6 \\
\hline
\end{tabular}
\caption{Asymmetric Call \& Put Options}
\label{tab:asym_call_&_put_options_params}
}
\hfill
\parbox[b]{.30\linewidth}{
\centering
\begin{tabular}{ |c|c|}
\hline
Parameter & Value  \\
\hline
$N_1$ & 10 \\
$K_1$ & 1.4\\
$N_2$ & 10 \\
$K_2$ & 0.6 \\
\hline
\end{tabular}
\caption{Symmetric Call \& Put Options}
\label{tab:sym_call_&_put_options_params}
}
\end{table}

\begin{table}[H]
\parbox[b]{.50\linewidth}{
\centering
\begin{tabular}{ |c|c|c|c|c|c|}
\hline
 & $i=1$ & $i=2$ & $i=3$ & $i=4$ & $i=5$ \\
\hline
$T_i^A$ & 0.2 & 0.4 & 0.6 & 0.8 & 1.0 \\
$B_i^A$ & 1.5 & 1.5 & 1.5 & 1.5 & 1.5 \\
$S_i^A$ & 0.1 & 0.1 & 0.1 & 0.1 & 0.1 \\
$C_i^P$ & 1.8 & 1.8 & 1.8 & 1.8 & 1.8 \\
\hline
$N^P$     & \multicolumn{5}{c|}{5} \\
$K^{PDI}$ & \multicolumn{5}{c|}{0.5} \\
$S^{PDI}$ & \multicolumn{5}{c|}{0.1} \\
\hline
\end{tabular}
\caption{Multi Coupons AutoCall}
\label{tab:multi_coupons_autocall_params}
}
\parbox[b]{.50\linewidth}{
\centering
\begin{tabular}{ |c|c|c|c|c|c|}
\hline
 & $i=1$ & $i=2$ & $i=3$ & $i=4$ & $i=5$ \\
\hline
$T_i^A$ & 0.2 & 0.4 & 0.6 & 0.8 & 1.0 \\
$B_i^A$ & 1.5 & 1.5 & 1.5 & 1.5 & 1.5 \\
$S_i^A$ & 0.1 & 0.1 & 0.1 & 0.1 & 0.1 \\
$C_i^P$ & 0.0 & 0.0 & 12.5 & 0.0 & 0.0 \\
\hline
$N^P$     & \multicolumn{5}{c|}{5} \\
$K^{PDI}$ & \multicolumn{5}{c|}{0.5} \\
$S^{PDI}$ & \multicolumn{5}{c|}{0.1} \\
\hline
\end{tabular}
\caption{Single Coupon AutoCall}
\label{tab:single_coupon_autocall_params}
}
\end{table}

\begin{table}
\parbox[t]{.45\linewidth}{
\centering
\begin{tabular}{ |c|c|}
\hline
Parameter & Value  \\
\hline
NumberOfBatchesForTraining & 100 \\
NBatchChangeProportion & 100 \\
NumberOfBatchesForEval & 10000 \\
NBatchSize & 1000 \\
LambdaConstrant & 0.001 \\
Contraint & 10.0 \\
LearningRate & 10.0 \\
\hline
\end{tabular}
\caption{Bachelier Call Option Run Parameters}
\label{tab:bachelier_call_option_run_params}
}
\parbox[t]{.45\linewidth}{
\centering
\begin{tabular}{ |c|c|}
\hline
Parameter & Value  \\
\hline
NumberOfBatchesForTraining & 100 \\
NBatchChangeProportion & 100 \\
NumberOfBatchesForEval & 10000 \\
NBatchSize & 1000 \\
LambdaConstrant & 0.01 \\
Contraint & 10.0 \\
LearningRate & 0.1 \\
\hline
\end{tabular}
\caption{Bachelier Asymmetric Call \& Put Options Run Parameters}
\label{tab:bachelier_asym_call_&_put_options_run_params}
}
\parbox[t]{.45\linewidth}{
\centering
\begin{tabular}{ |c|c|}
\hline
Parameter & Value  \\
\hline
NumberOfBatchesForTraining & 100 \\
NBatchChangeProportion & 100 \\
NumberOfBatchesForEval & 10000 \\
NBatchSize & 1000 \\
LambdaConstrant & 0.1 \\
Contraint & 10.0 \\
LearningRate & 0.03 \\
\hline
\end{tabular}
\caption{Bachelier Symmetric Call \& Put Options Run Parameters}
\label{tab:bachelier_sym_call_&_put_options_run_params}
}
\hfill
\parbox[t]{.45\linewidth}{
\centering
\begin{tabular}{ |c|c|}
\hline
Parameter & Value  \\
\hline
NumberOfBatchesForTraining & 100 \\
NBatchChangeProportion & 100 \\
NumberOfBatchesForEval & 10000 \\
NBatchSize & 1000 \\
LambdaConstrant & 0.01 \\
Contraint & 10.0 \\
LearningRate & 0.3 \\
\hline
\end{tabular}
\caption{Bachelier Multi Coupons AutoCall Run Parameters}
\label{tab:bachelier_multi_coupons_autocall_run_params}
}
\parbox[t]{.45\linewidth}{
\centering
\begin{tabular}{ |c|c|}
\hline
Parameter & Value  \\
\hline
NumberOfBatchesForTraining & 100 \\
NBatchChangeProportion & 100 \\
NumberOfBatchesForEval & 10000 \\
NBatchSize & 1000 \\
LambdaConstrant & 0.01 \\
Contraint & 10.0 \\
LearningRate & 0.03 \\
\hline
\end{tabular}
\caption{Bachelier Single Coupon AutoCall Run Parameters}
\label{tab:bachelier_single_coupon_autocall_run_params}
}
\end{table}

\begin{table}
\parbox[t]{.45\linewidth}{
\centering
\begin{tabular}{ |c|c|}
\hline
Parameter & Value  \\
\hline
NumberOfBatchesForTraining & 100 \\
NBatchChangeProportion & 100 \\
NumberOfBatchesForEval & 10000 \\
NBatchSize & 1000 \\
LambdaConstrant & 0.001 \\
Contraint & 10.0 \\
LearningRate & 0.3 \\
\hline
\end{tabular}
\caption{LV Call Option Run Parameters}
\label{tab:lv_call_option_run_params}
}
\parbox[t]{.45\linewidth}{
\centering
\begin{tabular}{ |c|c|}
\hline
Parameter & Value  \\
\hline
NumberOfBatchesForTraining & 100 \\
NBatchChangeProportion & 100 \\
NumberOfBatchesForEval & 10000 \\
NBatchSize & 1000 \\
LambdaConstrant & 0.01 \\
Contraint & 10.0 \\
LearningRate & 0.01 \\
\hline
\end{tabular}
\caption{LV Asymmetric Call \& Put Options Run Parameters}
\label{tab:lv_asym_call_&_put_options_run_params}
}
\parbox[t]{.45\linewidth}{
\centering
\begin{tabular}{ |c|c|}
\hline
Parameter & Value  \\
\hline
NumberOfBatchesForTraining & 100 \\
NBatchChangeProportion & 100 \\
NumberOfBatchesForEval & 10000 \\
NBatchSize & 1000 \\
LambdaConstrant & 0.1 \\
Contraint & 10.0 \\
LearningRate & 0.001 \\
\hline
\end{tabular}
\caption{LV Symmetric Call \& Put Options Run Parameters}
\label{tab:lv_sym_call_&_put_options_run_params}
}
\hfill
\parbox[t]{.45\linewidth}{
\centering
\begin{tabular}{ |c|c|}
\hline
Parameter & Value  \\
\hline
NumberOfBatchesForTraining & 100 \\
NBatchChangeProportion & 100 \\
NumberOfBatchesForEval & 10000 \\
NBatchSize & 1000 \\
LambdaConstrant & 0.01 \\
Contraint & 10.0 \\
LearningRate & 0.1 \\
\hline
\end{tabular}
\caption{LV Multi Coupons AutoCall Run Parameters}
\label{tab:lv_multi_coupons_autocall_run_params}
}
\parbox[t]{.45\linewidth}{
\centering
\begin{tabular}{ |c|c|}
\hline
Parameter & Value  \\
\hline
NumberOfBatchesForTraining & 100 \\
NBatchChangeProportion & 100 \\
NumberOfBatchesForEval & 10000 \\
NBatchSize & 1000 \\
LambdaConstrant & 0.01 \\
Contraint & 10.0 \\
LearningRate & 0.03 \\
\hline
\end{tabular}
\caption{LV Single Coupon AutoCall Run Parameters}
\label{tab:lv_single_coupon_autocall_run_params}
}
\end{table}

\begin{table}
\begin{tabular}{|c|c|c|c|c|c|c|c|c|c|c|}
\hline
$x_0$ & 
1 & 1.025 & 1.05 & 1.075 & 1.1 &
1.125 & 1.15 & 1.175 & 1.2 & 1.225\\
\hline
LearningRate & 
10 & 10 & 10 & 10 & 10 &
3 & 3 & 3 & 3 & 3
\\
\hline
$x_0$ & 
1.25 & 1.275 & 1.3 & 1.325 & 1.35 &
1.375 & 1.4 & 1.425 & 1.45 & 1.475\\
\hline
LearningRate & 
1 & 1 & 1 & 1 & 1 &
1 & 1 & 1 & 1 & 1
\\
\hline
$x_0$ & 
1.5\\
\cline{1-2}
LearningRate & 
1
\\
\cline{1-2}
\end{tabular}
\caption{Bachelier Call Learning Rates for Graph}
\label{tab:bachelier_call_learning_rates}
\end{table}

\begin{table}
\begin{tabular}{|c|c|c|c|c|c|c|c|c|c|c|}
\hline
$x_0$ & 
0.4 & 0.45 & 0.5 & 0.55 & 0.6 &
0.65 & 0.7 & 0.75 & 0.8 & 0.85\\
\hline
LearningRate & 
0.003 & 0.001 & 0.001 & 0.001 & 0.003 &
0.003 & 0.003 & 0.003 & 0.003 & 0.01
\\
\hline
$x_0$ & 
0.9 & 0.95 & 1 & 1.05 & 1.1 &
1.15 & 1.2 & 1.25 & 1.3 & 1.35\\
\hline
LearningRate & 
0.01 & 0.03 & 0.1 & 0.1 & 0.1 &
0.1 & 0.03 & 0.03 & 0.03 & 0.03
\\
\hline
$x_0$ & 
1.4\\
\cline{1-2}
LearningRate & 
0.03
\\
\cline{1-2}
\end{tabular}
\caption{Bachelier Asymmetric Calls and Puts Learning Rates for Graph}
\label{tab:bachelier_calls_and_puts_asym_learning_rates}
\end{table}

\begin{table}
\begin{tabular}{|c|c|c|c|c|c|c|c|c|c|c|}
\hline
$x_0$ & 
0.5 & 0.55 & 0.6 & 0.65 & 0.7 &
0.75 & 0.8 & 0.85 & 0.9 & 0.95\\
\hline
LearningRate & 
0.003 & 0.003 & 0.003 & 0.01 & 0.01 &
0.01 & 0.01 & 0.01 & 0.03 & 0.03
\\
\hline
$x_0$ & 
1 & 1.05 & 1.1 & 1.15 & 1.2 &
1.25 & 1.3 & 1.35 & 1.4 & 1.45\\
\hline
LearningRate & 
0.03 & 0.03 & 0.03 & 0.01 & 0.01 &
0.01 & 0.003 & 0.003 & 0.003 & 0.003
\\
\hline
$x_0$ & 
1.5\\
\cline{1-2}
LearningRate & 
0.003
\\
\cline{1-2}
\end{tabular}
\caption{Bachelier Symmetric Calls and Puts Learning Rates for Graph}
\label{tab:bachelier_calls_and_puts_sym_learning_rates}
\end{table}

\begin{table}
\begin{tabular}{|c|c|c|c|c|c|c|c|c|c|c|}
\hline
$x_0$ & 
0.1 & 0.15 & 0.2 & 0.25 & 0.3 &
0.35 & 0.4 & 0.45 & 0.5 & 0.55\\
\hline
LearningRate & 
0.3 & 0.3 & 0.3 & 0.3 & 0.3 &
0.1 & 0.1 & 0.03 & 0.03 & 0.03
\\
\hline
$x_0$ & 
0.6 & 0.65 & 0.7 & 0.75 & 0.8 &
0.85 & 0.9 & 0.95 & 1 & 1.05\\
\hline
LearningRate & 
0.03 & 0.01 & 0.03 & 0.03 & 0.03 &
0.03 & 0.03 & 0.03 & 0.03 & 0.03
\\
\hline
$x_0$ & 
1.1 & 1.15 & 1.2 & 1.25 & 1.3 & 
1.35 & 1.4 & 1.45 & 1.5 & 1.55 \\
\hline
LearningRate & 
0.03 & 0.03 & 0.03 & 0.03 & 0.03 &
0.03 & 0.03 & 0.03 & 0.03 & 0.03 \\
\hline
$x_0$ & 
1.6\\
\cline{1-2}
LearningRate & 
0.03
\\
\cline{1-2}
\end{tabular}
\caption{Bachelier Multi Coupons AutoCall Learning Rates for Graph}
\label{tab:bachelier_autocall_multi_learning_rates}
\end{table}

\begin{table}
\begin{tabular}{|c|c|c|c|c|c|c|c|c|c|c|}
\hline
$x_0$ & 
0.1 & 0.15 & 0.2 & 0.25 & 0.3 &
0.35 & 0.4 & 0.45 & 0.5 & 0.55\\
\hline
LearningRate & 
1 & 1 & 0.3 & 0.3 & 0.3 &
0.1 & 0.1 & 0.1 & 0.1 & 0.1
\\
\hline
$x_0$ & 
0.6 & 0.65 & 0.7 & 0.75 & 0.8 &
0.85 & 0.9 & 0.95 & 1 & 1.05\\
\hline
LearningRate & 
0.03 & 0.03 & 0.03 & 0.03 & 0.03 &
0.03 & 0.03 & 0.03 & 0.03 & 0.03
\\
\hline
$x_0$ & 
1.1 & 1.15 & 1.2 & 1.25 & 1.3 & 
1.35 & 1.4 & 1.45 & 1.5 & 1.55 \\
\hline
LearningRate & 
0.03 & 0.01 & 0.003 & 0.003 & 0.003 &
0.01 & 0.01 & 0.01 & 0.01 & 0.01 \\
\hline
$x_0$ & 
1.6\\
\cline{1-2}
LearningRate & 
0.01
\\
\cline{1-2}
\end{tabular}
\caption{Bachelier Single Coupon AutoCall Learning Rates for Graph}
\label{tab:bachelier_autocall_single_learning_rates}
\end{table}

\begin{table}
\begin{tabular}{|c|c|c|c|c|c|c|c|c|c|c|}
\hline
$x_0$ & 
0.5 & 0.55 & 0.6 & 0.65 & 0.7 &
0.75 & 0.8 & 0.85 & 0.9 & 0.95\\
\hline
LearningRate & 
10 & 10 & 10 & 10 & 10 &
3 & 3 & 3 & 3 & 3
\\
\hline
$x_0$ & 
1 & 1.05 & 1.1 & 1.15 & 1.2 &
1.25 & 1.3 & 1.35 & 1.4 & 1.45\\
\hline
LearningRate & 
1 & 1 & 0.3 & 0.3 & 0.3 &
0.3 & 0.3 & 0.3 & 0.3 & 0.3
\\
\hline
$x_0$ & 
1.5\\
\cline{1-2}
LearningRate & 
0.3
\\
\cline{1-2}
\end{tabular}
\caption{LV Call Learning Rates for Graph}
\label{tab:lv_call_learning_rates}
\end{table}

\begin{table}
\begin{tabular}{|c|c|c|c|c|c|c|c|c|c|c|}
\hline
$x_0$ & 
0.4 & 0.45 & 0.5 & 0.55 & 0.6 &
0.65 & 0.7 & 0.75 & 0.8 & 0.85\\
\hline
LearningRate & 
0.001 & 0.001 & 0.001 & 0.001 & 0.001 &
0.001 & 0.001 & 0.001 & 0.001 & 0.001
\\
\hline
$x_0$ & 
0.9 & 0.95 & 1 & 1.05 & 1.1 &
1.15 & 1.2 & 1.25 & 1.3 & 1.35\\
\hline
LearningRate & 
0.001 & 0.001 & 0.001 & 0.001 & 0.001 &
0.001 & 0.001 & 0.001 & 0.001 & 0.001
\\
\hline
$x_0$ & 
1.4\\
\cline{1-2}
LearningRate & 
0.001
\\
\cline{1-2}
\end{tabular}
\caption{LV Asymmetric Calls and Puts Learning Rates for Graph}
\label{tab:lv_calls_and_puts_asym_learning_rates}
\end{table}

\begin{table}
\begin{tabular}{|c|c|c|c|c|c|c|c|c|c|c|}
\hline
$x_0$ & 
0.5 & 0.55 & 0.6 & 0.65 & 0.7 &
0.75 & 0.8 & 0.85 & 0.9 & 0.95\\
\hline
LearningRate & 
0.0003 & 0.0003 & 0.0003 & 0.0003 & 0.0003 &
0.0003 & 0.0003 & 0.0003 & 0.0003 & 0.0003
\\
\hline
$x_0$ & 
1 & 1.05 & 1.1 & 1.15 & 1.2 &
1.25 & 1.3 & 1.35 & 1.4 & 1.45\\
\hline
LearningRate & 
0.0003 & 0.0003 & 0.0003 & 0.0003 & 0.0003 &
0.0003 & 0.0003 & 0.0003 & 0.0003 & 0.0003
\\
\hline
$x_0$ & 
1.5\\
\cline{1-2}
LearningRate & 
0.0003
\\
\cline{1-2}
\end{tabular}
\caption{LV Symmetric Calls and Puts Learning Rates for Graph}
\label{tab:lv_calls_and_puts_sym_learning_rates}
\end{table}

\begin{table}
\begin{tabular}{|c|c|c|c|c|c|c|c|c|c|c|}
\hline
$x_0$ & 
0.1 & 0.15 & 0.2 & 0.25 & 0.3 &
0.35 & 0.4 & 0.45 & 0.5 & 0.55\\
\hline
LearningRate & 
0.03 & 0.03 & 0.03 & 0.03 & 0.03 &
0.03 & 0.03 & 0.03 & 0.01 & 0.01
\\
\hline
$x_0$ & 
0.6 & 0.65 & 0.7 & 0.75 & 0.8 &
0.85 & 0.9 & 0.95 & 1 & 1.05\\
\hline
LearningRate & 
0.01 & 0.01 & 0.003 & 0.003 & 0.003 &
0.01 & 0.03 & 0.03 & 0.03 & 0.03
\\
\hline
$x_0$ & 
1.1 & 1.15 & 1.2 & 1.25 & 1.3 & 
1.35 & 1.4 & 1.45 & 1.5 & 1.55 \\
\hline
LearningRate & 
0.01 & 0.01 & 0.01 & 0.01 & 0.01 &
0.01 & 0.01 & 0.01 & 0.01 & 0.01 \\
\hline
$x_0$ & 
1.6\\
\cline{1-2}
LearningRate & 
0.01
\\
\cline{1-2}
\end{tabular}
\caption{LV Multi Coupons AutoCall Learning Rates for Graph}
\label{tab:lv_autocall_multi_learning_rates}
\end{table}

\begin{table}
\begin{tabular}{|c|c|c|c|c|c|c|c|c|c|c|}
\hline
$x_0$ & 
0.1 & 0.15 & 0.2 & 0.25 & 0.3 &
0.35 & 0.4 & 0.45 & 0.5 & 0.55\\
\hline
LearningRate & 
0.1 & 0.1 & 0.03 & 0.03 & 0.03 &
0.01 & 0.01 & 0.01 & 0.01 & 0.01
\\
\hline
$x_0$ & 
0.6 & 0.65 & 0.7 & 0.75 & 0.8 &
0.85 & 0.9 & 0.95 & 1 & 1.05\\
\hline
LearningRate & 
0.01 & 0.01 & 0.01 & 0.003 & 0.003 &
0.003 & 0.01 & 0.01 & 0.01 & 0.01
\\
\hline
$x_0$ & 
1.1 & 1.15 & 1.2 & 1.25 & 1.3 & 
1.35 & 1.4 & 1.45 & 1.5 & 1.55 \\
\hline
LearningRate & 
0.01 & 0.003 & 0.003 & 0.003 & 0.003 &
0.01 & 0.01 & 0.01 & 0.01 & 0.003 \\
\hline
$x_0$ & 
1.6\\
\cline{1-2}
LearningRate & 
0.003
\\
\cline{1-2}
\end{tabular}
\caption{LV Single Coupon AutoCall Learning Rates for Graph}
\label{tab:lv_autocall_single_learning_rates}
\end{table}

\begin{table}
\center
\begin{tabular}{|c|c|}
\hline
Diffusion and Payoff & BaseForAutomaticLambdaConstraint \\
\hline
Bachelier Call & 1 \\
\hline
Bachelier Asymmetric Calls and Puts & 0.3 \\
\hline
Bachelier Symmetric Calls and Puts & 0.3 \\
\hline
Bachelier Multi Coupons AutoCall & 0.3 \\
\hline
Bachelier Single Coupon Autocall & 0.3 \\
\hline
LV Call & 1 \\
\hline
LV Asymmetric Calls and Puts & 0.3 \\
\hline
LV Symmetric Calls and Puts & 0.3 \\
\hline
LV Multi Coupons Autocall & 0.3 \\
\hline
LV Single Coupon Autocall & 0.3 \\
\hline
\end{tabular}
\caption{Values of BaseForAutomaticLambdaConstraint}
\label{tab:baseforautomaticlambdaconstraint}
\end{table}

\section{Volatility Surfaces}

\begin{figure}[H]
  \begin{minipage}[t]{0.48\linewidth}
    \includegraphics[width=\linewidth]{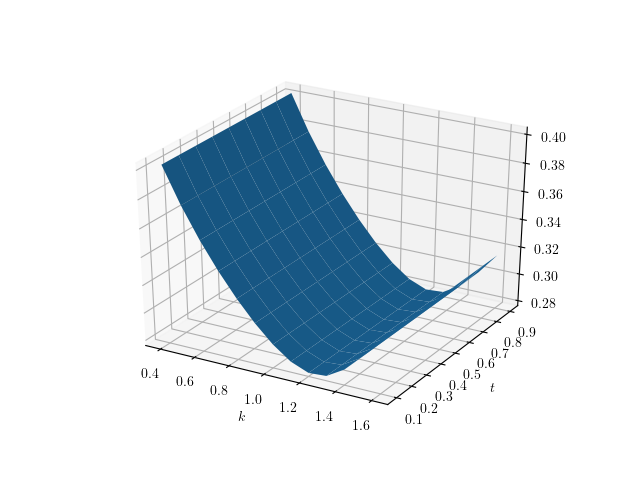}
    \caption{Implied Volatility Surface for Local Volatility Diffusion}
    \label{fig:imp_vol}
  \end{minipage}
  \begin{minipage}[t]{0.04\linewidth}
  \hfill
  \end{minipage}
  \begin{minipage}[t]{0.48\linewidth}
    \includegraphics[width=\linewidth]{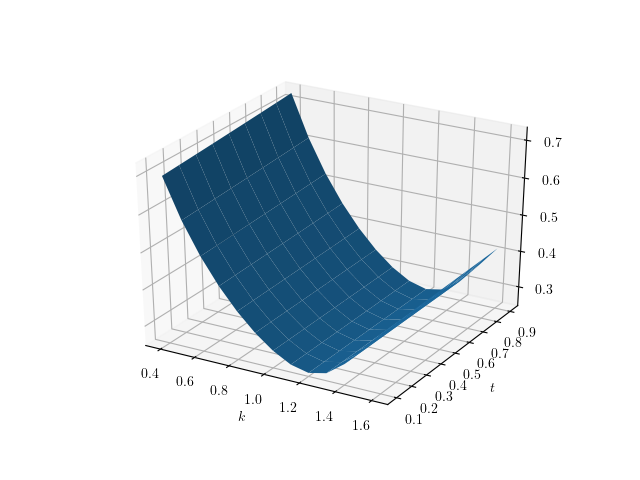}
    \caption{Local Volatility Surface for Local Volatility Diffusion}
    \label{fig:local_vol}
  \end{minipage}
\end{figure}

\end{document}